\documentclass[aps,prd,reprint,showpacs,nofootinbib,nobibnotes,amsmath,amssymb]{revtex4-1}
\usepackage{graphicx}
\usepackage{color}
\usepackage{longtable}
\allowdisplaybreaks[2]
\newcommand{\be}{\begin{equation}}\newcommand{\ee}{\end{equation}}
\newcommand{\bea}{\begin{eqnarray}}\newcommand{\eea}{\end{eqnarray}}

\newcommand{\F}{\phantom {1}}
\newcommand{\nn}{\nonumber}
\newcommand{\qqb}{{q\bar{q}}}

\newcommand{\figwidth}{8.3cm}

\begin{document}

\title{Precise determination of the three-quark potential in SU(3) lattice gauge theory}{}

\author{Yoshiaki Koma}
\email{koma@numazu-ct.ac.jp}
\affiliation{National Institute of Technology, Numazu College, Ooka 3600, Numazu 410-8501, Japan}

\author{Miho Koma}
\email{koma.miho@nihon-u.ac.jp}
\affiliation{Nihon University, College of International Relations, Mishima 411-8555, Japan}

\date{March 18, 2017}

\begin{abstract}
We investigate the static interquark potential for the three-quark system 
in SU(3) lattice gauge theory at zero temperature by using Monte Carlo simulations.
We extract the potential from the correlation function of the three Polyakov loops,
which are computed by employing the multilevel algorithm.
We obtain remarkably clean results of the three-quark potential for 
$O(200)$ sets of the three-quark geometries 
including  not only the cases  that  three quarks are put at the 
vertices of  acute,  right, and  obtuse triangles,
but also the extreme cases such that  three quarks are put in line.
We find several new interesting features of the three-quark potential
and then discuss its possible functional form.
\end{abstract}


\maketitle

\section{Introduction}
\label{sec:intro}

\par 
The static interquark potential for a three-quark system, 
the three-quark potential, is  one of the characteristic quantities
in quantum chromodynamics (QCD), which is relevant to 
spectroscopy of hadrons, especially, of baryons.
Therefore it is quite important to 
determine the functional form of the potential from the first principle,
clarify the properties of the three-quark system.
Since a direct interaction among three quarks, a three-body force,
is expected in QCD by virtue of SU(3) gauge symmetry,
it is also interesting to investigate how the three-quark potential 
is different from or the same as  the two-body quark-antiquark potential.

\par
The investigation of the static interquark potential generally requires a
nonperturbative method as the quarks  are strongly interacting  with each other
inside hadrons, and  Monte Carlo simulations of lattice QCD offer a powerful tool for this purpose.
The first lattice study of the three-quark potential goes 
back to the mid-1980s by Sommer and Wosiek~\cite{Sommer:1984xq,Sommer:1985da},
and by Thacker~{\it et\,al.}~\cite{Thacker:1987aq}.
The study was revisited around 2000 by several groups with improved 
numerical techniques and computer 
resources~\cite{Bali:2000gf,Takahashi:2000te,Takahashi:2002bw,Alexandrou:2001ip}.

\par
These latest results were, however, found to be inconsistent with each other.
Bali~\cite{Bali:2000gf} and  Alexandrou~{\it et\,al.}~\cite{Alexandrou:2001ip} 
claimed that the potential was described by the half of the 
sum of two-body potentials in the quark-antiquark system,
which is called the $\Delta$ area law, up to the interquark distance 
nearly 1\,fm.\footnote{Alexandrou~{\it et\,al.} updated their result with technical 
refinements and arrived at slightly different conclusion~\cite{Alexandrou:2002sn}.}
At the distances where the perturbation theory cannot be applied,
the $\Delta$ area law may suggest the formation of a $\Delta$-shaped 
color flux tube among the three quarks.
On the other hand, Takahashi~{\it  et\,al.}~\cite{Takahashi:2000te,Takahashi:2002bw},
claimed that the potential was described by
the sum of the two-body Coulombic terms and the three-body linear 
term, where the latter is proportional to the $Y$ distance 
with a junction at the Fermat-Torricelli point of a triangle spanned by the three quarks.
This may then be called the $Y$ area law, suggesting
 the formation of a $Y$-shaped color flux tube among the three quarks at long distances.
The confining feature of the $Y$ area law
can be  explained partly if the QCD vacuum possesses the property of 
dual superconductor~\cite{Kamizawa:1992np,Koma:2000hw},
which are explicitly demonstrated by lattice QCD simulations in the maximally 
Abelian gauge~\cite{Ichie:2002mi,Bornyakov:2004uv,Bornyakov:2004yg}.
Bissey~{\it et\,al.}~\cite{Bissey:2006bz}  investigated the profile of 
the non-Abelian action density in the three-quark system and found no $\Delta$-shaped 
flux-tube structure at long distance, but the structure was not always of $Y$ shape.
Several ideas were proposed to reconcile this situation based on 
the effective models~\cite{Caselle:2005sf,Dmitrasinovic:2009ma,Andreev:2015iaa,Andreev:2015riv}.
Perturbation theory may provide a guideline for solving the discrepancy~\cite{Brambilla:2009cd,Brambilla:2013vx},
if higher order contributions are property evaluated one after another.

\par
\begin{figure}[!t]
\includegraphics[width=7cm]{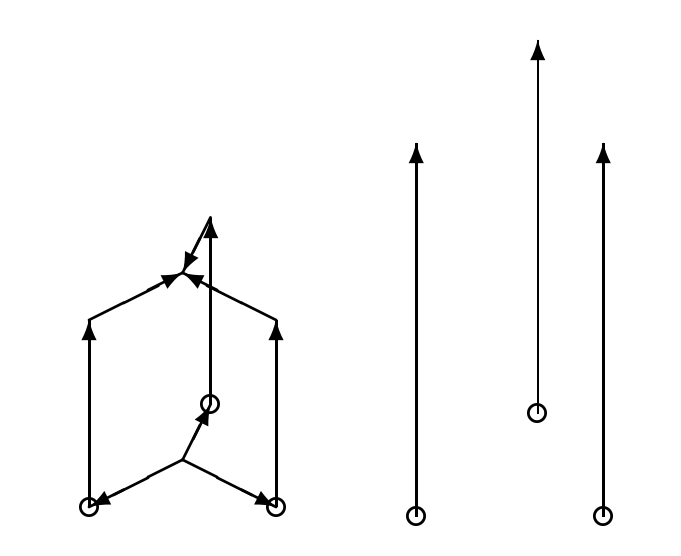}
\caption{The three-quark Wilson loop (left) and the  three-quark  PLCF (right).
While the Wilson loop is composed of the three temporal Wilson lines 
connected by the spatial Wilson lines with a junction,  
the PLCF is only composed of the three Polyakov loops, where a periodic boundary 
condition is imposed in the time direction.}
\label{fig:wilson_vs_PLCF}
\end{figure}

In this paper, we thus revisit the determination of the three-quark potential
in SU(3) lattice gauge theory at zero temperature.
All of the previous lattice results were obtained by using the Wilson loop as the three-quark source,
which is composed of the three temporal Wilson lines connected by the spatial
Wilson lines with a junction as illustrated  in Fig.~\ref{fig:wilson_vs_PLCF}~(left).
The potential and the flux-tube profile should not depend on the 
path of the spatial Wilson lines and the location of the junction, 
but some of the earlier results seem to be affected by them
especially when the temporal extent of the Wilson loop is not large enough.
Clearly, it is desirable to have the precise lattice data with less statistical and systematic errors
before evaluating the validity of the functional form.

\par
Our strategy is then to use the Polyakov loop correlation function~(PLCF) 
in the fundamental representation as the quark source.
In contrast to the Wilson loop, the PLCF is free from systematic effects caused by the 
spatial Wilson lines,
since the PLCF is composed only of the three Polyakov loops as illustrated 
in Fig.~\ref{fig:wilson_vs_PLCF}~(right),
where a periodic boundary condition is imposed in the time direction.
A severe problem,  which is why the PLCF has not been used so far 
for the zero temperature simulations, 
may be the smallness of the expectation value 
in contrast to the finite temperature case~\cite{Kaczmarek:1999mm,Cardoso:2011hh}, 
which means that  ordinary simulations are ineffective
as the signal is easily obscured by the statistical noise.
As we demonstrate in this paper, however,
this problem can be solved by employing 
the multilevel algorithm~\cite{Luscher:2001up,Luscher:2002qv}
with tuned simulation parameters~\cite{Koma:2014ffa}.
Another reason that  the PLCF has been  avoided may 
originate from a folklore that the potential from the PLCF can contain contributions
not only from the color-singlet state but also from
the color-adjoint state.
However, as we have demonstrated in SU(3) lattice gauge theory~\cite{Koma:2014ota},
all intermediate states of gluons equally contribute to the color-singlet potential,
at least as long as one uses the PLCF in the 
fundamental representation
 (see, an illuminating discussion in Ref.~\cite{Jahn:2004qr} for SU(2) lattice gauge theory).

\par
We obtain the  three-quark potential of 
$O(200)$ sets of the three-quark geometries
including not only the cases that three quarks are put at the 
vertices of acute, right, and  obtuse triangles,
but also the extreme cases such that  three quarks are put in line.
We find that most of the three-quark potentials from  triangle geometries
that the maximum inner angle is smaller than 120$^{\circ}$ can fall into one curve
as a function of the minimal length of lines connecting the three quarks,
which supports the $Y$-shaped flux-tube picture.
From the derivative of the potential, 
we observe that the string tension of the three-quark potential 
is the same as that of the quark-antiquark potential.
We also critically compare the three-quark potential 
to the half of the sum of the two-body quark-antiquark potential
and find a systematic deviation especially for larger triangle geometries, 
which brings us to a conclusion that
there is certainly a force  which cannot be
described by the superposition of the two-body forces.
We then discuss the functional form of the three-quark potential
and examine its scaling behavior with respect to the lattice spacing.

\par
This paper is organized as follows.
In Sec.~\ref{sec:procedure}, we describe 
how to compute the three-quark potential from the PLCF  with the multilevel algorithm.
We also classify various three-quark geometries.
In Sec.~\ref{sec:result}, we present our numerical results.
Section~\ref{sec:summary} is devoted to the summary of our findings.
Our preliminary results have been presented at Lattice 2013 in Mainz~\cite{Koma:2014ota}, at
Lattice 2014 in New York~\cite{Koma:2014ffa}, and
at Lattice 2015 in Kobe~\cite{Koma:2015brg}.

\section{Numerical procedures}
\label{sec:procedure}

In this section,  we describe  how to extract the three-quark potential from the PLCF
and how to implement the multilevel algorithm for computing the PLCF.
We then provide the definition of some practical distances and angles 
to classify various three-quark geometries
analyzed in the present study.

\subsection{The three-quark  potential from the PLCF}

\par
We perform 
simulations of SU(3) lattice gauge theory (lattice QCD  within the quenched approximation)
in four dimensions 
with the lattice volume $L^{3}\times T$ 
and the lattice spacing $a$ by 
imposing periodic boundary conditions in all space-time directions. 
The three-quark potential is extracted from the PLCF as follows.

\par
\begin{figure}[!t]
\includegraphics[width=8.8cm]{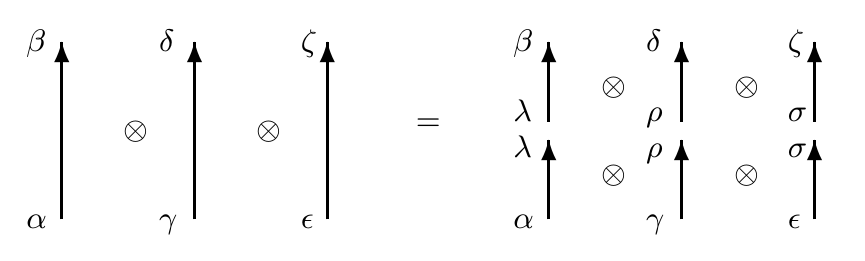}
\caption{A product of  two three-link correlators, 
where repeated Greek indices $\lambda$, $\rho$, 
and $\sigma$ are to be summed over from  one to three.}
\label{fig:fig2}
\end{figure}

We first define  a three-link correlator as
\bea
&&
\mathbb{T}(x_{0},\vec{x}_{1},\vec{x}_{2},\vec{x}_{3})_{\alpha \beta \gamma \delta\epsilon\zeta}\nn\\
&\equiv&  
U_{4}(x_{0},\vec{x}_{1})_{\alpha \beta} U_{4}(x_{0},\vec{x}_{2})_{\gamma \delta}
U_{4}(x_{0},\vec{x}_{3})_{\epsilon\zeta}\;,
\label{eqn:twolink}
\eea
which is a direct product  of 
three time-like link variables $U_{4}(x)$
placed at a time $x_{0}$ with  
spatial positions of three quarks,
$\vec{x}_{1}$, $\vec{x}_{2}$, and $\vec{x}_{3}$.
Greek indices $\alpha$, $\beta$, $\gamma$, $\delta$, $\epsilon$, $\zeta$ take the 
values from one to three, respectively.
Practically, a three-link correlator is a complex matrix with 
$3^{6}=729$ components.
The three-link correlator acts on a color state in the ${\bf 3}  \otimes {\bf 3}\otimes {\bf 3}$
representation of the SU(3) group $| n_{\alpha\beta\gamma}; \vec{x}_{1},\vec{x}_{2},\vec{x}_{3}\rangle$, which is
an eigenstate of the hamiltonian $\mathbb{H}$ defined by the transfer matrix  in the  temporal gauge,
$\mathbb{T} \equiv e^{-\mathbb{H}a}$, and then satisfies
\bea
&&
\mathbb{T}(x_{0},\vec{x}_{1},\vec{x}_{2},\vec{x}_{3})_{\alpha \lambda \beta \rho \gamma \sigma}
 | n_{\alpha\beta\gamma} ; \vec{x}_{1},\vec{x}_{2},\vec{x}_{3}\rangle \nn\\
&=&
 e^{-E_{n}(\vec{x}_{1},\vec{x}_{2},\vec{x}_{3})a}| n_{\lambda\rho\sigma}; \vec{x}_{1},\vec{x}_{2},\vec{x}_{3}\rangle \;,
\eea
where $n$ is the principal quantum number and 
repeated Greek indices $\alpha$, $\beta$, $\gamma$ are to be summed over from  one to three.
The energies $E_{n}(\vec{x}_{1},\vec{x}_{2},\vec{x}_{3})$ are positive 
and are  common to all of the $3^{3}=27$ color components of 
$| n_{\alpha\beta\gamma}; \vec{x}_{1},\vec{x}_{2},\vec{x}_{3}\rangle $.
The multiplication rule of two three-link correlators 
for  adjacent times at  $x_{0}$ and $x_{0}+a$ is
\bea
\!\!\!\!\!\!&&\!\!
\{  \mathbb{T}(x_{0},\vec{x}_{1},\vec{x}_{2},\vec{x}_{3})
\mathbb{T}(x_{0}+a,\vec{x}_{1},\vec{x}_{2},\vec{x}_{3}) \}_{\alpha \beta \gamma \delta\epsilon\zeta}\nn\\
\!\!\!\!\!\!&=&\!\!
\mathbb{T}(x_{0},\vec{x}_{1},\vec{x}_{2},\vec{x}_{3})_{\alpha \lambda \gamma \rho \epsilon \sigma}
\mathbb{T}(x_{0}+a,\vec{x}_{1},\vec{x}_{2},\vec{x}_{3})_{\lambda \beta \rho  \delta \sigma \zeta}\, .
\eea
With this multiplication rule a new and longer three-link correlator is created
as schematically shown in Fig.~\ref{fig:fig2}.

\par
We then construct the PLCF from the time-ordered product of the three-link correlators,
\bea
&&{\rm Tr}P(\vec{x}_{1}){\rm Tr}P(\vec{x}_{2}){\rm Tr}P(\vec{x}_{3})\nn\\*
&=&\{ \mathbb{T}(0,\vec{x}_{1},\vec{x}_{2},\vec{x}_{3})\mathbb{T}(a,\vec{x}_{1},\vec{x}_{2},\vec{x}_{3})
\nn\\*
&&
\quad \cdots \mathbb{T}(T-a,\vec{x}_{1},\vec{x}_{2},\vec{x}_{3})\}_{\alpha \alpha \gamma \gamma \epsilon\epsilon} \;.
\eea
By  inserting the complete set of eigenstates,
\be
{\bf 1}_{\alpha  \lambda  \beta \rho  \gamma\sigma} 
=\sum_{n} |n_{\alpha\beta\gamma}; \vec{x}_{1},\vec{x}_{2},\vec{x}_{3}\rangle
\langle 
n_{\lambda\rho\sigma};\vec{x}_{1},\vec{x}_{2},\vec{x}_{3}| \;,
\ee
at each time $x_{0}=0,a,...,T-a$, and by using the normalization condition,
\be
\langle n_{\alpha\beta\gamma};\vec{x}_{1},\vec{x}_{2},\vec{x}_{3}|
m_{\alpha\beta\gamma} ; \vec{x}_{1},\vec{x}_{2},\vec{x}_{3}\rangle = \delta_{nm} \;,
\ee
the expectation value of the PLCF is reduced to
\be
\langle {\rm Tr}P(\vec{x}_{1}){\rm Tr}P(\vec{x}_{2}) {\rm Tr}P(\vec{x}_{3})\rangle
  =  \sum_{n=0}^{\infty}  e^{-E_{n}(\vec{x}_{1},\vec{x}_{2},\vec{x}_{3})T} \;.
\label{eqn:plcf}
\ee
The ground state potential, $V_{3q}\equiv E_{0}$, is then extracted as
\bea
V_{3q}(\vec{x}_{1},\vec{x}_{2},\vec{x}_{3}) & =&
 -\frac{1}{T} \ln \langle  {\rm Tr}P(\vec{x}_{1}){\rm Tr}P(\vec{x}_{2}) {\rm Tr}P(\vec{x}_{3})\rangle \nn\\*
&& +  O(\frac{1}{T}e^{-(E_{1}-E_{0})T})\;,
\label{eqn:plcf-exract-E0}
\eea
where the terms of $O(e^{-(E_{1}-E_{0})T}/T)$ are  
always negligible at zero temperature.
Therefore, once the PLCF is computed accurately for a large temporal extent $T$,  
it is straightforward to extract the ground state potential.
For instance, if we refer to the value $a(E_{1}-E_{0})\sim 0.5$ at $\beta=6.00$ 
given by Takahashi and Suganuma~\cite{Takahashi:2004rw}, 
the order of magnitude of 
$O(e^{-(E_{1}-E_{0})T}/T)$ on a lattice with $T/a=24$, which is our numerical 
setting, is estimated as $O(e^{-0.5 \cdot 24}/24)= O(10^{-7})$, 
which is clearly negligible compared to $aV_{3q}$  at $\beta=6.00$.

\par
Note that if the sum of multiexponential functions in Eq.~\eqref{eqn:plcf} is forcibly
cast into a single exponential function, its exponent
may be understood as the minus of the free energy divided
by  corresponding temperature.
However, if the excited state contribution is negligible from the beginning
and the summation in Eq.~\eqref{eqn:plcf}
is represented only by the first term with $E_{0}$,
the energy we can extract following Eq.~\eqref{eqn:plcf-exract-E0}
is no longer the free energy but  just the ground state potential,
where the notion of temperature could be irrelevant.

\par
On the other hand,
if one uses the three-quark Wilson loop as in Fig.~\ref{fig:wilson_vs_PLCF} (left)
with a time extent $t$, 
the expectation value will be
\bea
&&\langle W(\vec{x}_{1},\vec{x}_{2},\vec{x}_{3},\{\vec{x}_{p}\},t) \rangle \nn\\*
 &=&  \sum_{n=0}^{\infty} w_{n}(\vec{x}_{1},\vec{x}_{2},\vec{x}_{3},\{\vec{x}_{p}\},t)
 e^{-E_{n}( \vec{x}_{1},\vec{x}_{2},\vec{x}_{3})\,t} \;,
\label{eqn:wilson}
\eea
and
the ground state potential is  then extracted as
\bea
V_{3q}(\vec{x}_{1},\vec{x}_{2},\vec{x}_{3})  
&=&
-\frac{1}{t}  \ln \langle W  (\vec{x}_{1},\vec{x}_{2},\vec{x}_{3},\{\vec{x}_{p}\},t)\rangle \nn\\*
&&
+ \frac{1}{t} \ln  w_{0}+ O (\frac{1}{t} e^{-(E_{1}-E_{0})t})\;.
\label{eqn:wilson-exract-E0}
\eea
The crucial difference from Eq.~\eqref{eqn:plcf-exract-E0}
is the presence of the nontrivial second term, since the weight factor $\{w_{n}\}$ 
is dependent not only  on the temporal extent of the Wilson loop $t$ but also 
on the path of the spatial Wilson lines $\{\vec{x}_{p}\}$.
One may adopt smearing techniques~\cite{Albanese:1987ds}
to the spatial links to achieve a better overlap with the ground state, $w_{0}  \to 1$,
and then the second term can be dismissed.
However, 
the achievement
is usually incomplete especially when the interquark distance becomes larger.
A possible way to overcome the incompleteness of  the smearing 
may be to combine this technique and  the variational method, 
although it requires a further careful look on the 
validity for the choice of the variational basis~\cite{Alexandrou:2002sn,Takahashi:2004rw}.
Moreover, the  terms  of  $O(e^{-(E_{1}-E_{0})t} / t)$
are not easily suppressed, since $t$ cannot be large practically.
There is also  limitation such as $t <T/2$ due to the periodic 
boundary condition in the time direction.
Thus, in order to identify the first term of the r.h.s.~of Eq.~\eqref{eqn:wilson-exract-E0} as
the ground state potential, these problems must be solved.
Otherwise the contamination from the excited states cannot be avoided
and the resulting potential may be overestimated, since 
the second term is usually negative.

\subsection{The multilevel algorithm for the PLCF}
\label{subsect:multilevel}

\par
A severe problem of the PLCF
is that the expectation values of the PLCF are extremely small at zero temperature,
so that they are  immediately obscured by the statistical noise.
By using the multilevel algorithm~\cite{Luscher:2001up,Luscher:2002qv}, however, 
it is possible to overcome the problem.
The idea  is to compute  a desired correlation function, which 
may have an extremely small expectation value, 
from the product of relatively large sub-lattice averages of its components 
(in our case it corresponds to the product of $\mathbb{T}$s within a sub-lattice), 
where the sub-lattices are defined by dividing the lattice volume into several layers along the time direction.
During the computation of the sub-lattice averages, 
the spatial links at the sub-lattice boundaries are kept intact.
The computation of the correlation function in this way 
 is regarded as the hierarchical functional integral method and
is supported by the  transfer matrix formalism of quantum field theory.

\par
In order to make efficient use of the multilevel algorithm,
it is important to choose  the following two parameters appropriately.
One is the number of time slices in a sub-lattice, $N_{\rm tsl} =T/(aN_{\rm sub})$,
where $N_{\rm sub}$ is the number of sub-lattices.
The other is the number of internal
updates for the sub-lattice averages,  $N_{\rm iupd}$.

\par
Let us explain the reason how and why  the multilevel algorithm works well
by looking at a simple case that the lattice volume is divided 
into two sub-lattices
at the time slice $x_{0}=0$ and $x_{0}=T/2$ ($N_{\rm sub}=2$ and $N_{\rm tsl} =T/(2a)$).
We may prepare the product of the three-link correlators
$ \{\mathbb{T}(0,\vec{x}_{1},  \vec{x}_{2},  \vec{x}_{3})\cdots 
\mathbb{T}(T/2-a,\vec{x}_{1},  \vec{x}_{2},  \vec{x}_{3})\}_{\alpha\beta\gamma \delta \epsilon\zeta}$ 
and $\{\mathbb{T}(T/2,\vec{x}_{1},  \vec{x}_{2},  \vec{x}_{3})\cdots
 \mathbb{T}(T-a,\vec{x}_{1},  \vec{x}_{2},  \vec{x}_{3})\}_{\alpha\beta\gamma \delta \epsilon\zeta}$, 
and construct the PLCF by
\bea
\!\!\!\!\!\!&&{\rm Tr}\,P(\vec{x}_{1}){\rm Tr}\,P\,(\vec{x}_{2}){\rm Tr}\,P\,(\vec{x}_{3})\nn\\
\!\!\!\!\!\!&=&
[  \mathbb{T}(0,\vec{x}_{1},  \vec{x}_{2},  \vec{x}_{3})\cdots 
\mathbb{T}(\frac{T}{2}\!-\!a,\vec{x}_{1},  \vec{x}_{2},  \vec{x}_{3}) ]_{\alpha \lambda \gamma \rho \epsilon\sigma}\nn\\
\!\!\!\!\!\!&&
\times  [ \mathbb{T}(\frac{T}{2},\vec{x}_{1},  \vec{x}_{2},  \vec{x}_{3})\cdots 
\mathbb{T}(T\!-\!a,\vec{x}_{1},  \vec{x}_{2},  \vec{x}_{3}) ]_{\lambda \alpha \rho  \gamma \sigma \epsilon}\,, 
\label{eqn:plcf-sublattice-ave}
\eea
where $[\cdots ]$ represent taking the sub-lattice averages (this is not yet an expectation value).
Fixing the spatial links at the sub-lattice boundaries may
correspond to inserting two fixed sources
$|\phi_{1} \rangle = \sum_{n} a_{n}|n_{\alpha\beta\gamma} ;\vec{x}_{1},  \vec{x}_{2},  \vec{x}_{3}\rangle $
and $|\phi_{2} \rangle = \sum_{n} b_{n}|n_{\alpha\beta\gamma};\vec{x}_{1},  \vec{x}_{2},  \vec{x}_{3}\rangle $
at $x_{0}=0$ and $x_{0}=T/2$,  respectively,
where $\{ a_{n} \}$ and $\{ b_{m} \}$ are 
unknown complex values but satisfy $|a_{n}|^{2}=|b_{n}|^{2}=1$ for arbitrary $n$.
Then, Eq.~\eqref{eqn:plcf-sublattice-ave} is evaluated,  omitting arguments of the 
spatial vectors  for simplicity,  as
\bea
&&
{\rm Tr}\,P(\vec{x}_{1}){\rm Tr}\,P\,(\vec{x}_{2}){\rm Tr}\,P\,(\vec{x}_{3})\nn\\*
 &=&
 \left [  |\phi_{1}\rangle \langle \phi_{1}  | \mathbb{T}(0)\cdots  \mathbb{T}(\frac{T}{2}- a)  
 \right  ]_{\alpha \lambda \gamma \rho \epsilon\sigma}\nn\\*
&&\times
\left   [  |\phi_{2} \rangle  \langle \phi_{2} |  \mathbb{T}(\frac{T}{2} ) \cdots \mathbb{T}(T- a)  
 \right ]_{\lambda \alpha \rho  \gamma \sigma \epsilon} \nn\\*
 &=&
 \sum_{n,m} a_{m}^{*}a_{n} 
 |n_{\alpha  \gamma  \epsilon}  \rangle
\langle m_{\lambda  \rho \sigma}| e^{-E_{m}\frac{T}{2}} 
\nn\\*
&&\times
  \sum_{n',m'} b_{m'}^{*}b_{n'}
   |n'_{\lambda  \rho   \sigma }\rangle \langle m'_{ \alpha   \gamma  \epsilon} |e^{-E_{m'}\frac{T}{2}}\nn\\*
 &=&
\sum_{n}^{}b_{n}^{*}a_{n}e^{-E_{n} \frac{T}{2}} \cdot 
\sum_{m}^{} a_{m}^{*}b_{m}e^{-E_{m} \frac{T}{2}}  \;.
\label{eqn:lowest}
\eea
If we take the average for a large number of  different fixed sources 
at $x_{0}=0$ and $x_{0}=T/2$ of other independent gauge configurations,
we obtain the expectation value of the PLCF
as in Eq.~\eqref{eqn:plcf}, since
inserting many fixed sources corresponds to inserting the complete set.

\par
It is worth noting that if $T/2$ is large enough such that the contribution from the 
terms of $O(e^{-(E_{1}-E_{0})(T/2)})$ is negligible, 
which is  the case at zero temperature limit, Eq.~\eqref{eqn:lowest} further reduces to 
\bea
{\rm Tr}P(\vec{x}_{1}){\rm Tr}P\,(\vec{x}_{2}){\rm Tr}P\,(\vec{x}_{3})
&=& 
|a_{0}|^{2} |b_{0}|^{2}  e^{-E_{0}(\vec{x}_{1},  \vec{x}_{2},  \vec{x}_{3}) T}\nn\\*
&=&
 e^{-E_{0}(\vec{x}_{1},  \vec{x}_{2},  \vec{x}_{3}) T}  \; ,
\label{eqn:plcf-selection}
\eea
where $|a_{0}|^{2}= |b_{0}|^{2} =1$.
This means that the ground state potential $E_{0}$ can be extracted 
from one gauge configuration.
Of course, one should be careful that
the weight of one gauge configuration in the multilevel algorithm
is  different from that in ordinary simulations.
It is important to notice that 
each color component of the intermediate 
states equally contributes to the exponential decay 
of the PLCF (the degeneracy is just $3^{6}=729$ in this example),
and there is no dominant contribution to the potential from a particular color component, 
which implies that contributions from the color-singlet and color-adjoint states are the same.
Eq.~\eqref{eqn:plcf-selection}  also indicates that
it is possible to obtain the same gauge invariant potential even from the gauge variant PLCF
constructed  by  selected partial intermediate states~\cite{Koma:2014ota}.
In this sense,  gauge invariance of the PLCF is 
desirable to maximize the number of internal color statistics (the degeneracy), 
which may help to obtain stable values with a smaller $N_{\rm iupd}$.
On the other hand, if the terms of $O(e^{-(E_{1}-E_{0})(T/2)})$ 
in Eq.~\eqref{eqn:lowest} are not small enough,
the ground state potential  cannot be extracted
since the PLCF always suffer from contamination of excited states.
In such a case, increasing the number of independent gauge
configurations (statistics) does not help.
Instead,  the larger temporal lattice volume is needed from the beginning.

\par
This example tells us that it is crucial to find an appropriate~$N_{\rm tsl}$ so that 
the terms of $O(e^{-(E_{1}-E_{0})(a N_{\rm tsl})})$ are  small enough.
In our experience with the standard SU(3) Wilson gauge action,
there is a critical minimal length of $aN_{\rm tsl}$ to obtain the ground state potential.
We find the value
$a N_{\rm tsl} = 0.36\sim 0.37\,{\rm [fm]}$~\cite{Koma:2006si,Koma:2006fw,Koma:2009ws},
which corresponds to 
$N_{\rm tsl}=3$ at $\beta \simeq 5.85$, 
$N_{\rm tsl}=4$ at $\beta \simeq 6.00$, 
and $N_{\rm tsl}=6$ at $\beta \simeq 6.30$.
Then, the appropriate $N_{\rm iupd}$ is chosen
by looking at the convergence history of the PLCF as a function of  $N_{\rm iupd}$.
If one is interested in the behavior of the potential at long distances, the larger 
$N_{\rm iupd}$ is needed.

\subsection{The classification of various three-quark geometries}

\begin{figure*}[!t]
\includegraphics[width=17cm]{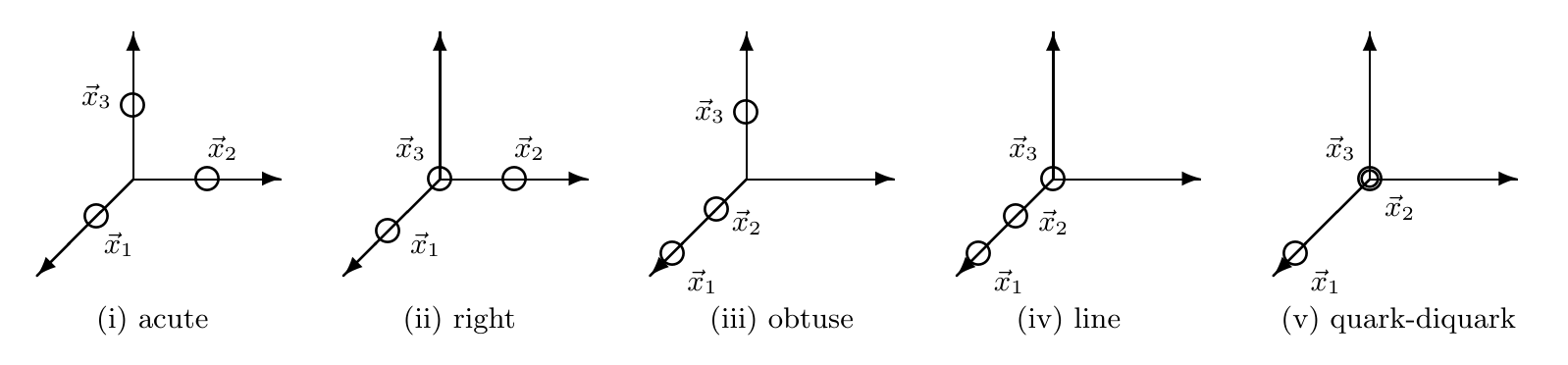}
\caption{The three-quark geometries investigated in our numerical simulations.
The circles represent the spatial location of  quarks.
Three Polyakov loops are put at the vertices of 
 (i) acute, (ii)  right, (iii)  obtuse triangles, and are also put in (iv) line, 
and are put to be (v) the quark-diquark system.}
\label{fig:howtoputPLCF}
\end{figure*}

\par
We compute the PLCF composed of the three Polyakov loops, 
${\rm Tr}P(\vec{x}_{1}){\rm Tr}P(\vec{x}_{2}) {\rm Tr}P(\vec{x}_{3})$, 
where the spatial locations of the Polyakov loops, $\vec{x}_1$, $\vec{x}_2$, and 
$\vec{x}_3$ correspond to those of  three quarks in the three-dimensional space, respectively.
As shown in Fig.~\ref{fig:howtoputPLCF}, 
there are 
five types of  three-quark geometries: 
three quarks are put at the vertices of  acute (ACT), right (RGT), 
obtuse (OBT) triangles, and are put in line (LIN).
As a special case, two of three quarks are put at the same location,
which corresponds to a quark-diquark system (QDQ).

\par
\begin{figure}[!t]
\includegraphics[width=5.5cm]{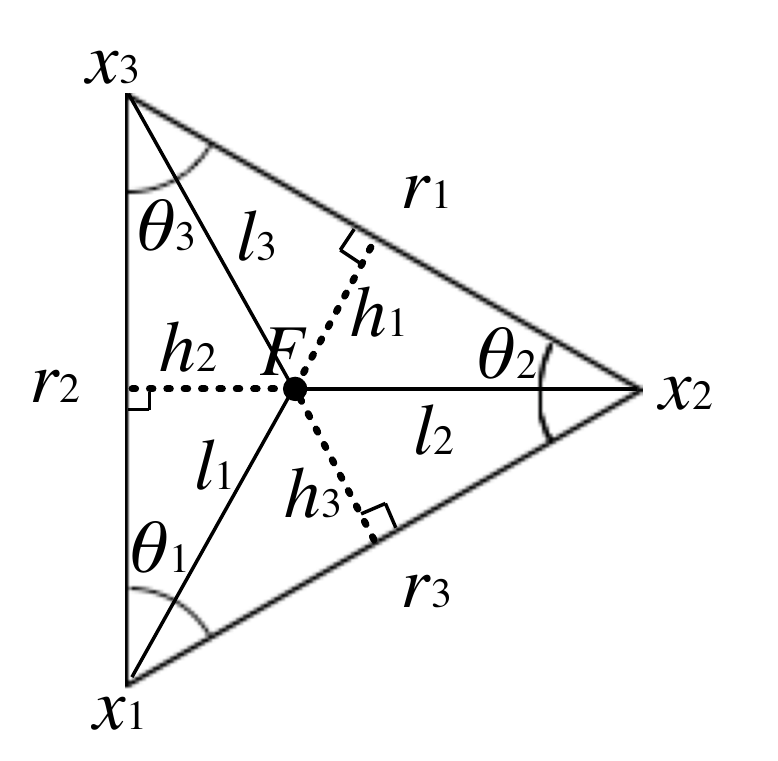}
\caption{The definition of distances and angles of a triangle used in our analyses:
$x_{i}$ denotes the location of $i$\,th quark,  $F$ the Fermat-Torricelli point,
so that $\angle x_{1}Fx_{2} = \angle x_{2}Fx_{3}=\angle x_{3}Fx_{1}=120^{\circ}$.
$h_{i}$ denotes the distance between $F$  and  each  side.
$\Delta$ and $Y$ are given by $\Delta = r_{1}+r_{2}+r_{3}$ 
and $Y = l_{1} +l_{2}+l_{3}$, respectively.}
\label{fig:triangle-h}
\end{figure}

These three-quark geometries
 can be classified by the value of
the maximum inner angle of a triangle 
\bea
\theta_{\rm max}
&=&
\max (\theta_1,\theta_2,\theta_3)\nn\\*
&=& 
\cos^{-1} \left ( \frac{r_{\rm max}  (r_1^2+r_2^2+r_3^2 - 2 r_{\rm max}^2)}{2r_1 r_2 r_3} \right )\,,
\label{eq:thetamax}
\eea
where 
\be
r_{1}=|\vec{x}_2-\vec{x}_3|, \quad r_{2}=|\vec{x}_3-\vec{x}_1|, \quad r_{3}=|\vec{x}_1-\vec{x}_2|\;,
\label{eqn:inter-quark-distance}
\ee
are  interquark distances and $r_{\rm max}=\max (r_1,r_2,r_3)$  (see, Fig.~\ref{fig:triangle-h}).
Acute triangles satisfy $\theta_{\rm max} < 90^{\circ}$, which 
contain equilateral and isosceles triangles in our study.
Right triangles are the case $\theta_{\rm max} = 90^{\circ}$.
Obtuse triangles are further classified into two types
depending on $\theta_{\rm max}$,
obtuse-narrow (OBTN) triangles for $90^{\circ} <\theta_{\rm max}  < 120^{\circ}$ and 
obtuse-wide (OBTW) triangles for $120^{\circ} \leqq \theta_{\rm max}  < 180^{\circ}$.

\par
In contrast to the classification of the three-quark geometries,
the parametrization of the three-quark  potential is not straightforward.
This is due to the fact that the potential can depend not only on the location of three 
quarks, $\vec{x}_{1}$, $\vec{x}_{2}$, and $\vec{x}_{3}$,
but also on the structure of the flux tube spanned among the three quarks,
which is unknown a priori because of the nonperturbative  feature of the QCD vacuum.
Therefore, the determination of the functional form of the potential is
nothing but the finding of appropriate distances that can 
capture the systematic behavior of the potential.
Such distances should be symmetric under 
the permutation of the quark positions.

\par 
The simplest distance is then given by
the sum of  interquark distances in Eq.~\eqref{eqn:inter-quark-distance},
\be
\Delta = r_1+ r_2 + r_3 \; .
\label{eq:Delta}
\ee
Another possible distance is given by
the minimal total length of lines connecting the three quarks 
via the Fermat-Torricelli point of a triangle,
\be
Y=\sqrt{\frac{r_{1}^{2}+r_{2}^{2}+r_{3}^{2}+4\sqrt{3}S}{2}}\;, 
\label{eq:Y}
\ee
where
$S$ is the area of the triangle given by Heron's formula,
\be
s = \frac{1}{2}\Delta\;,\quad
S = \sqrt{s(s-r_{1})(s-r_{2})(s-r_{3})}\;.
\ee
Note that the distance between the Fermat-Torricelli point and each vertex   is
\be
l_{i} = Y-\frac{1}{Y} (r_{i}^{2}+\frac{4S}{\sqrt{3}})\;.
\ee
For finding the location of the Fermat-Torricelli point the method 
presented in \cite{Bicudo:2008yr,Ay:2009zp} may be useful.

\par
In fact, these two types of distances, $\Delta$ and $Y$, were often used to examine the 
behavior of the potential in the earlier studies.
We also follow them in our analyses of the three-quark potential.
In terms of the minimal length of connected lines,  
$Y$  is reduced to 
\be
\Lambda = \Delta - r_{\rm max} \;,
\label{eq:Lambda}
\ee
when $\theta_{\rm max} \geqq 120^{\circ}$.
It is then convenient to introduce a combined distance of $Y$ and $\Lambda$ 
classified by $\theta_{\rm max}$ as
\be
L_{\rm str}=\biggl\{
\begin{array}{cl}
Y&\quad (\theta_{\rm max} < 120^{\circ})\\
\Lambda & \quad(\theta_{\rm max}\geqq 120^{\circ})\\
\end{array}
\biggr.\,.
\label{eq:L}
\ee

\par
For a detailed comparison with the quark-antiquark potential,
we also use a reduced interquark distance $R$ defined by 
\be
\frac{1}{R}=\frac{1}{r_{1}} + \frac{1}{r_{2}} +\frac{1}{r_{3}} \; ,
\label{eq:defR}
\ee
and an averaged distance between the Fermat-Torricelli
point and three sides of a triangle (see, Fig.~\ref{fig:triangle-h}),
\be
h=\frac{1}{3}(h_{1}+h_{2}+h_{3})\;,
\label{eqn:distance-h}
\ee
where 
\be
h_{i} = \frac{2 S_{i}}{r_{i}} = \frac{\sqrt{3} l_{j}l_{k}}{2r_{i}}
\quad (i,j,k: {\rm cyclic}) \,.
\ee
Note that $S_{i}=(1/2)  l_{j}l_{k} \sin 120^{\circ}$ is
the area of the triangle spanned by three points $x_{j}$, $x_{k}$, and $F$.

\par
As we will see in the next section, these distances 
are quite useful to see
the systematic behaviors of the three-quark potential,
although it may be possible to define other symmetric distances~\cite{Dmitrasinovic:2009ma}.

\section{Numerical results}
\label{sec:result}

\par
In this section, we present results of our lattice Monte Carlo simulations.
We emphasize that the crucial difference of our simulations 
from the earlier ones of other groups is that we 
compute the three-quark potential from the PLCF,
which could provide us with numerical results with less systematic effects 
than that from the Wilson loop. 

\par
We carried out Monte Carlo simulations
using the standard Wilson gauge action in SU(3) lattice gauge theory.
The basic simulation parameters are summarized in Table~\ref{tbl:simulation}.
The lattice spacing was determined by the Sommer scale $r_{0}=0.50\,{\rm [fm]}$~\cite{Necco:2001xg}.
One Monte-Carlo update consisted of one heat-bath and five over-relaxation steps.
The gauge coupling $\beta$ and the lattice volume $L^{3}\times T$ 
were chosen to make maximal use of the multilevel algorithm
within our computer resource.

\par
In fact, we once investigated the quark-antiquark  potential from the PLCF at $\beta=6.00$ 
for various lattice volumes including $16^{4}$, $20^{4}$, and $20^{3}\times 40$ using the 
multilevel algorithm~\cite{Koma:2006fw,Koma:2009ws}, 
and found no noticeable dependence on the temporal size $T$.
Therefore, our results of  the three-quark potential 
on the $24^{4}$ lattice at $\beta=6.00$ are also expected to be independent of $T$,
and can be regarded as those at  zero temperature.
Note that the lattice volume $24^{4}$ at $\beta=5.85$ and $\beta=6.30$
approximately corresponds to $32^{4}$ and $16^{4}$ at $\beta=6.00$, respectively.

\par
The numbers of internal updates $N_{\rm  iupd}$ and of 
the gauge configurations $N_{\rm cnf}$ are dependent on the observables,
which will be described in the following subsections individually.
The special attention is paid to the data from one gauge configuration at $\beta=6.00$,
which has no statistical error.
On the other hand, the data with statistical errors are from a certain number of  gauge configurations,
and the corresponding errors are estimated by the standard jackknife method.
When we perform  the $\chi^{2}$ fit, we  mainly use the data with statistical errors.

\begin{table}[t]
\caption{The basic simulation parameters used in this study.
The numbers of internal updates $N_{\rm  iupd}$ and of 
the gauge configurations $N_{\rm cnf}$ are dependent on the observables.}
\centerline{
\begin{tabular}{ccccc}
\hline\hline
$\beta =6/g^{2}$ & $(L/a)^{3} (T/a)$ & $a$\,[fm] 
&$N_{\rm sub}$ &$N_{\rm tsl}$  \\
\hline
5.85 & $24^{4}$ & 0.123 & 8 & 3 \\ 
6.00 & $24^{4}$ & 0.093 & 6 & 4 \\
6.30 &$24^{4}$ & 0.059 & 4 & 6 \\
\hline\hline
\end{tabular}}
\label{tbl:simulation}
\end{table}

\subsection{The potential from one gauge configuration}
\label{subsec:oneconf}

\par
\begin{figure}[!t]
\includegraphics[width=\figwidth]{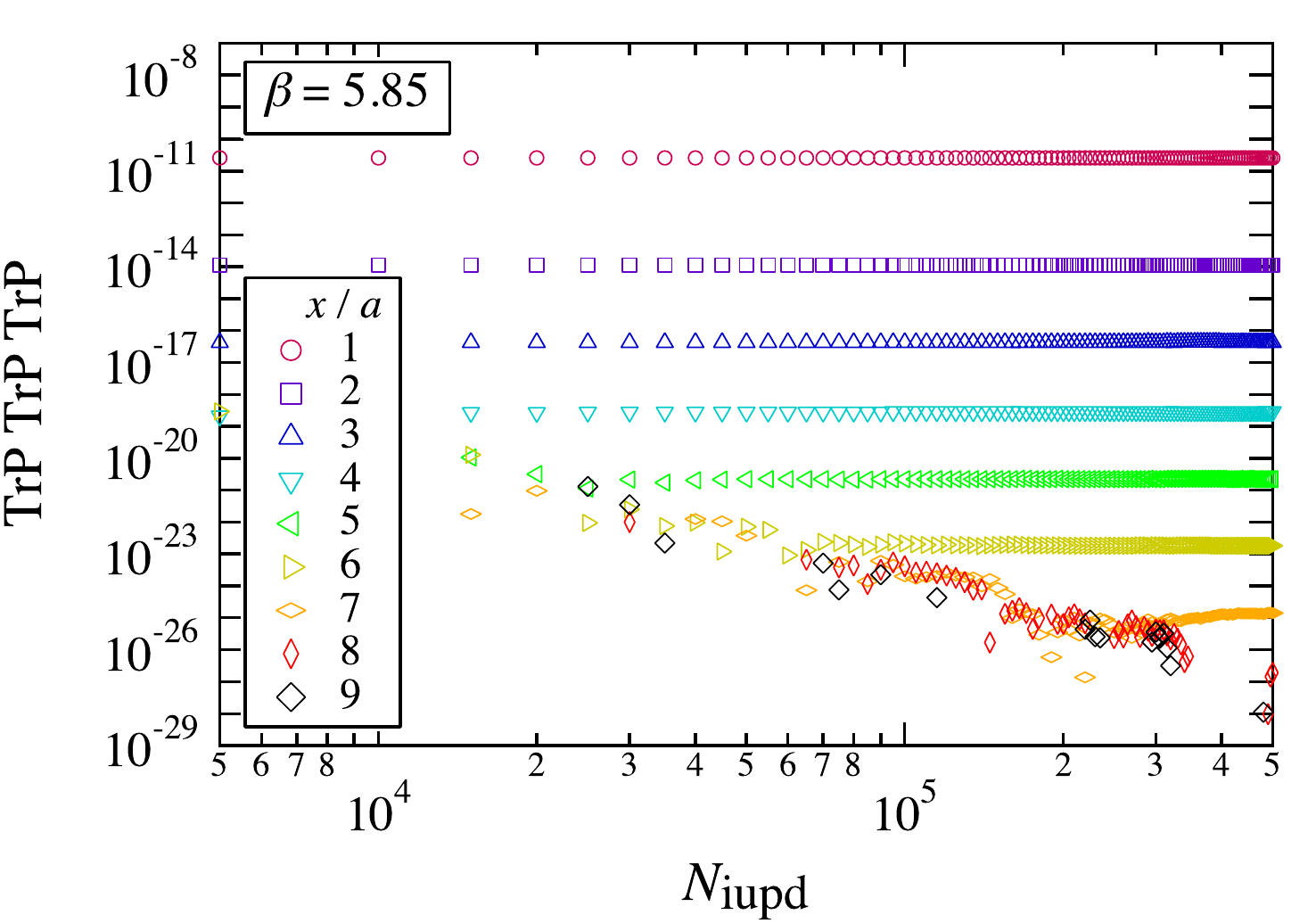}\\
\includegraphics[width=\figwidth]{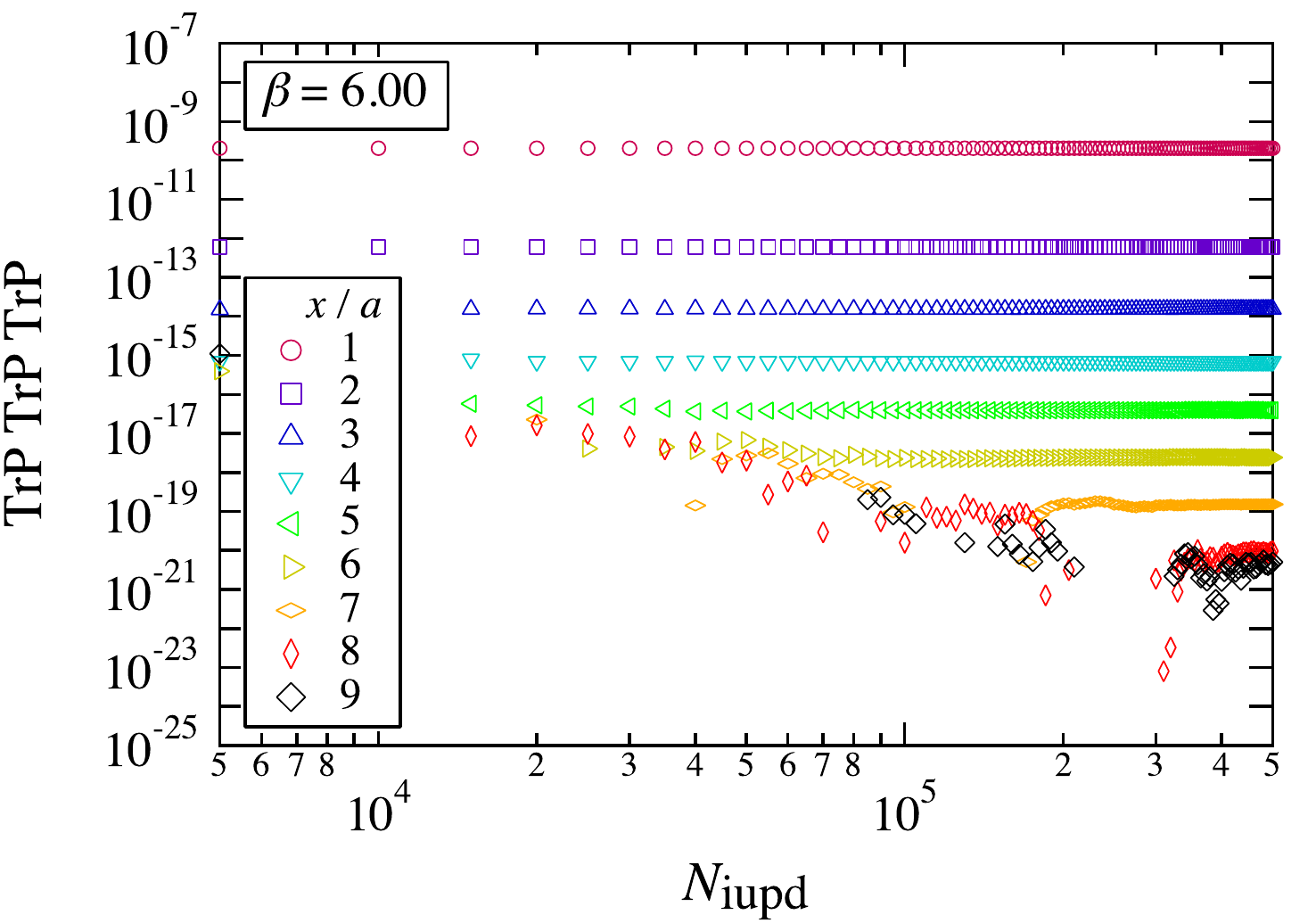}\\
\includegraphics[width=\figwidth]{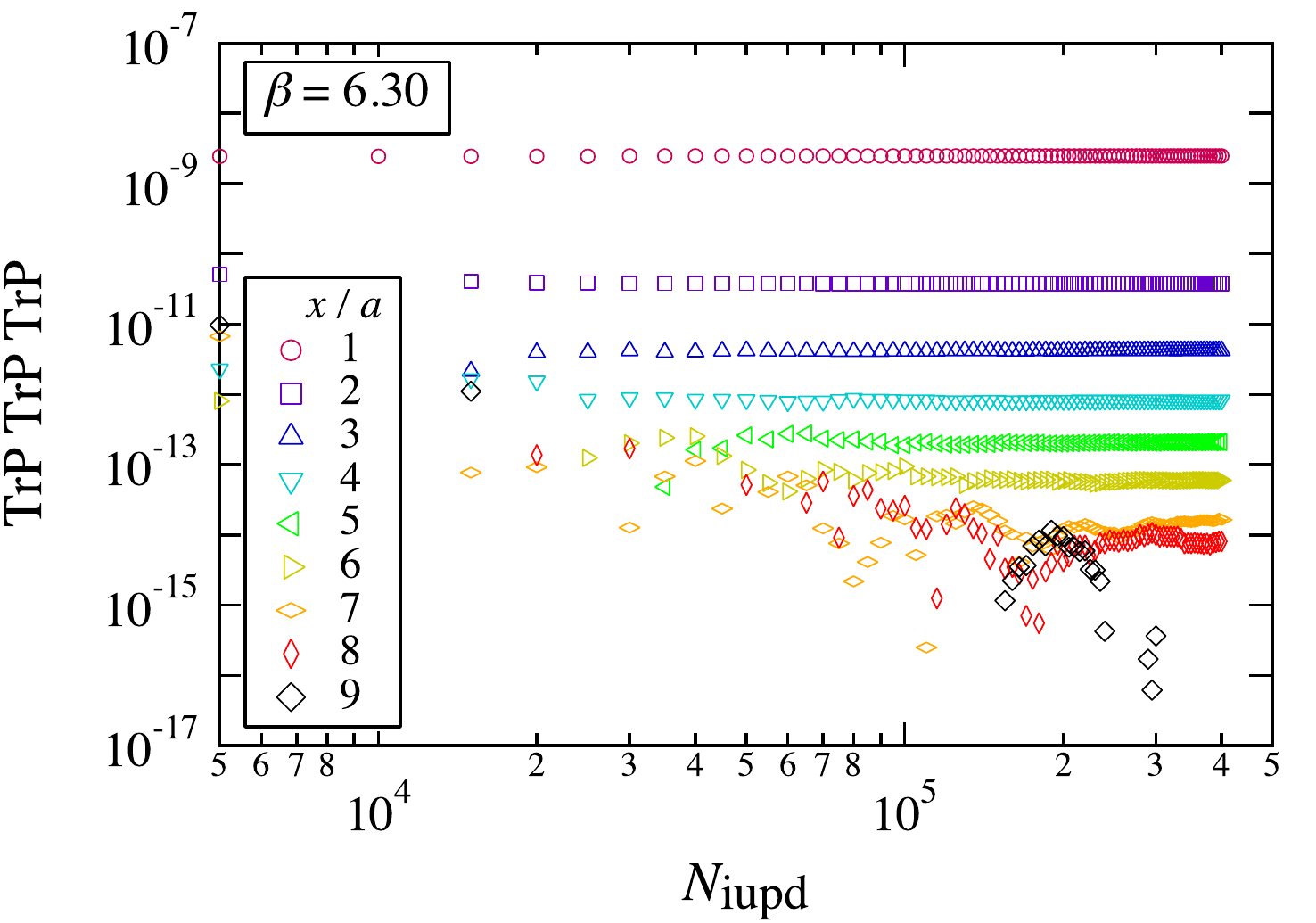}
\caption{Convergence histories of the PLCF  of the equilateral triangle geometries 
at $\beta=5.85$ (upper),  $\beta=6.00$ (middle), and  $\beta=6.30$ (lower) 
as a function of $N_{\rm iupd}$,
where quarks are placed at $(x,0,0)$, $(0,x,0)$, and $(0,0,x)$.}
\label{fig:hist-plcf}
\end{figure}

At the beginning, we demonstrate that the three-quark potential can be obtained from one gauge configuration
by tuning the parameters of the multilevel algorithm as explained in Sec.~\ref{subsect:multilevel}.
In Fig.~\ref{fig:hist-plcf}, we plot typical convergence histories of the PLCF 
for the equilateral triangle configurations of one gauge configuration 
at $\beta=5.85$ ($N_{\rm tsl}=3$), $\beta=6.00$ ($N_{\rm tsl}=4$), and $\beta=6.30$ ($N_{\rm tsl}=6$)
as a function of  the number of internal updates~$N_{\rm iupd}$.
We find that the fluctuation of the PLCF is washed out 
and the values become stable as we increase~$N_{\rm iupd}$.
Required $N_{\rm iupd}$ for convergence depends on the size of the triangle.

\par
\begin{figure}[t]
\includegraphics[width=\figwidth]{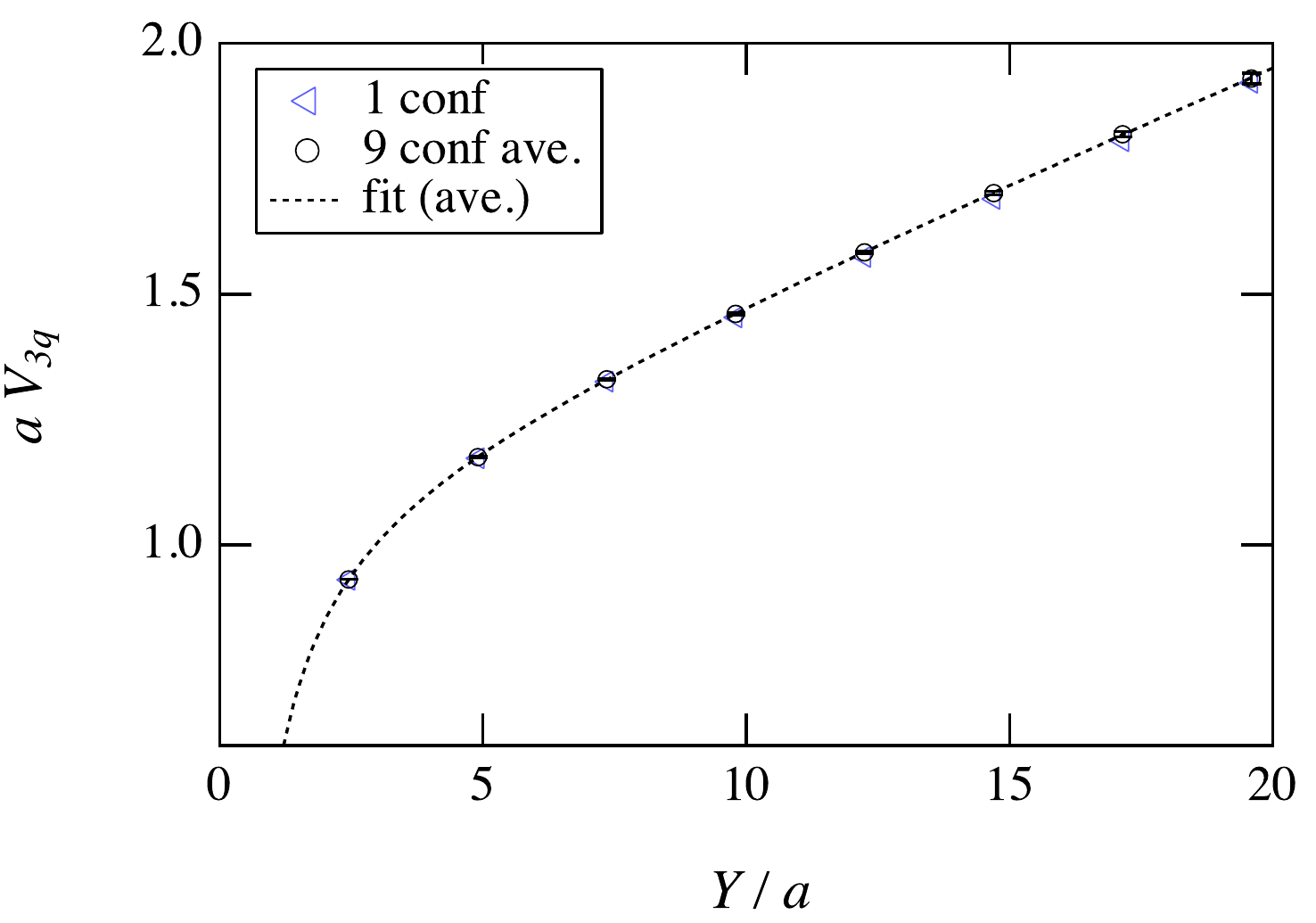}
\caption{The three-quark potentials of the equilateral triangle geometries at $\beta=6.00$
obtained from one gauge configuration and from the average of 9 gauge configurations as a function of~$Y$. 
The dotted line represents the fit curve to the averaged potential. }
\label{fig:ave-vs-oneconf}
\end{figure}
\begin{figure}[t]
\includegraphics[width=\figwidth]{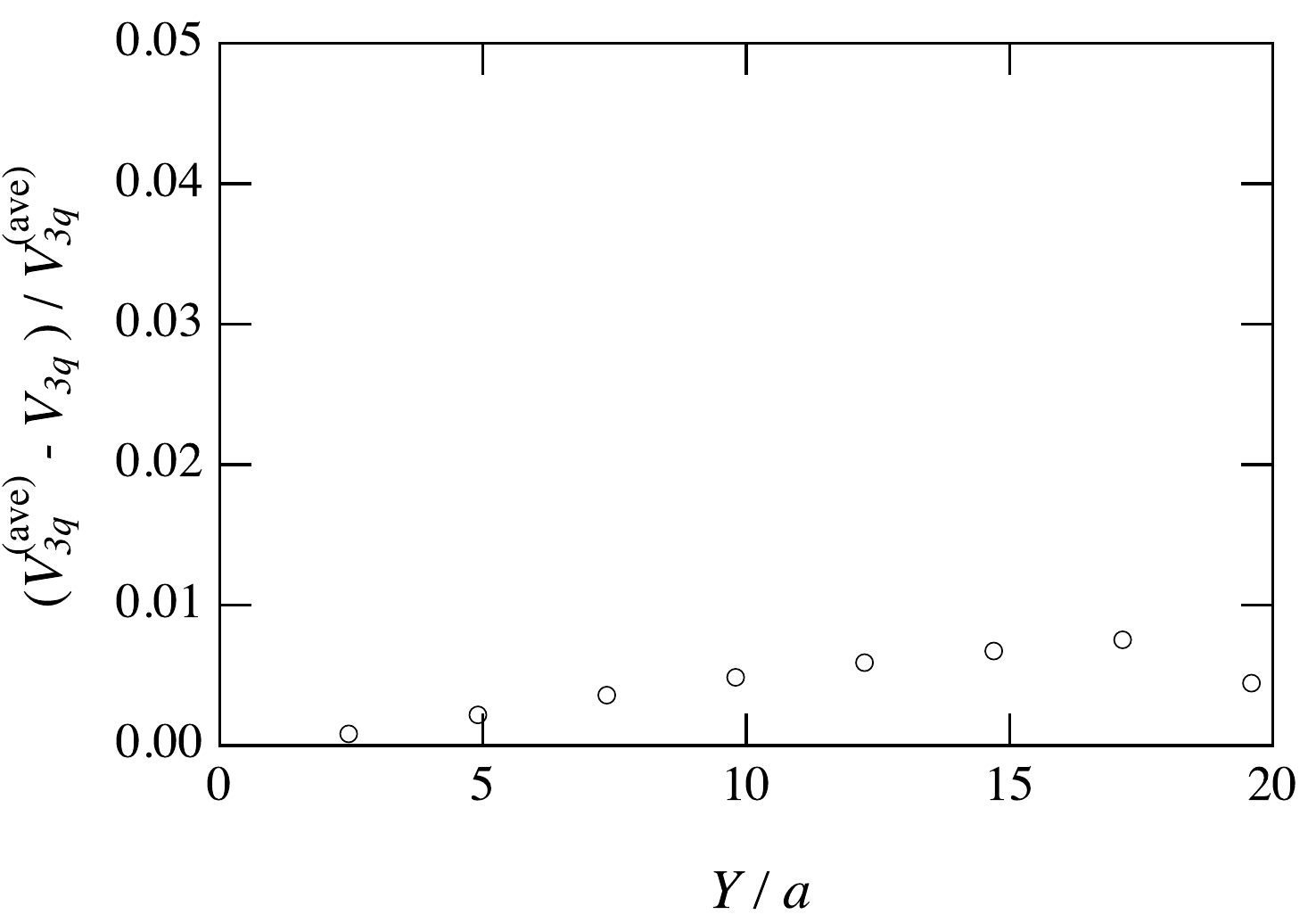}
\caption{The relative error between the two potentials  in Fig.~\ref{fig:ave-vs-oneconf},
$(V^{\rm (ave)}_{3q}-V_{3q})/V^{\rm (ave)}_{3q}$.}
\label{fig:ave-oneconf}
\end{figure}

Once the PLCF becomes stable, it can be regarded as the 
expectation value as in Eq.~\eqref{eqn:plcf-selection},
and then the potential is computed by Eq.~\eqref{eqn:plcf-exract-E0}.
In Fig.~\ref{fig:ave-vs-oneconf}, we show the potential at $\beta=6.00$ from one gauge configuration
at $N_{\rm iupd}=500000$ as a function of~$Y$ defined in Eq.~\eqref{eq:Y}.
The potential is also compared to that from 
the average of $N_{\rm cnf}=9$ independent gauge configurations
(the number 9 is just the maximum gauge configurations 
that we obtained within our available computer resources).
The numerical error of the average is estimated by the standard jackknife method.
It is obvious that the potential is determined accurately,
where the potential from one gauge configuration already represents the average.
The level of agreement between the two potentials  can be quantified by evaluating the
relative error $(V^{\rm (ave)}_{3q}-V_{3q})/V^{\rm (ave)}_{3q}$,
which is plotted in  Fig.~\ref{fig:ave-oneconf}.
This figure shows that  the relative error is gradually increasing at large $Y$, 
which is one of the origin of the statistical error of the average.
However it is only 0.8~\% even at a quite long distance $Y/a \simeq 18$.

\par
The fit of the averaged potential to an empirical functional form, 
\be
V_{3q}^{\rm (Y)} = - \frac{A_{3q}^{\rm (Y)}}{Y} + \sigma_{3q} Y +\mu_{3q}\;,
\ee
yields $A_{3q}^{\rm (Y)} = 0.662(6)$, $ \sigma_{3q} a^{2}=0.0446(3)$, and $\mu_{3q} a = 1.092(3)$,
which describes the behavior of the data nicely as shown in Fig.~\ref{fig:ave-vs-oneconf}.
It is interesting to note that  the coefficient in front of~$Y$,
which we may call the three-quark string tension $\sigma_{3q}$,
is consistent with the string tension of the quark-antiquark potential $\sigma_{\qqb}$.
In addition, the constant term  $\mu_{3q}$ is approximately $(3/2)\mu_{\qqb}$,
where $\mu_{\qqb}$ is  the constant term of the
quark-antiquark potential (discussed later in Sec.~\ref{subsec:functional-form}).

\subsection{The potential of various three-quark geometries}
\label{sect:various-pot3q}

\begin{table}[t]
\caption{The number of various three-quark 
geometries investigated from one gauge configuration at $\beta=6.00$ (total 221).
The potential data are summarized  in Tables in Appendix~\ref{sect:alldata}.}~\\
\label{tbl:3quark-class}
\centerline{
\begin{tabular}{lcc}
\hline\hline
Classification (abbr.) &Count  & Table \\
\hline
acute (ACT) & 68   & \ref{tbl:datatable_act}\\
right (RGT) & 43 & \ref{tbl:datatable_rgt}\\
obtuse-narrow (OBTN)  & 27 &\ref{tbl:datatable_obn} \\
obtuse-wide (OBTW) & 28 & \ref{tbl:datatable_obw} \\
line (LIN) &32 & \ref{tbl:datatable_lin}\\
quark-diquark (QDQ) & 23& \ref{tbl:datatable_qdq} \\
\hline\hline
\end{tabular}}
\end{table}

\begin{figure}[t]
\includegraphics[width=\figwidth]{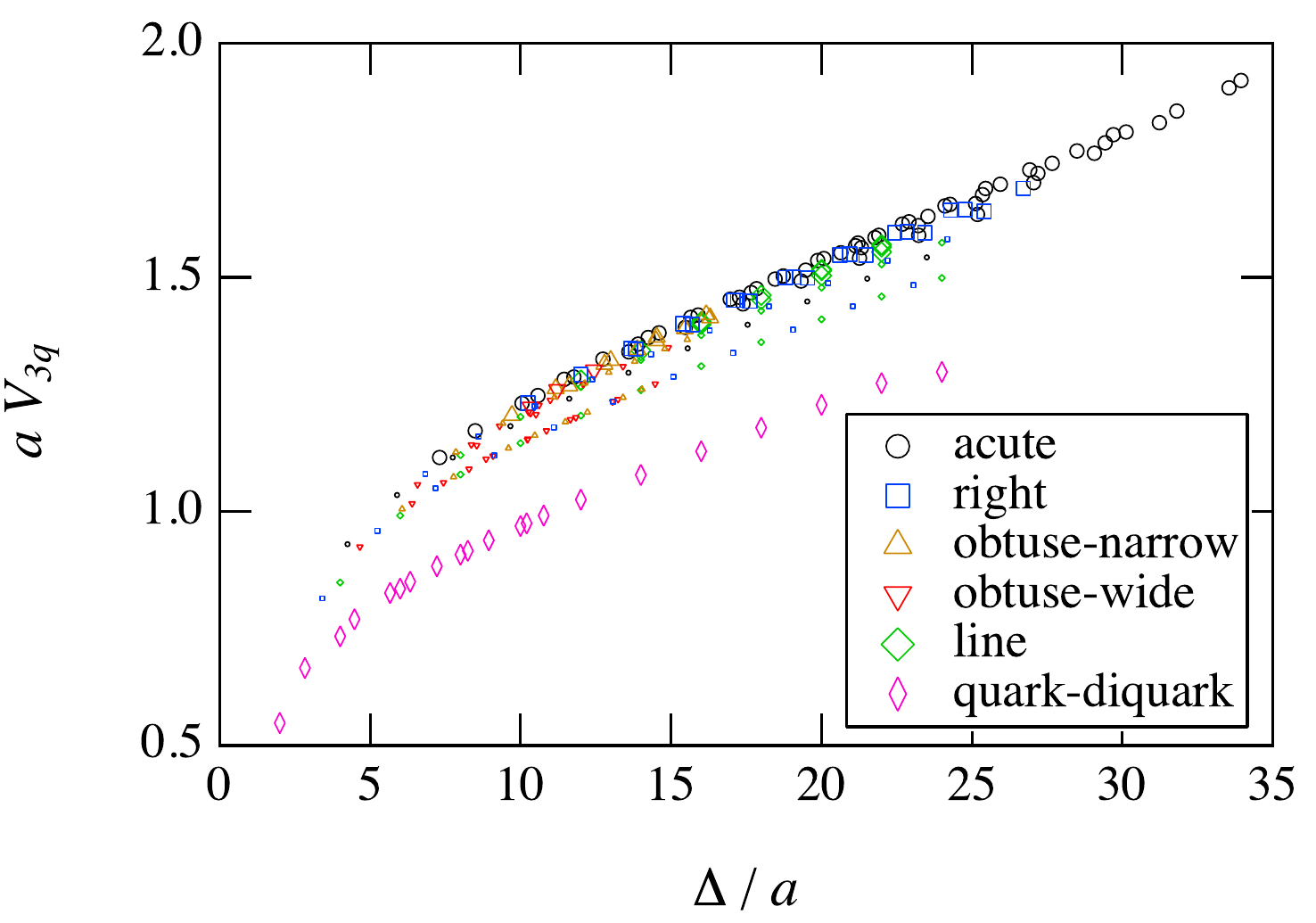}
\includegraphics[width=\figwidth]{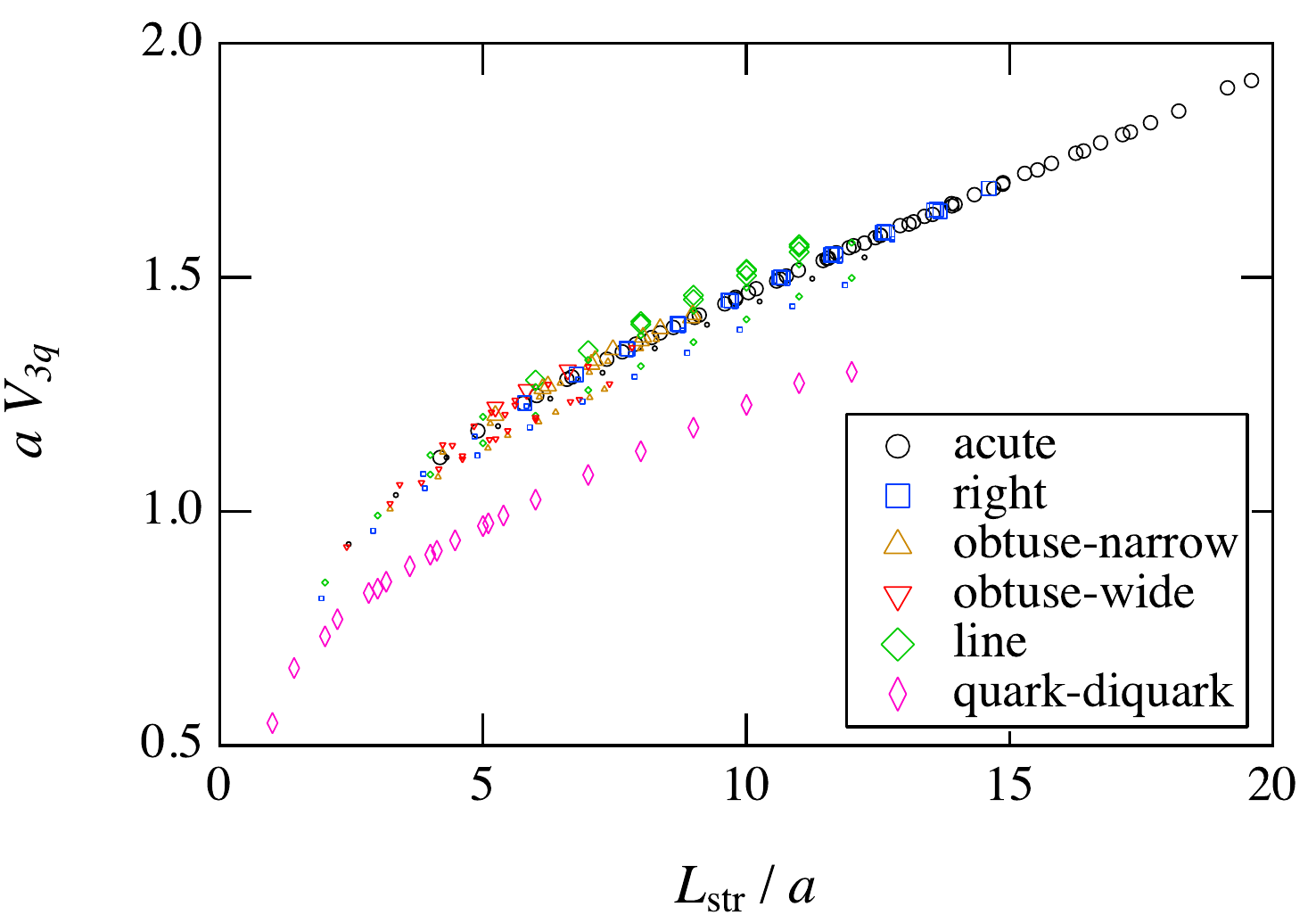}
\caption{The three-quark potential against 
$\Delta$ defined in Eq.~\eqref{eq:Delta} (upper) and $L_{\rm str}$ defined in Eq.~\eqref{eq:L} (lower).
Smaller markers denote  data from the geometries for $r_{\rm min}/a \leqq 2$, 
which may suffer from lattice cutoff effects.}
\label{fig:pot-b600-all}
\end{figure}
\begin{figure}[t]
\includegraphics[width=\figwidth]{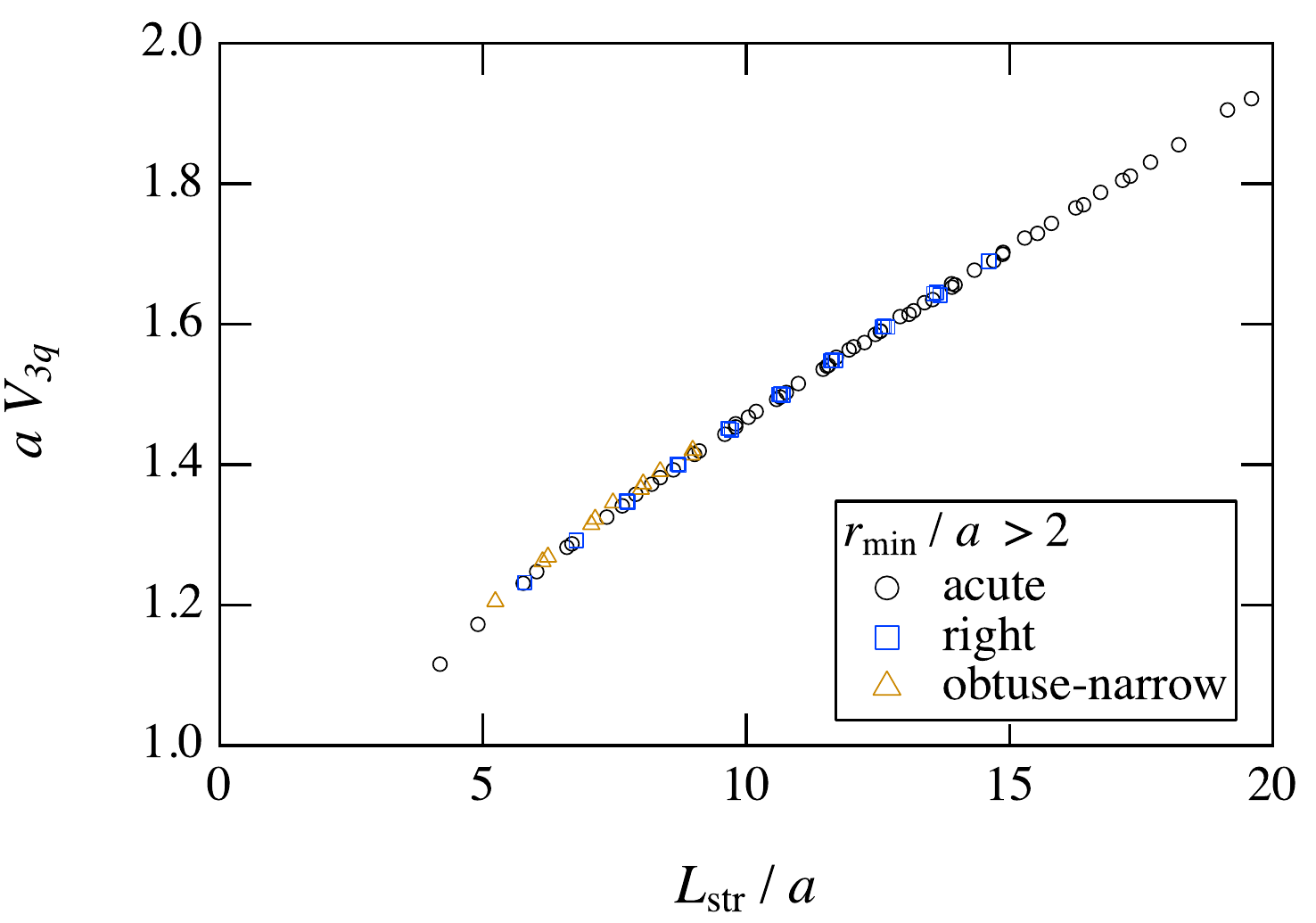}
\caption{The three-quark potential of acute,  right,
and  obtuse-narrow geometries  restricted to $r_{\rm min}/a > 2$ against~$L_{\rm str}$.}
\label{fig:pot-b600-all-r2}
\end{figure}

\par 
We then extend the computation of the potential to various  three-quark geometries
as in Fig.~\ref{fig:howtoputPLCF} by using the same one gauge configuration at $\beta=6.00$.
In total we investigate 221 three-quark geometries
as summarized in Table~\ref{tbl:3quark-class}.
We list all of these potential data within the significant digits in Appendix~\ref{sect:alldata}.

\par
In Fig.~\ref{fig:pot-b600-all}, 
we plot all the potential data against  $\Delta$ and $L_{\rm str}$ 
defined in Eqs.~\eqref{eq:Delta} and~\eqref{eq:L}.
At glance, we find that neither  $\Delta$ nor $L_{\rm str}$ 
can provide a universal functional form of the three-quark potential, 
since all the data do not fall into one curve with these distances.
However, if we look at the potential of  ACT, RGT, and OBTN against~$L_{\rm str}$ 
carefully and restrict the data only for $r_{\rm min}/a >2$,
where $r_{\rm min}=\min (r_{1},r_{2},r_{3})$,
by assuming that they do not suffer from severe lattice cutoff effects,
we observe an excellent linearly-rising behavior
as explicitly shown in Fig.~\ref{fig:pot-b600-all-r2}.
As we will demonstrate in the following analysis, 
the corresponding string tension is consistent with
that of the quark-antiquark potential.

\par
\begin{figure}[t]
\includegraphics[width=\figwidth]{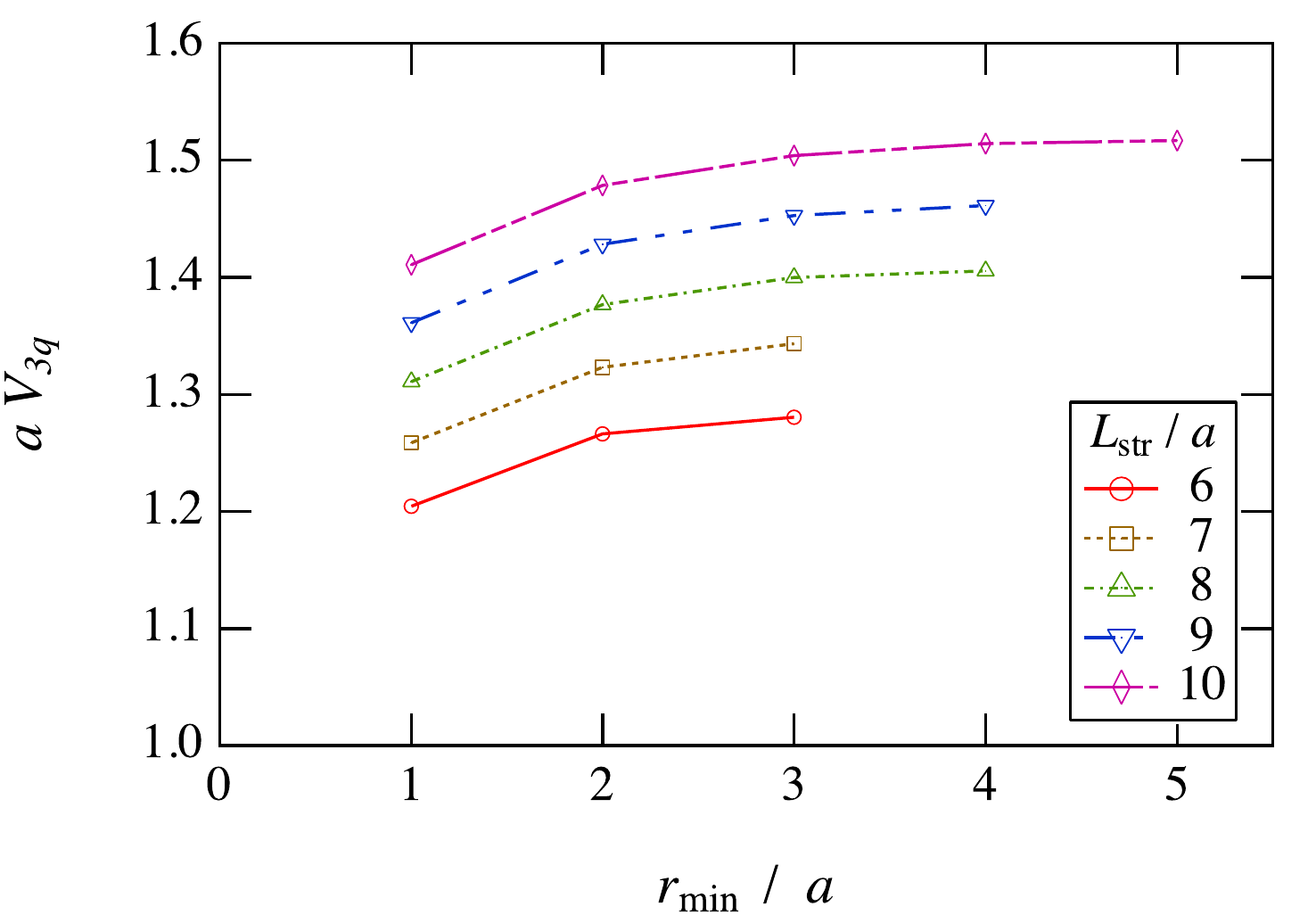}
\caption{The three-quark potential of line geometries against~$r_{\rm min}$.}
\label{fig:line3qvsrmin}
\end{figure}

The behavior of the potentials of  OBTW and LIN
is not as clear as that of  ACT and RGT. 
They are not on a simple function of~$L_{\rm str}$.
In particular, as explicitly shown in Fig.~\ref{fig:line3qvsrmin} for LIN, 
the potentials are dependent also on $r_{\rm min}$.
The increasing behaviors as a function of $r_{\rm min}$ 
are similar with each other,
which in turn imply that the three quarks in LIN prefers to be QDQ.
As clarified later in Sec.~\ref{subsec:h}, the potential for LIN is
well described by the half of the sum of the quark-antiquark potential,
so that the interval between the potentials for various $L_{\rm str}$ in Fig.~\ref{fig:line3qvsrmin}
is approximately given by $\sigma_{\qqb}\delta L_{\rm str}$, where 
$\delta L_{\rm str}$ is the difference of $L_{\rm str}$.

\begin{figure}[t]
\includegraphics[width=\figwidth]{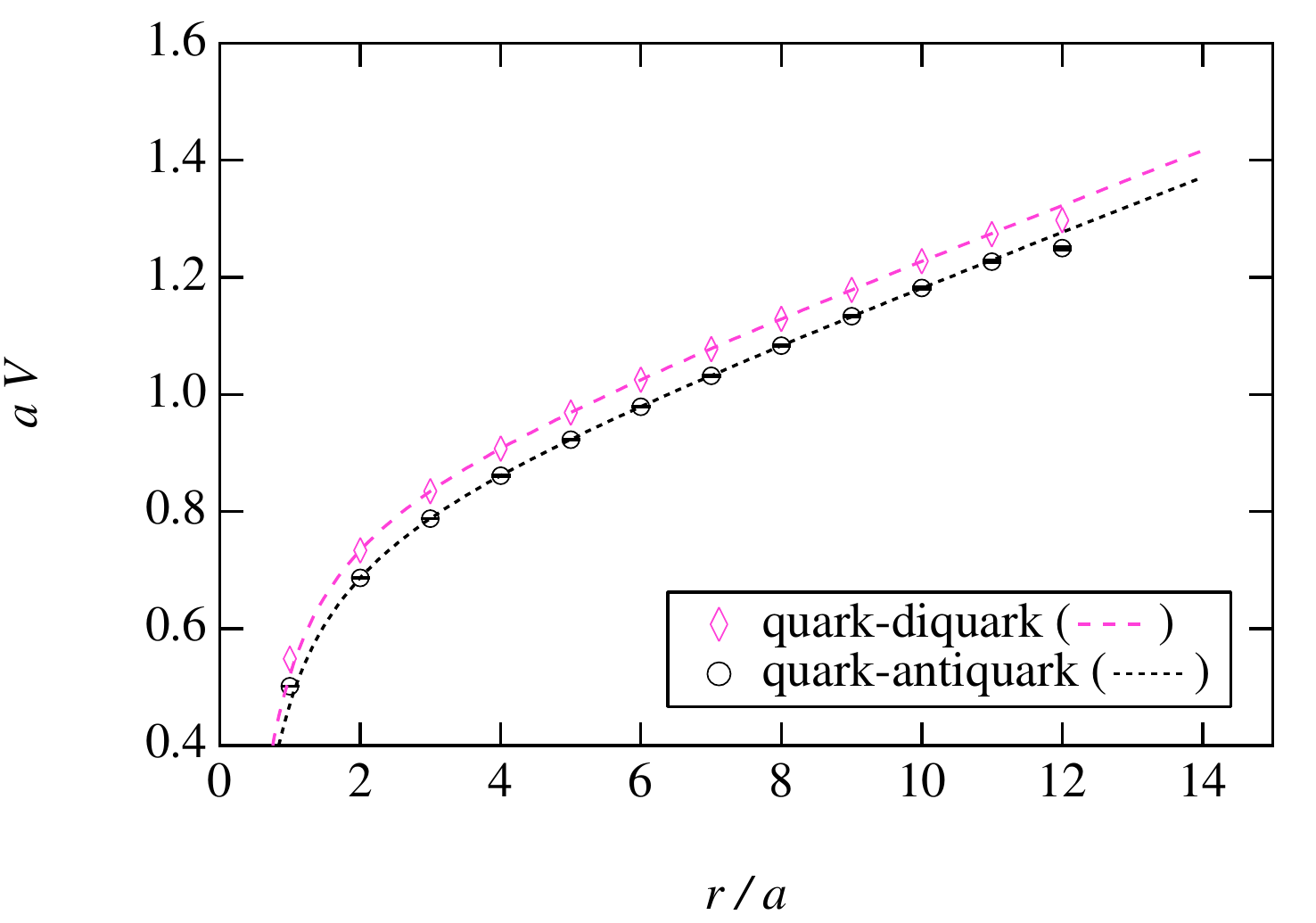}
\caption{The quark-diquark and quark-antiquark potentials against $r$.
The dashed and dotted lines are the fit curves to Eq.~\eqref{eqn:fit-vqdq}, respectively.}
\label{fig:pot_qdiq_qqbar}
\end{figure}

\begin{figure}[t]
\includegraphics[width=\figwidth]{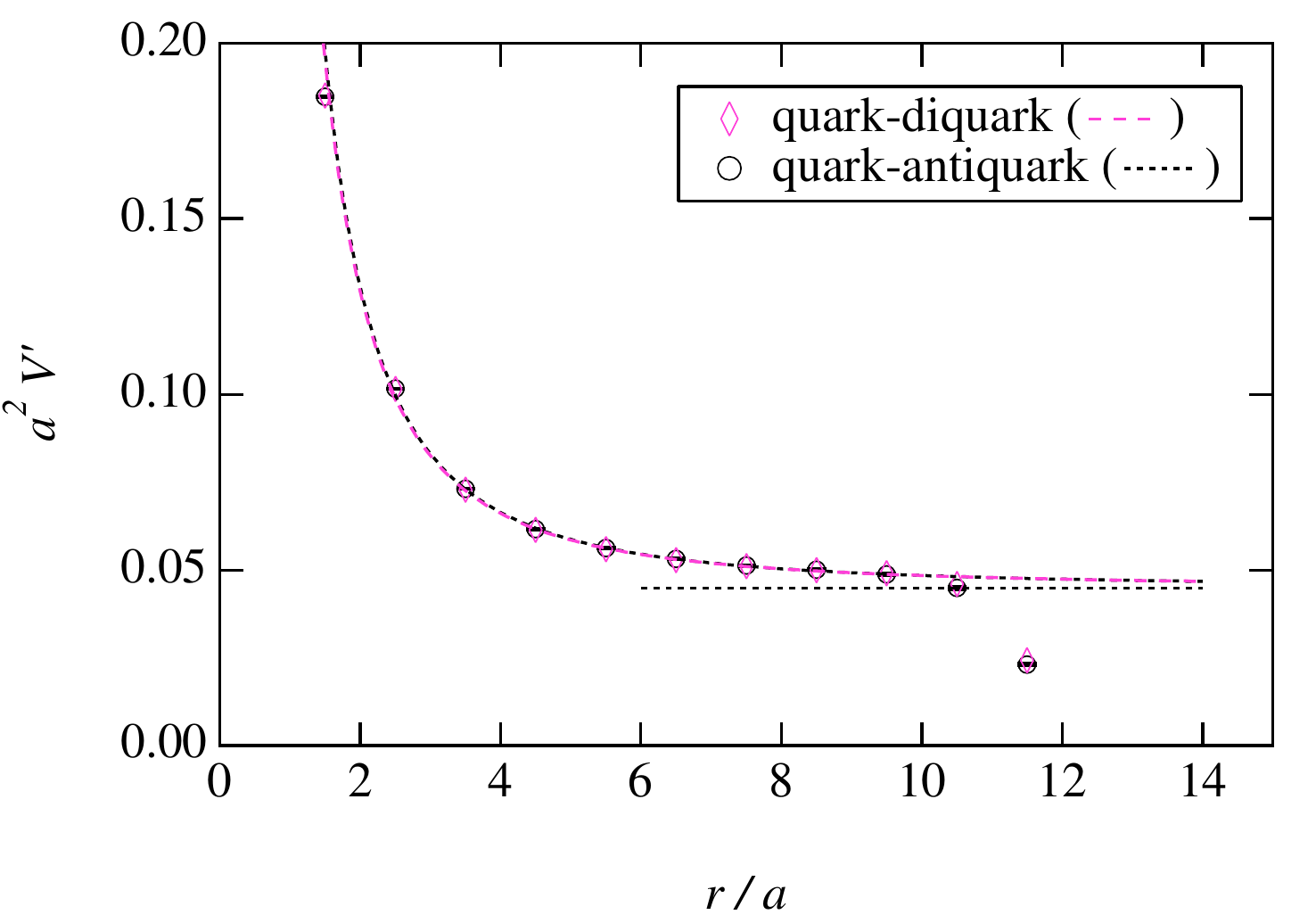}
\caption{The derivatives of the quark-diquark and quark-antiquark potentials with respect to $r$.
The horizontal dotted line corresponds to the string tension of the quark-antiquark 
potential $\sigma_{\qqb}a^{2}=0.0449$ (see, Table~\ref{tbl:fit-potqqbar}).}
\label{fig:frc_qdiq_qqbar}
\end{figure}

\par
The potential of QDQ exhibits a different behavior from that of other three-quark geometries.
In Fig.~\ref{fig:pot_qdiq_qqbar}, we plot the QDQ potential (we have selected only on-axis data),
which is then compared to the quark-antiquark potential with the same distance 
between a quark and an antiquark $r$ (the raw data
of the quark-antiquark potential are 
summarized in Table~\ref{tbl:potqqbar_scale} in Appendix~\ref{sect:qqbarpot}). 
It appears that the two potentials are almost the same.
Fitting the two potentials to the functional form,
\be
V =-\frac{A}{r} +\sigma \,  r + \mu\;,
\label{eqn:fit-vqdq}
\ee
leads to the values 
$A_{qdq}=0.336(2)$, $\sigma_{qdq}a^{2}=0.0449(1)$, and $\mu_{qdq}a = 0.811(1)$
for the QDQ potential\footnote{The errors of the QDQ data, which are absent because of
 ${\displaystyle N_{\rm cnf}\!=\!1}$,
are estimated from the residuals in the fitting process, and are 
used to evaluate the errors of the fit parameters.}, while
$A_{\qqb}=0.340(2)$, $\sigma_{\qqb}a^{2}=0.0449(2)$, and $\mu_{\qqb}a = 0.766(1)$
for the quark-antiquark potential  (see, Table~\ref{tbl:fit-potqqbar} in Appendix~\ref{sect:qqbarpot}), where 
subscripts are added to distinguish the fit results;
two potentials are consistent with each other except for the constant shift, 
$a(\mu_{qdq}-\mu_{\qqb})=0.045(3)$.

\par
We also compute the derivatives of the QDQ and the 
quark-antiquark potentials with respect to $r$, 
which are shown in Fig.~\ref{fig:frc_qdiq_qqbar}.
We find that the two results completely agree with each other,
including a systematic effect caused by finite volume.
These results mean that reduction of 
the representation in SU(3) color group,
${\bf 3}\otimes (\bar{\bf 3} \oplus {\bf 6}) \Rightarrow  {\bf 3} \otimes \bar{\bf 3}$,
is realized nonperturbatively.
Note that a similar result is obtained in Ref.~\cite{Bissey:2009gw}
by using the $T$-shaped
three-quark Wilson loop, where the distance between the two quarks
for the diquark is set to two lattice steps.

\subsection{The string tension of the flux tube}
\label{subsec:st}

The analysis in Sec.~\ref{sect:various-pot3q} indicates 
that all of the three-quark potentials of various three-quark 
geometries cannot be parametrized by a unique distance simultaneously.
However,  if we look at the three-quark potential plotted in Fig.~\ref{fig:pot-b600-all} 
optimistically, especially the plot against $L_{\rm str}$,
there seems to be a common increasing behavior with different constant shifts.
This may probably be due to the fact that somehow
a common type of flux tube is formed among three quarks to minimize 
the total energy of the three-quark system,
while the difference of the constants
originates only from  short distance effects among the three quarks,
which persists even if one of the three quarks is located at a distance.
We thus investigate the derivative of the potential 
with respect to $L_{\rm str}$
so that the short distance effects can be removed  from the potential.

\begin{figure}[t]
\includegraphics[width=\figwidth]{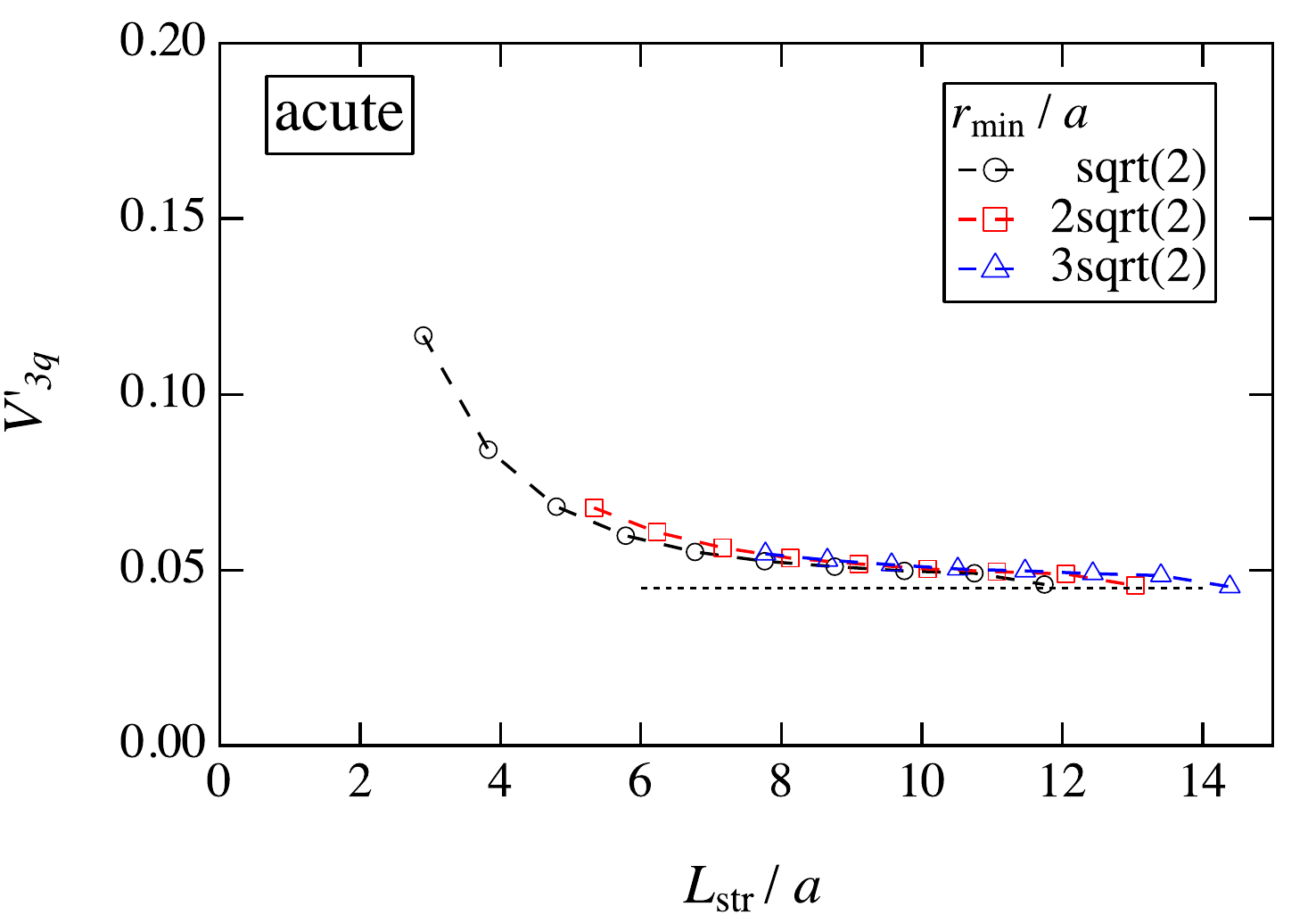}\\
\includegraphics[width=\figwidth]{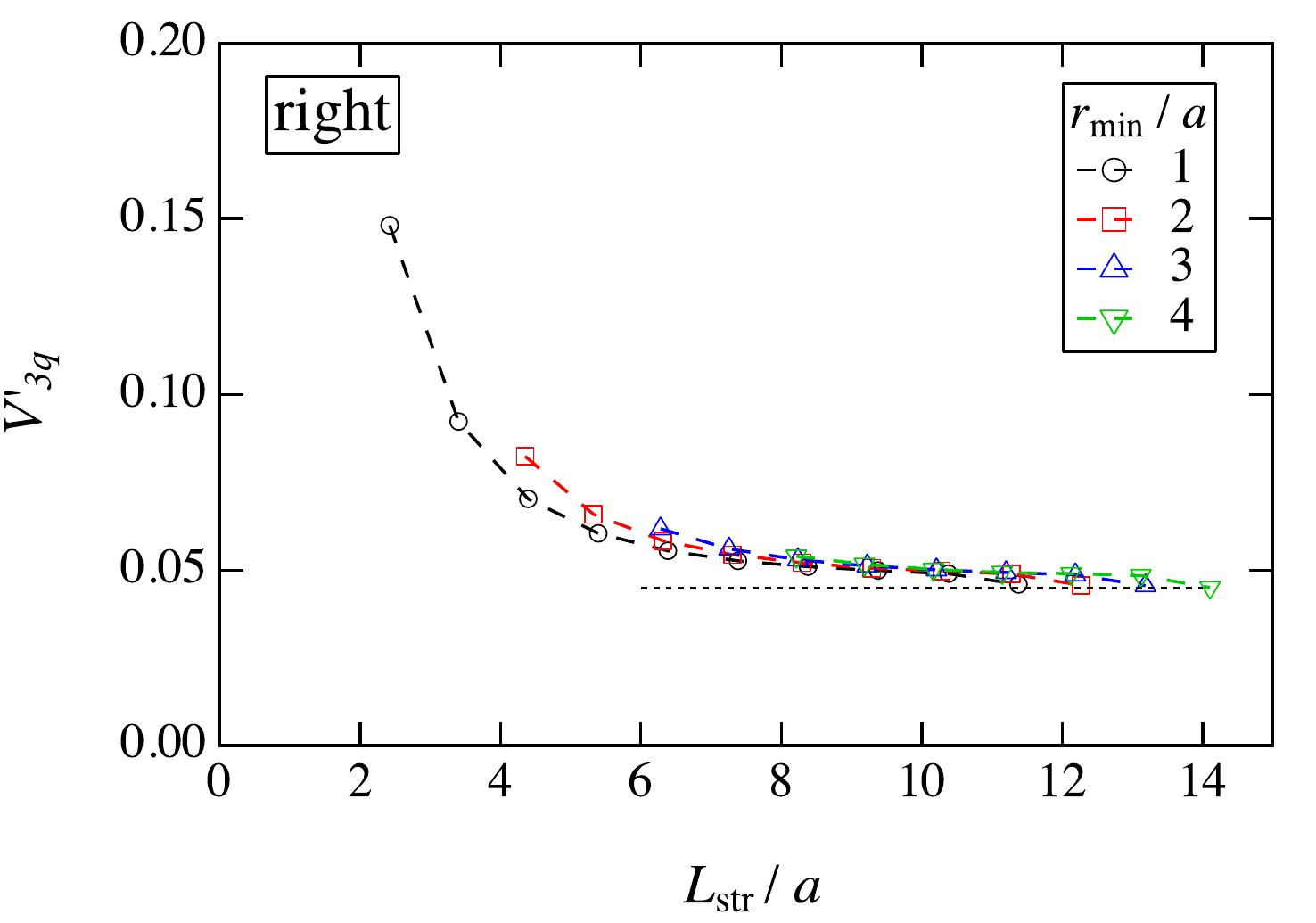}\\
\includegraphics[width=\figwidth]{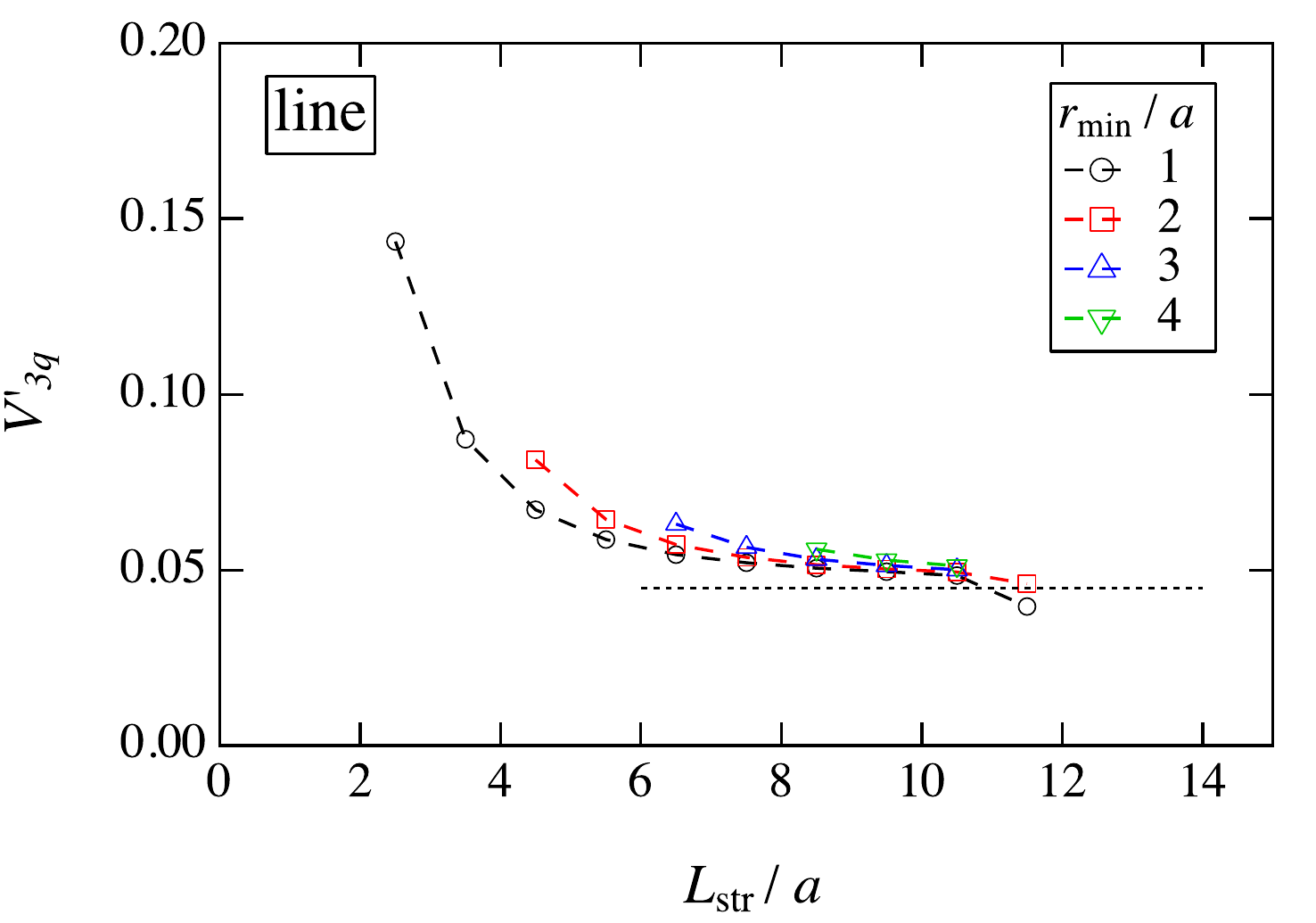}
\caption{The derivatives of the three-quark potential with
respect to $L_{\rm str}$ for acute  (upper), right  (middle), and line  (lower) geometries.
The dotted line in each plot corresponds to the string tension of the quark-antiquark 
potential $\sigma_{\qqb}a^{2}=0.0449$ (see, Table~\ref{tbl:fit-potqqbar}).}
\label{fig:frc-b600-act}
\end{figure}

\par
Let us first focus on the potential of the isosceles triangle geometries
within ACT, where the two of three quarks are placed at
$\vec{x}_{1}=(x,0,0)$ and $\vec{x}_{2}=(0,x,0)$,
and the remaining third quark is placed at  $\vec{x}_{3}=(0,0,z)$ with $z\geqq x$.
In this case,  $L_{\rm str}$ is identical to $Y$
and the distance between the Fermat-Torricelli point and 
$\vec{x}_{1}$ and~$\vec{x}_{2}$, respectively, is the same
 $l_{1}=l_{2}=(\sqrt{6}/3)x$ (see, Fig.~\ref{fig:triangle-h}).
Therefore, pulling the third quark (changing $z$) 
with the fixed first and second quarks does 
not affect the location of the Fermat-Torricelli point,
which just affects the increase of the energy 
between the Fermat-Torricelli point and the third quark,
where $l_{3}=\sqrt{z^{2} +x^{2}/2} - x/ \sqrt{6}$.
We then compute the derivative of the potential with respect to $Y$ for 
several fixed values of $x$,
\be
V'_{3q} = \frac{V_{3q}(x,z+\delta z) - V_{3q}(x,z)}{ \delta Y }    \;.
\ee
In Fig.~\ref{fig:frc-b600-act} (upper), we plot the result 
for  one gauge configuration at $\beta=6.00$
with the classification in terms of the 
distance between the first and second quarks,
$r_{\rm min} =\sqrt{2}\, x$, where $x/a=1$, $2$, and $3$ (in this case, $\delta Y = \delta l_{3}$).
We find that all the derivatives  behave quite similarly 
and approach a constant value at long distance.
Remarkably, the  constant value is nothing but the string tension in the
quark-antiquark system, $\sigma_{\qqb}a^{2}=0.0449$  (see, Table~\ref{tbl:fit-potqqbar}).
Since the third quark is chosen arbitrarily among the three,
this result also supports a picture of the  $Y$-shaped flux-tube formation.
This feature agrees with that  was pointed out by Takahashi~{\it et\,al.}~\cite{Takahashi:2000te,Takahashi:2002bw}
based on the~$\chi^{2}$ fit to the potential data with the $Y$~{\it Ansatz}.

\par
We then pay attention to the potentials of RGT,
where $\vec{x}_{2}=(0,y,0)$ and $\vec{x}_{3}=(0,0,0)$,
and the remaining first quark is placed at  $\vec{x}_{1}=(x,0,0)$, where $x\geqq y$.
In this case,  although the location of the  Fermat-Torricelli point 
is slightly dependent on changing $x$,
it becomes insensitive to $x$ when $x\gg y$.
The derivative is then defined by
\be
 V'_{3q} =  \frac{V_{3q}(x+\delta x, y)  - V_{3q}(x,y)}{ \delta Y } 
 \;.
\ee
In Fig.~\ref{fig:frc-b600-act} (middle),
we plot the result for the same one gauge configuration
with the classification in terms of the distance between the second and third quarks,
$r_{\rm min} =y$, where $y/a =1 \sim 4$.
We find that all the derivatives approach the constant value,
 $\sigma_{\qqb}a^{2}=0.0449$, at long distance: 
the tendency is quite the same as that for ACT.

\par
We finally examine the potentials of LIN,
where $\vec{x}_{1}=(x_{1},0,0)$,  $\vec{x}_{2}=(x_{2},0,0)$, and $\vec{x}_{3}=(0,0,0)$,
which is an extreme case that there is probably no chance to form a junction of the flux tube.
For a fixed distance $r_{\rm min}=x_{2}$ ($0 <   x_{2} < x_{1}/2$), the derivative is then defined by
\be
 V'_{3q} =  \frac{V_{3q}(x_{1}+\delta x_{1},x_{2})  - V_{3q}(x_{1},x_{2})}{ \delta x_{1} }  \;.
\ee
In Fig.~\ref{fig:frc-b600-act} (bottom),
we plot the result for the same one gauge configuration as a function of $L_{\rm str}=x_{1}$.
We again find that all the derivatives approach the constant value,
 $\sigma_{\qqb}a^{2}=0.0449$, at long distance.

\par
These results strongly indicate that
the energy of the flux tube per unit length,
the string tension, is common to that of the quark-antiquark potential
regardless of the 
geometry of three quarks.

\subsection{Detailed comparison with the two-body quark-antiquark potential}
\label{subsec:h}

\par
In the earlier studies of the three-quark potential,
there was a claim that the potential was described by the half of the 
sum of two-body potentials in the quark-antiquark system~\cite{Bali:2000gf, Alexandrou:2001ip}.
However, our results in Sec.~\ref{subsec:st},
indicating the presence of the flux-tube junction in larger ACT and RGT,
clearly contradict the earlier claim.
Thus we critically compare the three-quark potential 
with the quark-antiquark potential.
In fact, this is possible only when the both potentials
are determined accurately up to long distance, 
otherwise one cannot distinguish the difference between them
since it is not so apparent in practice as we will see below.

\begin{figure}[t]
\includegraphics[width=\figwidth]{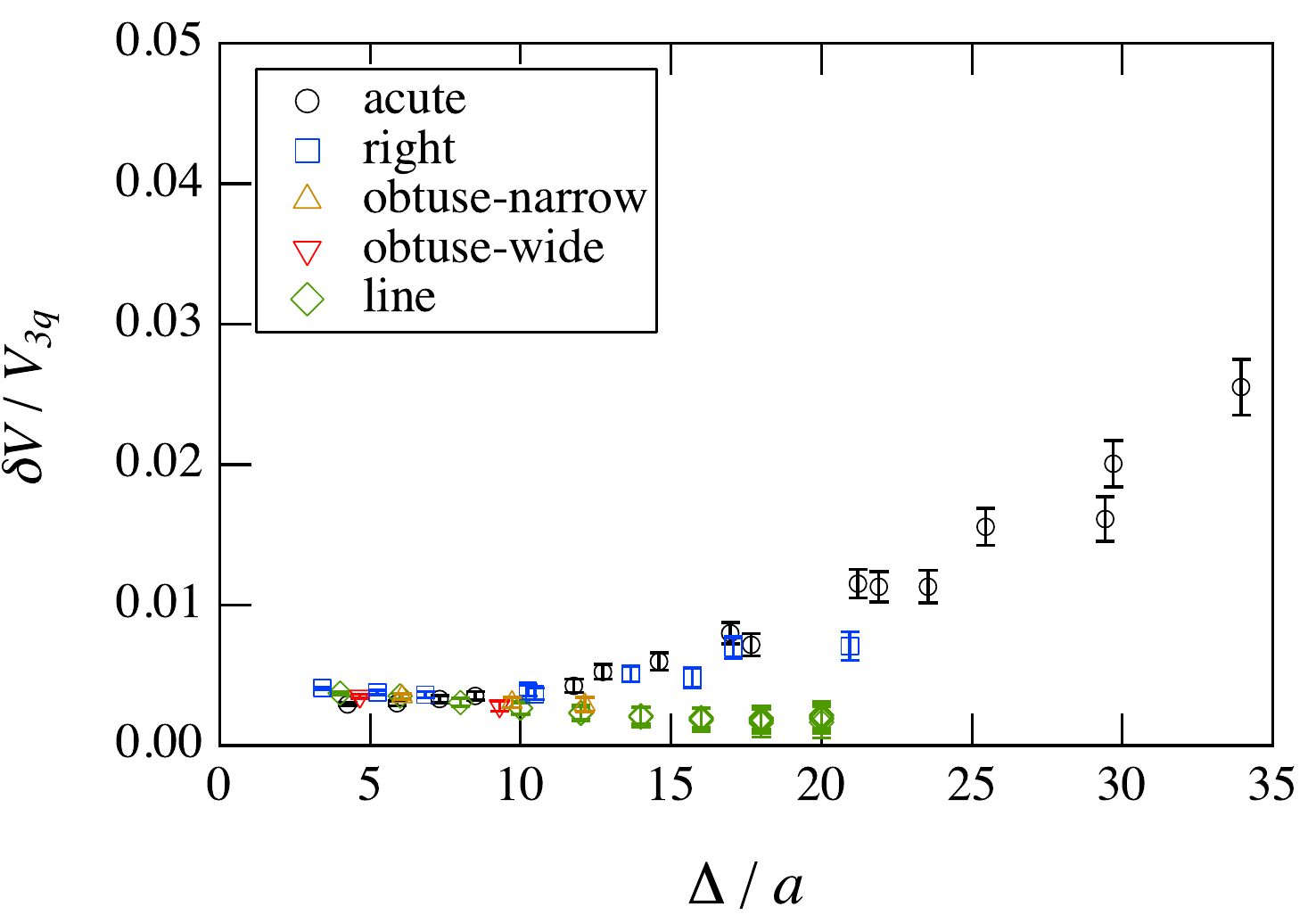}
\caption{The relative error
between three-quark potential and the half of the
sum of the quark-antiquark potential, $\delta V/V_{3q}$ against $\Delta$  
defined in Eq.~\eqref{eq:Delta}.
As the three-quark potential is from one gauge configuration with no statistical error,
the error bars in this plot are purely from that of the quark-antiquark potential from 
$N_{\rm cnf}=20$ (see, Tables~\ref{tbl:potqqbar_scale} and 
\ref{tbl:potqqbar-b600}).}
\label{fig:rratio_vs_delta}
\includegraphics[width=\figwidth]{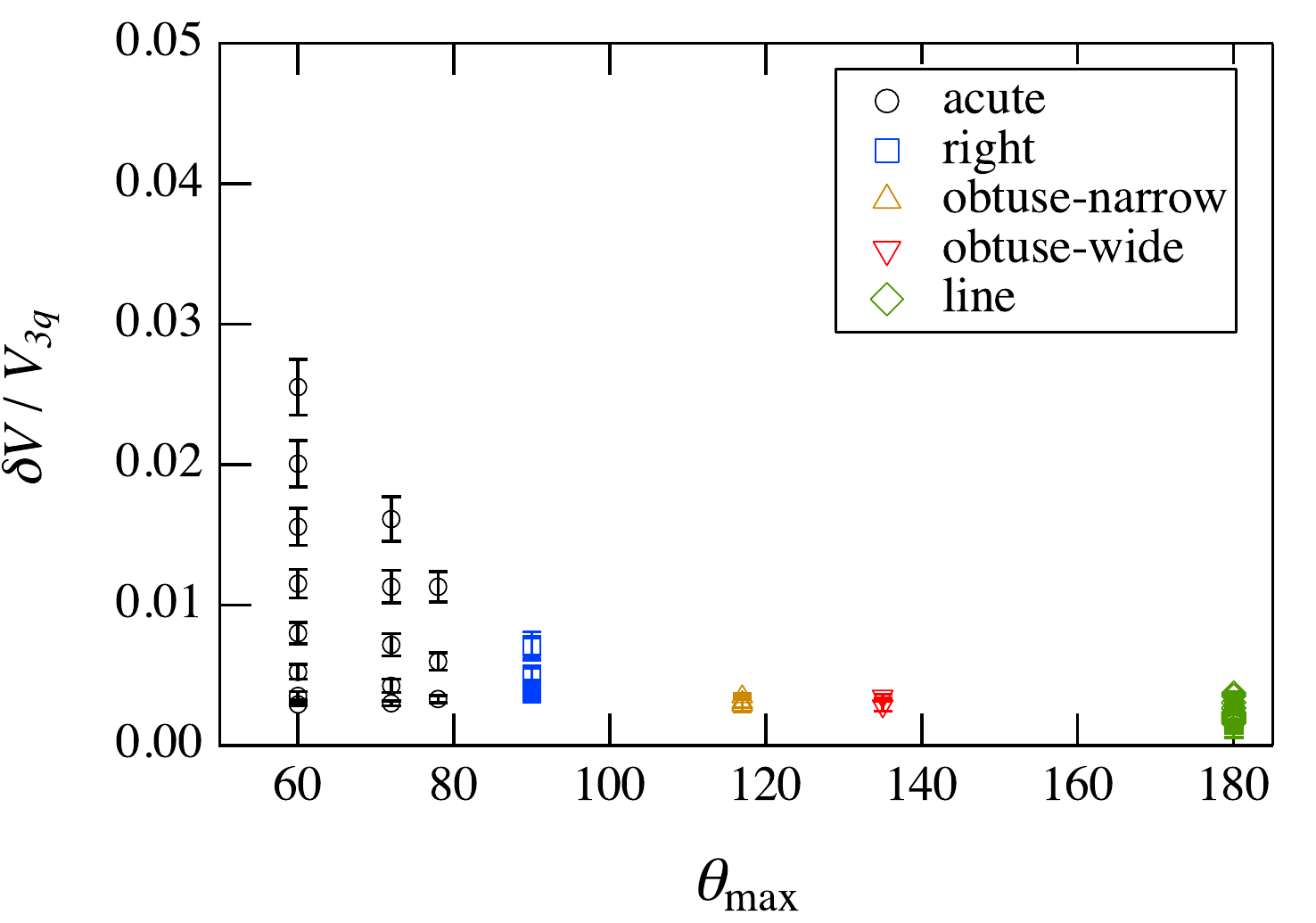}
\caption{The same plot as in Fig.~\ref{fig:rratio_vs_delta}, but 
against $\theta_{\rm max}$ defined in Eq.~\eqref{eq:thetamax}.}
\label{fig:rratio_vs_theta}
\end{figure}

\par
In Fig.~\ref{fig:rratio_vs_delta}, we plot the relative error 
between the three-quark potential and the
half of the sum of the quark-antiquark potentials at $\beta=6.00$,
\be
\frac{\delta V}{V_{3q}} =\frac{1}{V_{3q}} \left (V_{3q} (\vec{x}_{1},\vec{x}_{2},\vec{x}_{3})
- \frac{1}{2}\sum_{i=1}^{3} V_{\qqb}(r_{i}) \right )\;,
\label{eqn:pot3q-sumqqbar}
\ee
against~$\Delta$ defined in Eq.~\eqref{eq:Delta},
where we have selected only 55 three-quark geometries
out of 221 ones.
The selection is to perform the subtraction in Eq.~\eqref{eqn:pot3q-sumqqbar}
by using only the available raw data of the quark-antiquark potential 
as summarized in Tables~\ref{tbl:potqqbar_scale} 
and~\ref{tbl:potqqbar-b600} in Appendix~\ref{sect:qqbarpot}.
We have avoided the use of the fit function of 
the quark-antiquark potential in Eq.~\eqref{eqn:fit-vqdq},
which may cause unwanted systematic effects especially at short distance.

\par
We observe that the relative error is almost constant about 0.003 at $\Delta /a <  10$,
which may be understood as zero within the systematic error due to the use of 
different gauge configurations,\footnote{Note, however, that our further analysis 
using exactly the same $N_{\rm cnf}=200$ gauge configurations 
for the three-quark and the quark-antiquark potentials at $\beta=6.00$ 
with $N_{\rm iupd}=10000$
indicates that this constant still remains finite about 0.002, which
may be due to lattice cutoff effects. 
These potentials are not presented in this paper as the number of geometries is
very limited, but the data are available on request.}
where the three-quark potential is from 
one gauge configuration with $N_{\rm iupd}=500000$,
while the quark-antiquark potential is from $N_{\rm cnf}=20$ gauge configurations
with $N_{\rm iupd}=100000$.
On the other hand, the difference becomes apparent at $\Delta /a >  10$
especially for ACT and RGT, both of which show a similar increasing behavior 
up to 0.03 at $\Delta /a = 35$.
Note that the relative error 0.03 is already large enough compared to
that  from the gauge configuration dependence as demonstrated in Fig.~\ref{fig:ave-oneconf}.
In Fig.~\ref{fig:rratio_vs_theta}, we also plot the same 
relative error in Eq.~\eqref{eqn:pot3q-sumqqbar} against $\theta_{\rm max}$
defined in Eq.~\eqref{eq:thetamax}.
Clearly, some of ACT and RGT data (correspond to larger $\Delta$)
exhibit a large deviation from zero.

\begin{figure}[!t]
\includegraphics[width=\figwidth]{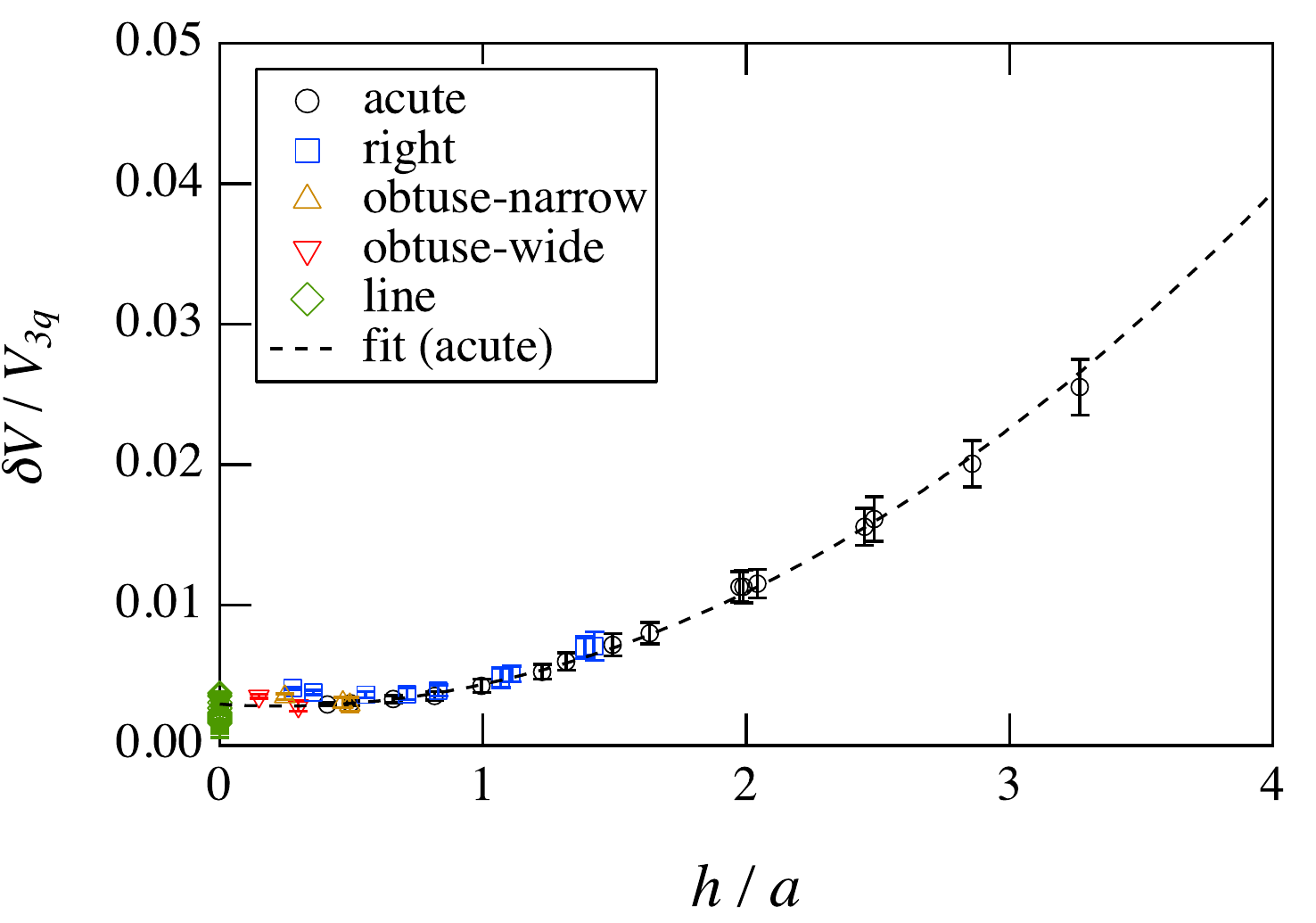}
\caption{The same plot as in Fig.~\ref{fig:rratio_vs_delta}, but 
against $h$ defined in Eq.~\eqref{eqn:distance-h}.
The fit curve is given by Eq.~\protect{\eqref{eq:deltaV}}.}
\label{fig:rratio_vs_h}
\end{figure}

\par
The above two results seem to indicate that the large deviation in ACT and RGT 
originates from the formation of the flux-tube junction, since
only these configurations are possible to form a proper junction at larger triangle.
On the other hand, even in ACT and RGT, the flux-tube junction
cannot be formed properly when the size of the triangle is insufficient.
It seems that there is a critical size of the triangle to form a proper flux-tube junction.
Based on this expectation, in Fig.~\ref{fig:rratio_vs_h},
we then plot the relative error in Eq.~\eqref{eqn:pot3q-sumqqbar}
against the distance $h$ defined in Eq.~\eqref{eqn:distance-h},
the averaged distance between the
Fermat-Torricelli point and three sides of a triangle.
Note that $h=0$ for LIN, while $h>0$ for the other triangles.
We find that the relative errors of 
all the three-quark potential are well parametrized by $h$.
The dashed line corresponds to a fit curve for the ACT data,
which is an  empirical quadratic function given by
\be
\frac{\delta V}{V_{3q}}= c_{0} +c_{1}(h/a)+c_{2}(h/a)^{2} \;,
\label{eq:deltaV}
\ee
where $c_{0}=0.0029(3)$, $c_{1}=-0.0011(7)$, and $c_{2}=0.0026(3)$.
The relative errors seem to start increasing around $h/a \sim 0.6$.

\par
The difference between the  three-quark potential and 
the half of the sum of the quark-antiquark potentials
signals an existence of three-body force among the three quarks.
Our results in this subsection indicate that 
the difference is not so drastic when the size of the triangle
is small or $\theta_{\rm max}$ approaches 180$^{\circ}$,
while it shows up gradually for larger ACT and RGT.
In other words, it seems that the emergence of the three-body effects
depends on whether 
the interquark distance among three quarks
is large enough to
form a flux-tube junction inside the triangle.

\subsection{The functional form of the three-quark potential}
\label{subsec:functional-form}

The analysis in Sec.~\ref{subsec:st} suggests that
the long distance part of the three-quark potential can be described by the term
$\sigma_{3q} L_{\rm str}$,
where the coefficient $\sigma_{3q}$
is common to that of the quark-antiquark potential $\sigma_{\qqb}$.
In addition, a naive fit result of the three-quark potential of the equilateral triangle  geometries
in Sec.~\ref{subsec:oneconf} as well as the analysis in Sec.~\ref{subsec:h}
have indicated that the potential also contains a constant term,
which is approximately
$3/2$ of that in the quark-antiquark potential $\mu_{\qqb}$.
The presence of such a constant term in the three-quark potential is 
quite natural in terms of the self-energy of quarks, 
which will be proportional to the number 
of quarks involved in the system.

\par
\begin{figure}[t]
\includegraphics[width=\figwidth]{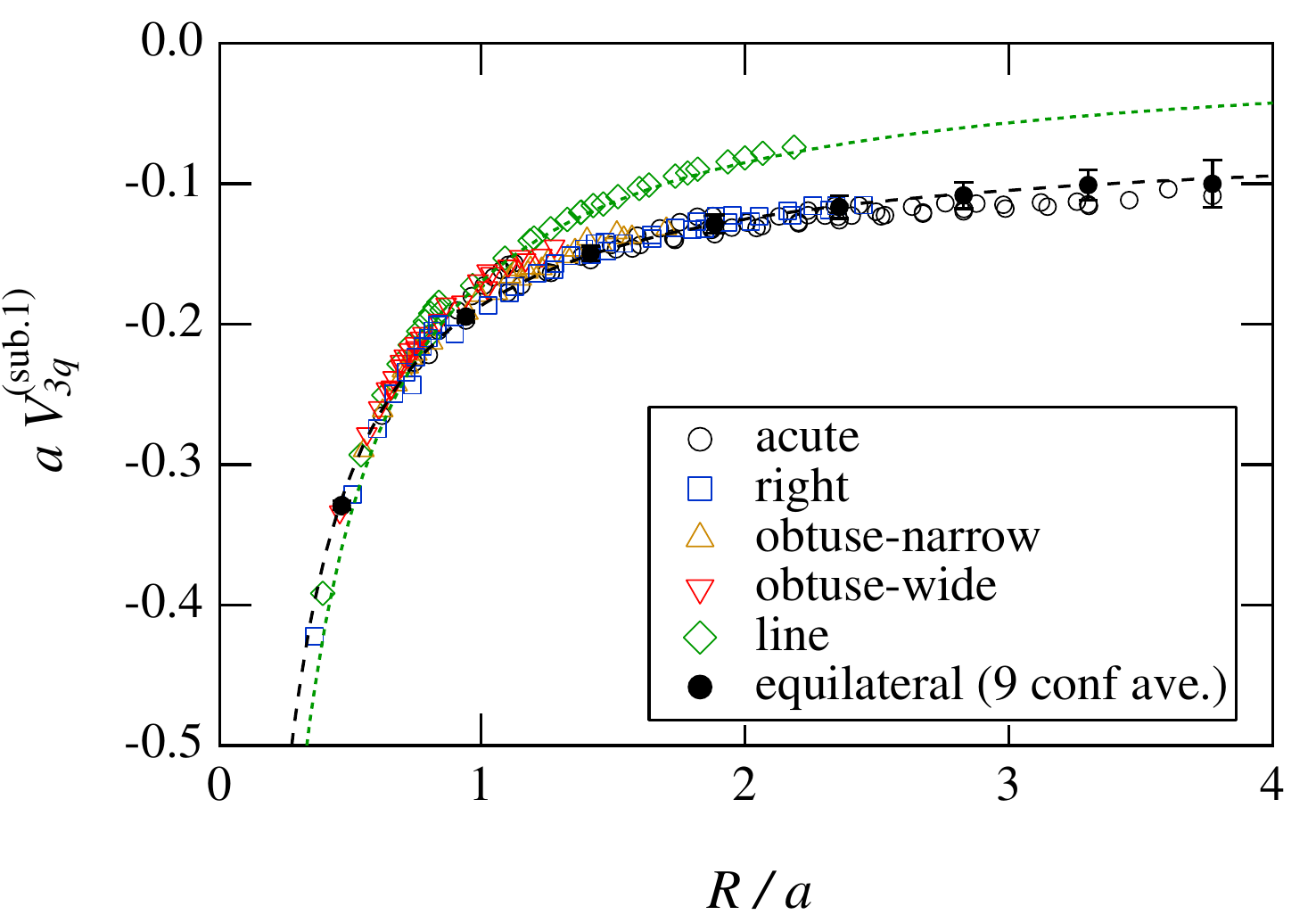}
\caption{The three-quark potential after subtracting the confinement and constant terms 
$V_{3q}^{\rm (sub.1)}$ in Eq.~\eqref{eq:vr} against the reduced distance $R$ defined in Eq.~\eqref{eq:defR}.
The open symbols are the three-quark potential from one gauge configuration at $\beta=6.00$,
and the filled circles with error bars are from 9 gauge configurations
of the equilateral triangle geometries.
The dotted line corresponds to $ -A_{\qqb}/(2R)$,
and the dashed line the fit curve given by Eq.~\eqref{eq:vr-form}.}
\label{fig:vsR_all}
\end{figure}

We then investigate the behavior of the rest of the 
potential by subtracting the confinement term $\sigma_{\qqb}L_{\rm str}$ 
and an expected constant term $(3/2)\mu_{\qqb}$ from the three-quark potential,
\be
V_{3q}^{\rm (sub.1)} = 
V_{3q} - (\sigma_{\qqb}L_{\rm str} + \frac{3}{2}\mu_{\qqb})\; ,
\label{eq:vr}
\ee
which will be useful to clarify the short distance part of the potential.
In Fig.~\ref{fig:vsR_all}, we plot Eq.~\eqref{eq:vr} against $R$ defined in Eq.~\eqref{eq:defR},
where the three-quark potential used in this analysis is 
the same as that used in the previous subsections at $\beta=6.00$.
From the quark-antiquark potential we take the values
$\sigma_{\qqb}a^{2} =0.0449(2)$ and $\mu_{\qqb}a =0.766(1)$
(see, Table~\ref{tbl:fit-potqqbar} in Appendix~\ref{sect:qqbarpot}).
Remarkably, we find two systematic 
curves:\footnote{Although we attempted  to plot  $V_{3q}^{\rm (sub.1)}$ with $\Delta$ and $L_{\rm str}$, 
we failed to see any systematic behaviors.}
one is mostly for LIN and the other is for triangle 
geometries 
although both seem to overlap at $R/a < 1$.
The data of LIN are well described by a function $-A_{\qqb}/(2R)$
(the dotted line in the figure), 
where $A_{\qqb}=0.340(2)$ is also from Table~\ref{tbl:fit-potqqbar},
which may not be surprising after the analysis in Sec.~\ref{subsec:h}.
Since $L_{\rm str}= \Delta / 2$ for LIN, 
the difference of the potential from the half of the sum of the 
quark-antiquark potential is at most 0.3~\% relative 
error as explicitly shown in Fig.~\ref{fig:rratio_vs_delta}.
What we should pay attention to is then the behavior of the other data for triangle 
geometries,
which clearly deviates from the function $ -A_{\qqb}/(2R)$.
Moreover, it seems that the curve approaches a negative constant value at large $R$.
We then perform an empirical $\chi^{2}$ fit to the functional form, 
\be
V_{3q}^{\rm (R)} = -\frac{A_{3q}^{\rm (R)}}{R} + \tilde{\mu}_{3q}\,,
\label{eq:vr-form}
\ee 
which yields $A_{3q}^{\rm (R)} = 0.126(3)$ and $\tilde{\mu}_{3q} a= - 0.062(4)$
with $\chi^{2}/N_{\rm df} =0.04$, where the averaged potential
of the equilateral triangle configuration with error bars is taken into account in the fit.
It turns out that Eq.~\eqref{eq:vr-form} nicely captures 
the increasing behavior of $V_{3q}^{\rm (sub.1)}$ (the dashed line in the figure).
We find that $A_{3q}^{\rm (R)}$ is significantly smaller than $A_{\qqb}/2=0.170$ 
about 26~\%, namely,
\be
A_{3q}^{\rm (R)} =\dfrac{A_{\qqb}}{2}(1 - 0.259 )\; ,
\ee
and $\tilde{\mu}_{3q}$ has a negative value as expected.
The value $0.259$ may be further tuned by performing a more sophisticated fit.
The absolute value of $\tilde{\mu}_{3q}$ is one order of magnitude 
smaller than $\mu_{\qqb}$, but still seems to be finite.

\par
\begin{figure}[t]
\includegraphics[width=\figwidth]{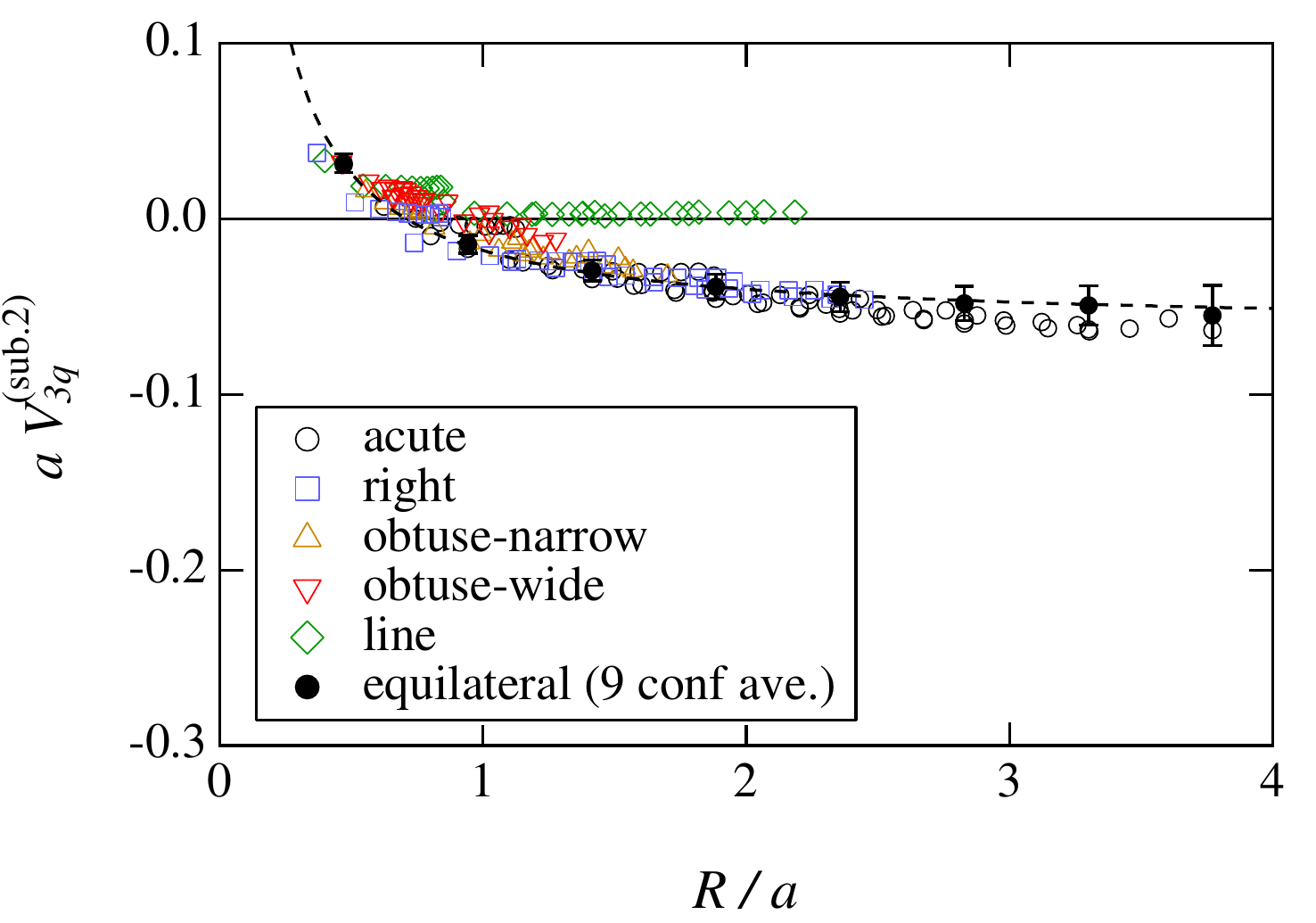}
\caption{The same plot as in Fig.~\ref{fig:vsR_all}, but the term $-A_{\qqb}/(2R)$
is further subtracted from the three-quark potential ($V_{3q}^{\rm (sub.2)}$ in Eq.~\eqref{eq:vr-v2}).
The dashed line corresponds to $0.259A_{\qqb}/(2R)+ \tilde{\mu}_{3q}$.}
\label{fig:pot_remnant_h0}
\end{figure}

One may suspect at this stage that
the existence of $\tilde{\mu}_{3q}$ just reflects a lattice artifact at $\beta=6.00$,
however, the following analysis in Sec.~\ref{subsec:scaling}
shows a kind of scaling behavior on $\tilde{\mu}_{3q}$, 
which implies that $\tilde{\mu}_{3q}$ reflects a physical effect.
Since the negative shift of the energy at large $R$ appears only for triangle
geometries, 
we speculate that it just represents an energy reduction due to 
formation of the flux-tube junction.
Of course, if this is the case, the energy reduction could depend on the size of the triangle.
This feature seems to be incorporated effectively by rewriting Eq.~\eqref{eq:vr-form} as
\be
V_{3q}^{\rm (R)} = -\dfrac{A_{\qqb}}{2R} + 
\left ( 0.259\cdot \dfrac{A_{\qqb}}{2R} + \tilde{\mu}_{3q} \right )\, ,
\label{eq:vr-form-v2}
\ee
where the first term purely represents two-body interaction between quarks
while the terms inside parenthesis are interpreted as the three-body junction effect.
In Fig.~\ref{fig:pot_remnant_h0}, we plot 
\be
V_{3q}^{\rm (sub.2)} = V_{3q} - (- \dfrac{A_{\qqb}}{2R} + \sigma_{\qqb}L_{\rm str} + \frac{3}{2}\mu_{\qqb} ) \;,
\label{eq:vr-v2}
\ee
against $R$.
Although it is not obvious in Fig.~\ref{fig:vsR_all} 
whether OBTW belongs to LIN or triangle category, the plot
in Fig.~\ref{fig:pot_remnant_h0}
seems to support the latter.
The change of sign from positive to negative values around $R/a \sim 0.7$ 
may reflect the fact that forming a simple $Y$-shaped junction
is rather costly for these smaller $R$,
implying that a different type of flux structure is realized.
It would be quite interesting to investigate the energy density 
of  various three-quark systems with the PLCF.

\par
To summarize, the functional form of the three-quark potential 
for triangle geometries
 is effectively parametrized by 
\be
V_{3q}  =
 - \frac{A_{3q}^{\rm (R)}}{R} + \sigma_{\qqb} L_{\rm str} + \frac{3}{2}\mu_{\qqb}+ \tilde{\mu}_{3q}\;.
\label{eq:finalform-case1}
\ee
On the other hand, the functional form for LIN can be described 
by the half of the sum of the quark-antiquark potential,
\be
V_{3q} =\dfrac{1}{2}\displaystyle{\sum_{i=1}^{3}} V_{\qqb} (r_{i}) \; ,
\label{eq:finalform-case2}
\ee
up to the tiny 0.3~\% relative error as shown in Sec.~\ref{subsec:h}.
We cannot answer here whether or not both functional forms can change continuously depending 
on the movement of quarks, which should be clarified in future study.
It is certainly important for this purpose to see the potential of OBTW at long distance in detail.
The functional form in Eq.~\eqref{eq:finalform-case1} may be similar to that proposed by
Takahashi~{\it et\,al.}~\cite{Takahashi:2000te,Takahashi:2002bw} for triangle 
geometries,
where the potential was parametrized by 
${\displaystyle V_{3q}= -A_{\qqb}/(2R)+\sigma_{\qqb} Y + {\it constant}}$.
However, our result shows a noticeable difference from it
in terms of the junction term in Eq.~\eqref{eq:vr-form-v2}.

\subsection{Scaling test of the three-quark potential}
\label{subsec:scaling}

\par
So far we have concentrated on analyzing the potential at  $\beta=6.00$,
and have found the possible functional form of the potential  as in
Eqs.~\eqref{eq:finalform-case1} and~\eqref{eq:finalform-case2}.
We finally examine whether the functional form is still valid
even if the lattice spacing decreases or increases.
We then computed the  three-quark potential at $\beta=5.85$ and~$6.30$.
We only pay attention to the potential of  equilateral triangle 
geometries
because of the limited computer resources,
however, it will provide some hints concerning the scaling behavior of the potential.
The number of gauge configurations to compute the final
expectation values are not so many, 
but the statistical errors are highly suppressed by employing the multilevel algorithm
with the tuned parameters,
where $N_{\rm cnf} =8$ at $\beta=5.85$  with $N_{\rm iupd} =500000$,
and $N_{\rm cnf} =6$ at $\beta=6.30$ with $N_{\rm iupd} =400000$.
Typical convergence histories of the PLCF  of the equilateral triangle 
geometry
at $\beta=5.85$ and  $\beta=6.30$ are already shown in Fig.~\ref{fig:hist-plcf}.

\begin{figure}[!t]
\includegraphics[width=\figwidth]{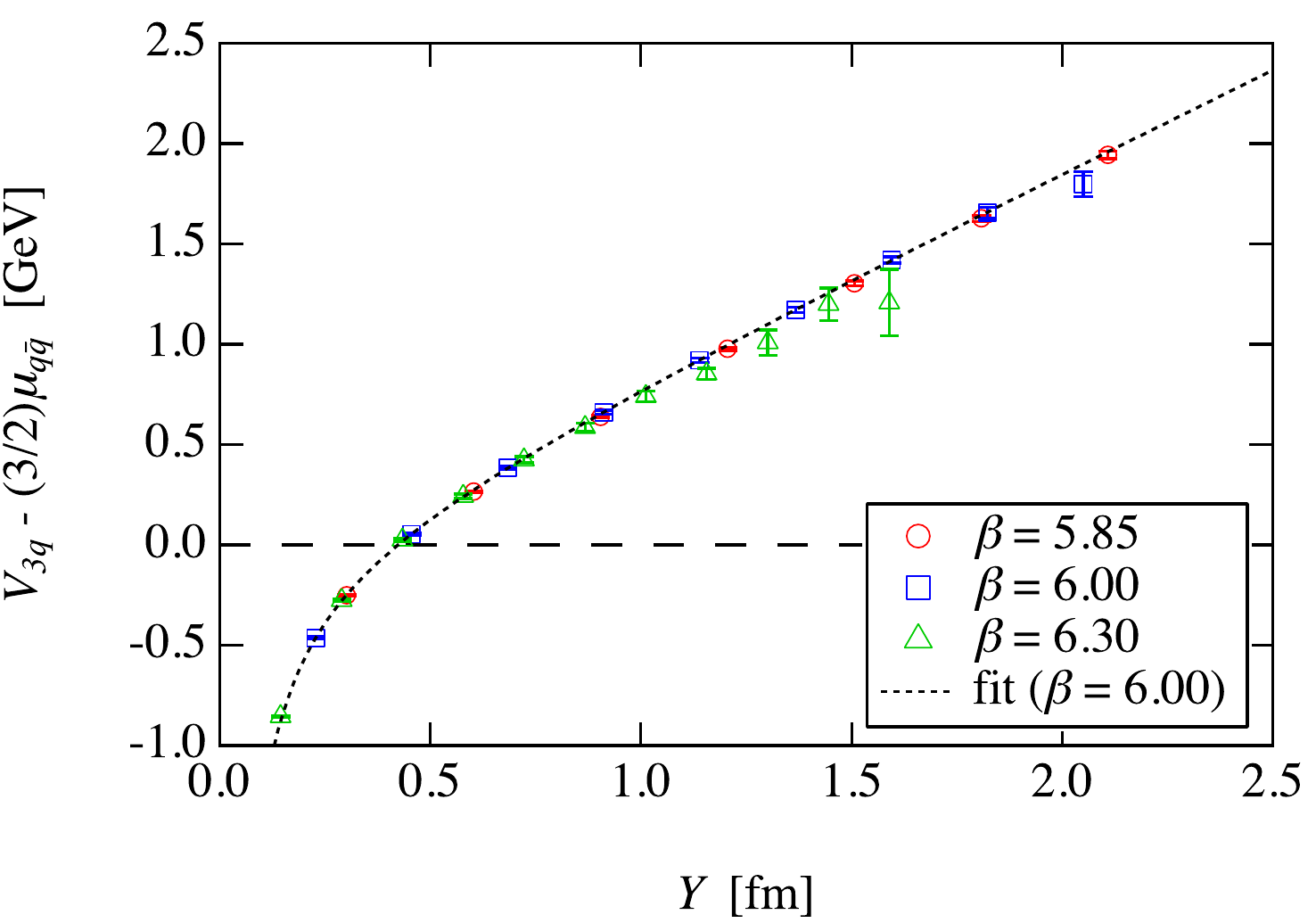}
\caption{The three-quark potential  as a function of $Y$ in physical unit, where the 
constant $(3/2)\mu_{\qqb}$ is subtracted.
The dotted line corresponds to the fit curve to Eq.~\eqref{eqn:fit-v3q-y}, which yields
$A_{3q}^{\rm (Y)}\hbar c = 0.131(4)\,{\rm [GeV\,fm]}$,  $\sigma_{3q} = 1.02(1)\,{\rm [GeV/fm]}$,
and $\tilde{\mu}_{3q} = -0.12(2)\,{\rm [GeV]}$ with $\chi^{2}/N_{\rm df}=0.02$.} 
\label{fig:pot3q_scale}
\includegraphics[width=\figwidth]{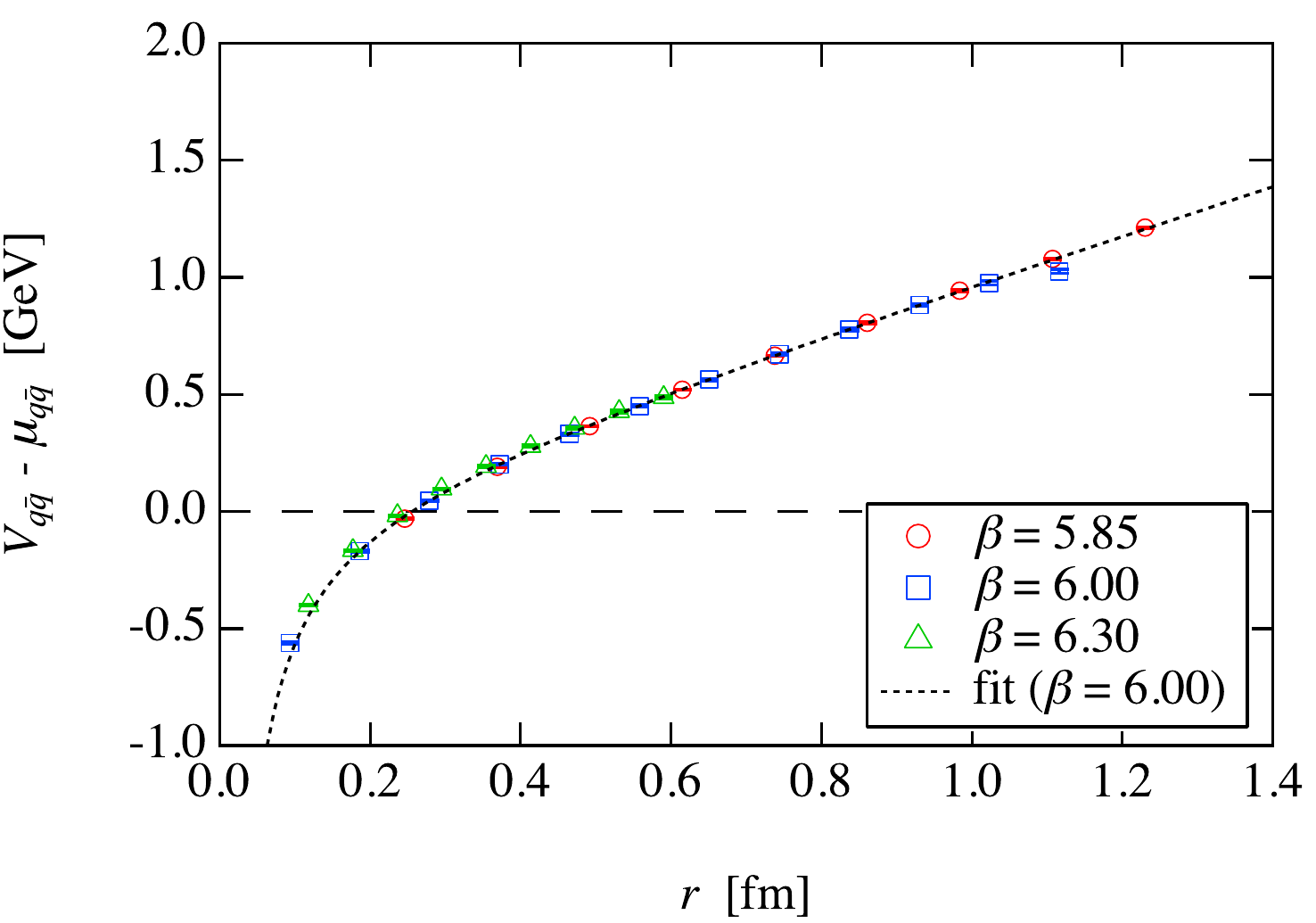}
\caption{The quark-antiquark potential as a function of $r$ in physical unit, 
where the constant $\mu_{\qqb}$ is subtracted.
The dotted line is the fit curve to 
$V_{\qqb}(r) =-A_{\qqb}\hbar c/r  +\sigma_{\qqb}\,  r $,
which yields
$A_{\qqb}\hbar c =0.0671(6)\,{\rm [GeV\,fm]}$,  $\sigma_{\qqb} = 1.025(4)\,{\rm [GeV/fm]}$
with $\chi^{2}/N_{\rm df}=0.013$.}
\label{fig:potqqbar_scale}
\end{figure}

\begin{figure}[t]
\includegraphics[width=\figwidth]{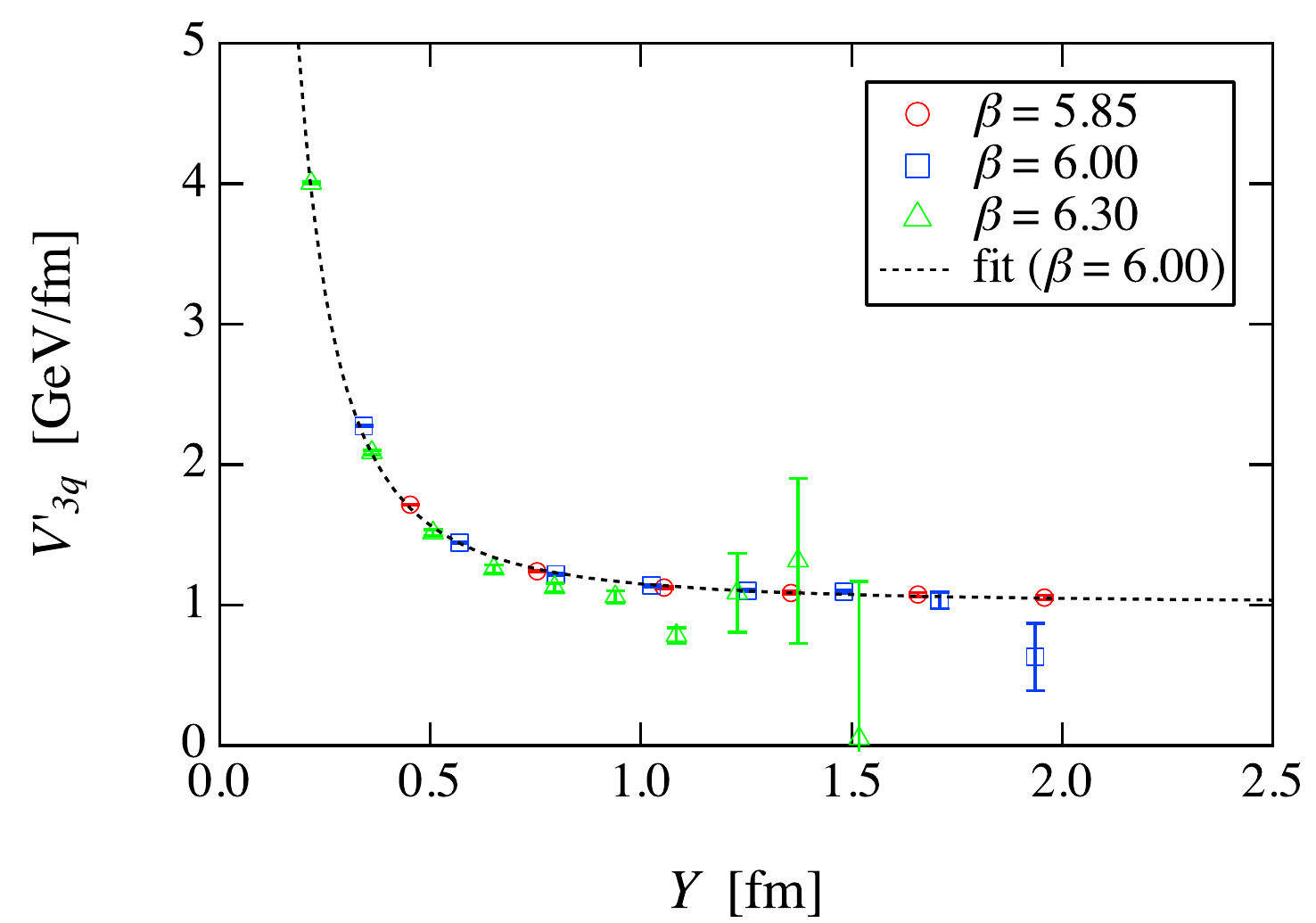}
\caption{The derivative of the potential with respect to $Y$ in physical unit.
The dotted line corresponds to the fit curve to the derivative of Eq.~\eqref{eqn:fit-v3q-y} with respect to $Y$, which yields
$A_{3q}^{\rm (Y)}\hbar c = 0.139(4)\,{\rm [GeV\,fm]}$,  $\sigma = 1.013(7)\,{\rm [GeV/fm]}$ with $\chi^{2}/N_{\rm df} = 2.0$.}
\label{fig:frc3q_ave_sc}
\end{figure}

\par
In Fig.~\ref{fig:pot3q_scale}, we plot the potential as a function of $Y$,
where physical scales are introduced according to  Table~\ref{tbl:simulation}
 (the raw data and the fit results  are summarized in 
 Tables~\ref{tbl:pot3q_regular_scale} and~\ref{tbl:fit-pot3q-regular}).
In addition, the constant term $(3/2)\mu_{\qqb}\,a$
 is subtracted before we introduce the physical scale for each $\beta$ value.
The constant term $\mu_{\qqb}$  is extracted from  the quark-antiquark potential 
as summarized in Table~\ref{tbl:fit-potqqbar},
which reflects the self-energy of a quark and an antiquark,
and diverges in the continuum limit $a\to 0$.
We find that the potential beautifully  falls into one curve, 
\be
V_{3q}^{\rm (Y)} =-\frac{A_{3q}^{\rm (Y)}\hbar c}{Y} 
+ \sigma_{3q} Y + \tilde{\mu}_{3q}
 \;,
\label{eqn:fit-v3q-y}
\ee
indicating a scaling behavior with respect to 
the lattice spacing.\footnote{When we put three quarks at $(x,0,0)$, $(0,x,0)$, and $(0,0,x)$, 
the distance between the two of three quarks is $r= \sqrt{2}x$
and $Y=\sqrt{6}r = 2\sqrt{3}x$.
On the other hand, $R =\sqrt{2}x/3$, and then $R=\sqrt{3} Y/9 =0.19245 Y$.
Thus, the function $1/Y$ and $1/R$ is the same 
for equilateral triangles up to a multiplicative factor.}
The quark-antiquark potential used in this analysis also exhibits  a
good scaling behavior as shown 
in Fig.~\ref{fig:potqqbar_scale} (the raw data and the fit results  
are summarized in Tables~\ref{tbl:potqqbar_scale} and~\ref{tbl:fit-potqqbar}
 in Appendix~\ref{sect:qqbarpot}).
Comparison of the potentials at various $\beta$ values may be affected by 
the way of subtraction of the constant $\mu_{3q}$.
This uncertainty can however be avoided by 
looking at the derivative of the potential with respect to $Y$,
which is independent of the constant $\mu_{3q}$.
The result is  plotted in Fig.~\ref{fig:frc3q_ave_sc} and
the data fall into a curve given by the derivative of Eq.~\eqref{eqn:fit-v3q-y} with respect to~$Y$.

\par
\begin{figure}[t]
\includegraphics[width=\figwidth]{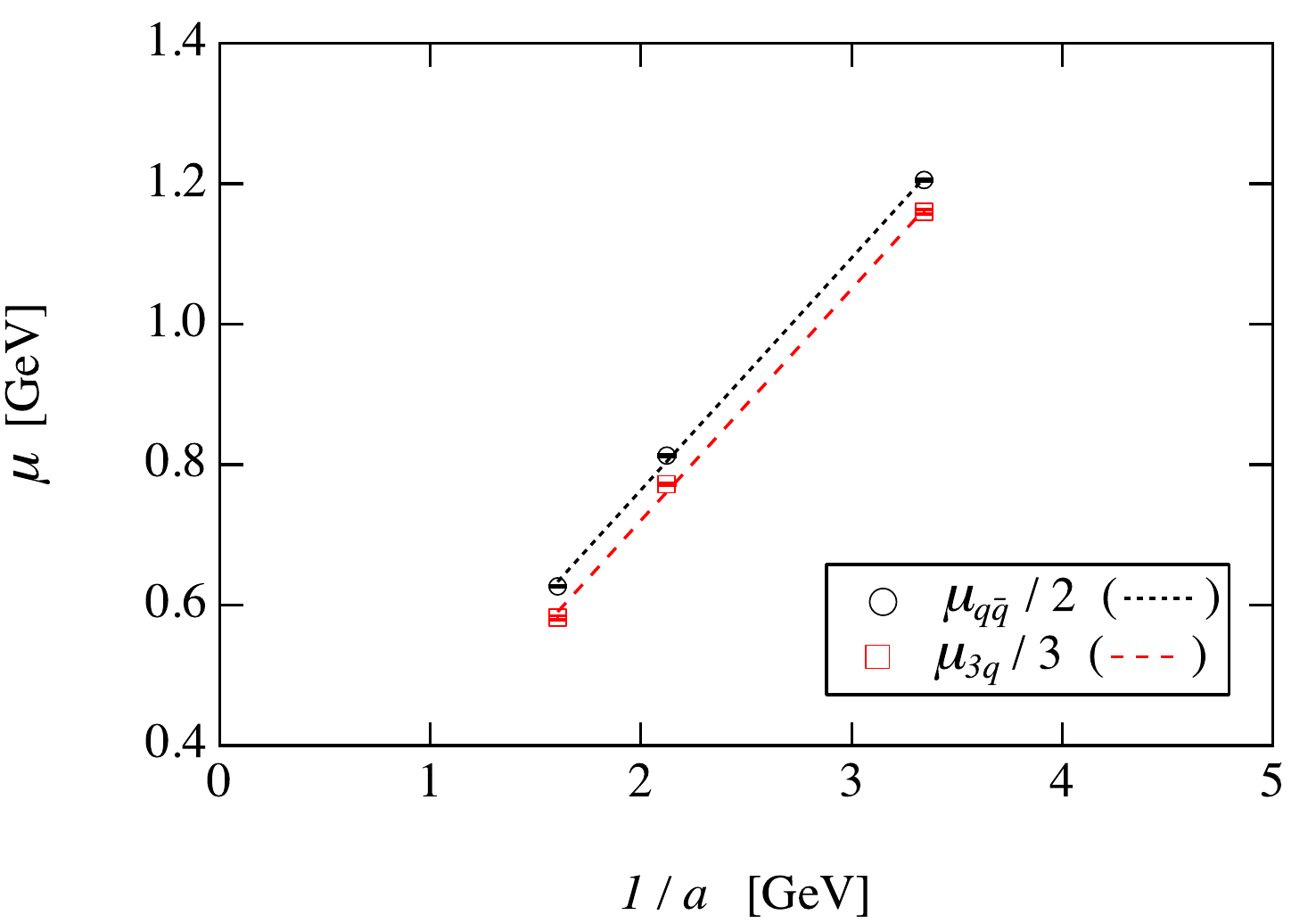}
\caption{The scale dependence of the constant term 
of the quark-antiquark and  three-quark potentials per one quark.
A naive fit to a linear function yields the slope $0.33$ for both cases.}
\label{fig:mu_onequark_diff_sc}
\end{figure}
\begin{figure}[t]
\includegraphics[width=\figwidth]{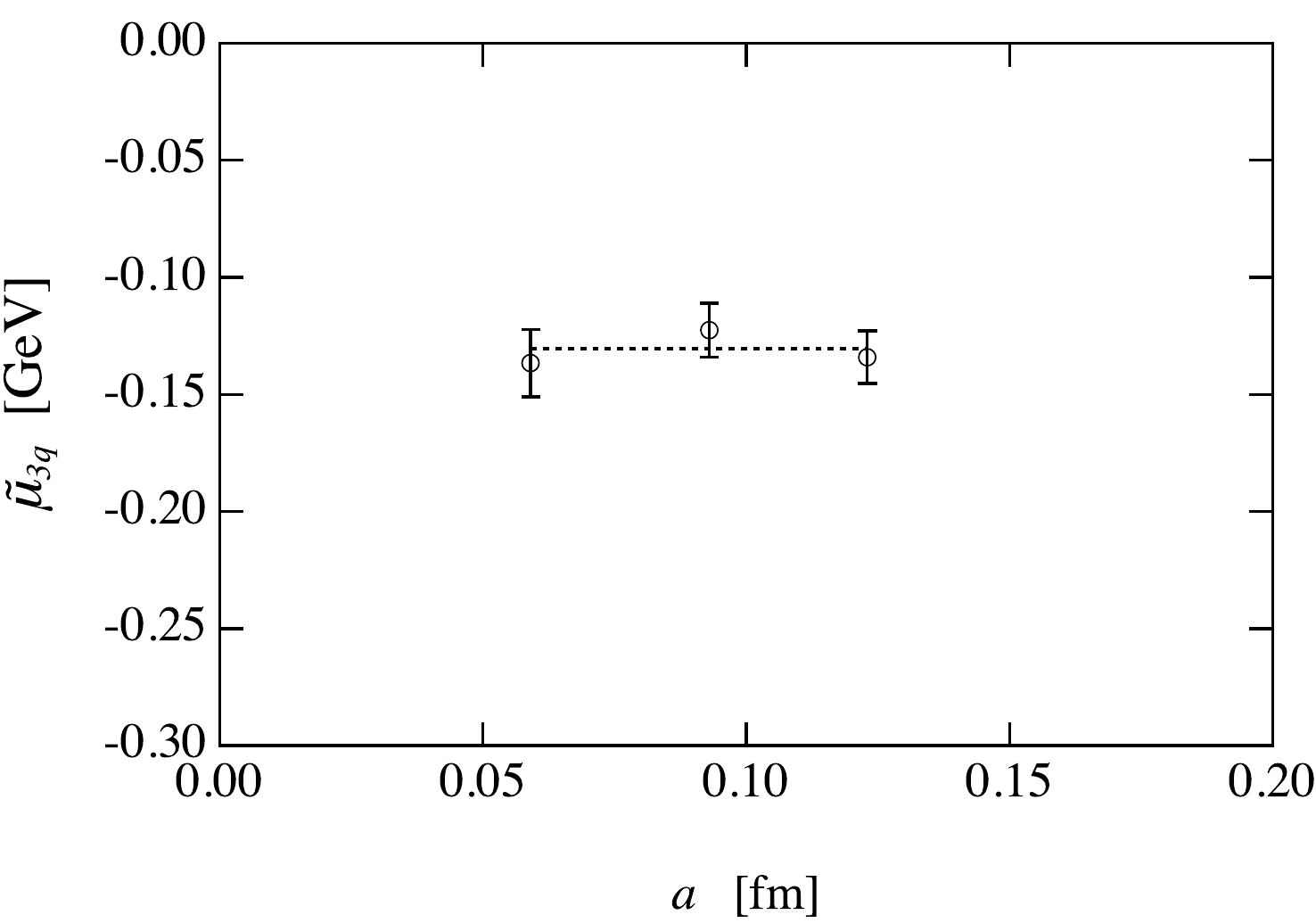}
\caption{The scaling behavior of the 
remnant of the constant term of the three-quark potential $\tilde{\mu}_{3q}$
in Eq.~\eqref{eqn:mu3q-remnant}.
A naive fit to a constant value  yields $\tilde{\mu}_{3q} = -0.131(7)\,{\rm [GeV]}$.}
\label{fig:mu_diff_sc_v2}
\end{figure}

We then look at the scaling behavior of the constant term in 
the three-quark potential.
Although it has already been suggested in Sec.~\ref{subsec:functional-form}
that the constant~$\mu_{3q}$ consists of 
two contribution $(3/2)\mu_{\qqb}$ and the remnant~$\tilde{\mu}_{3q}$,
\be
\mu_{3q} =\frac{3}{2} \mu_{\qqb} +  \tilde{\mu}_{3q} \;,
\label{eqn:mu3q-remnant}
\ee
we simply compare the behavior of $\mu_{\qqb}/2$ and $\mu_{3q}/3$ in physical unit,
which is plotted in Fig.~\ref{fig:mu_onequark_diff_sc} as a function of the inverse of the lattice spacing~$1/a$.
If the constant originates only from the divergent self-energies of quarks,
both should coincide with each other.
As can be seen, the increasing behavior is the same as $1/a$ increases, however, 
there is a difference by a constant, which seems to be independent of~$a$.
Indeed, as we explicitly show in Fig.~\ref{fig:mu_diff_sc_v2},
the remnant of the constant $\tilde{\mu}_{3q}$ seems to exhibit a scaling behavior against~$a$.
As already discussed in Sec.~\ref{subsec:functional-form},
$\tilde{\mu}_{3q}$ is absent for LIN and appears only
when the three quarks form a triangle.
Therefore, it would be quite reasonable to  understand that this
is caused by the formation of the flux-tube junction, which
can reduce the total energy of the three-quark system.

\begin{figure}[t]
\includegraphics[width=\figwidth]{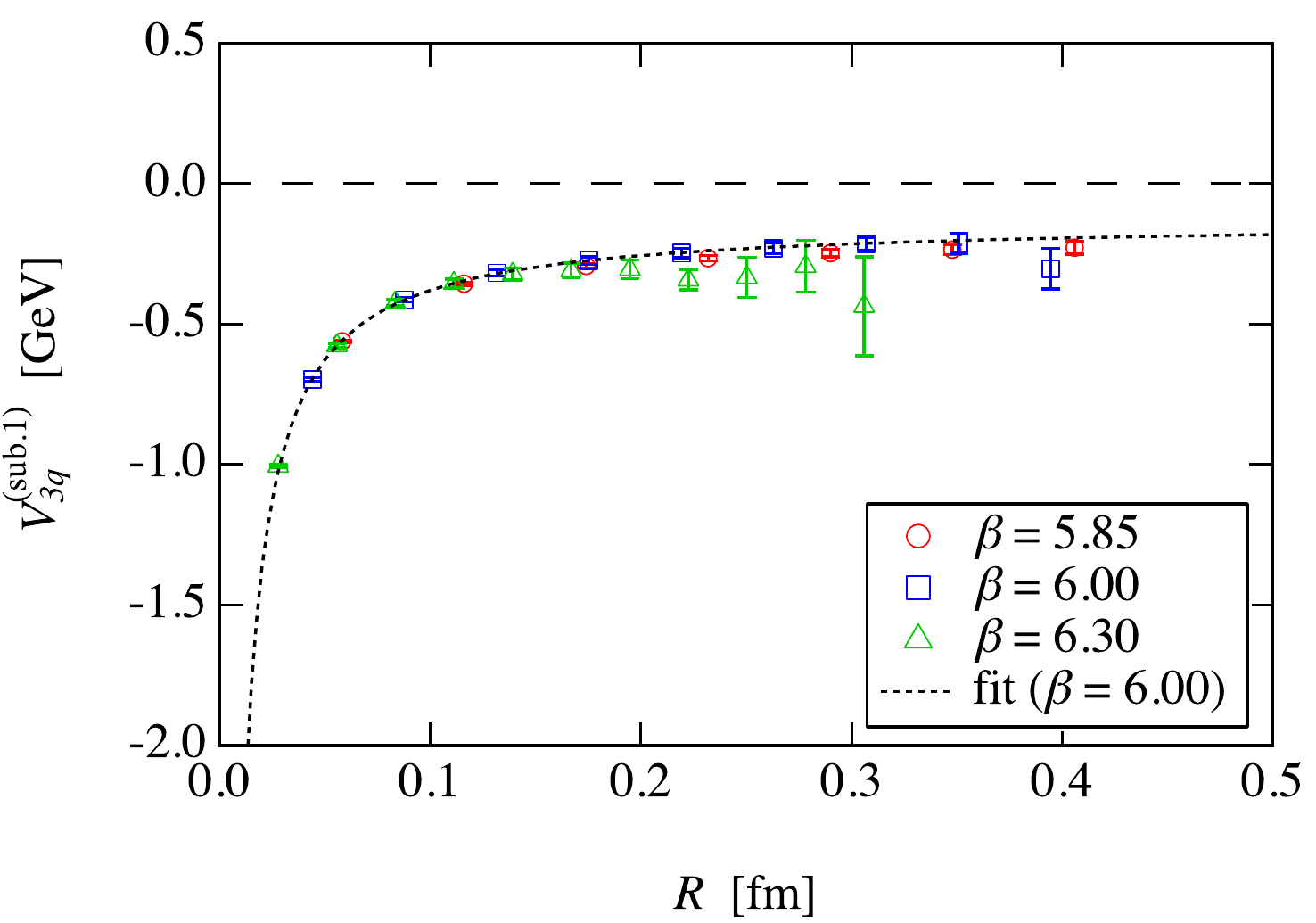}
\caption{The three-quark potential as a function of 
$R$ in physical  unit 
after subtracting the confinement  and divergent constant terms
expected from the quark-antiquark potential, $V_{3q}^{\rm (sub.1)}$.
The dotted line corresponds to the fit curve to 
$V_{3q}^{\rm (R)} =-A_{3q}^{\rm (R)}\hbar c/R +\tilde{\mu}_{3q}$ in Eq.~\eqref{eq:vr-form},
which yields $A_{3q}^{\rm (R)}\hbar c =0.0248(5)\,{\rm [GeV\,fm]}$
and $\tilde{\mu}_{3q} =-0.132(8)\,{\rm [GeV]}$ with 
$\chi^{2}/N_{\rm df} = 0.04$.}
\label{fig:pot3q_scale_subt}
\end{figure}

\par
Finally, in Fig.~\ref{fig:pot3q_scale_subt},
we plot the three-quark potential as a function of  $R$ in physical unit 
after subtracting the confinement and divergent constant terms
expected from the quark-antiquark potential, $V_{3q}^{\rm (sub.1)}$ in Eq.~\eqref{eq:vr}.
The behavior of the potential is nicely described by the functional form
in Eq.~\eqref{eq:vr-form}.
Although we cannot exclude the possibility of 
other parameterizations
 with different distances,
such as $Y$, which is different from $R$ only by a  multiplicative factor
for the equilateral triangle 
geometries,
this plot and the detailed analysis of the potential at $\beta=6.00$ in 
Fig.~\ref{fig:vsR_all} indicate that the parametrization with  $R$  
also works well for other three-quark geometries
at $\beta=5.85$ and~$6.30$.

\section{Summary}
\label{sec:summary}

\par
We have investigated the static interquark potential 
for the three-quark system, the  three-quark potential,
in SU(3) lattice gauge theory at zero temperature by using Monte Carlo simulations.
The crucial difference of our study from earlier ones by other groups
is that we have used the Polyakov loop correlation function (PLCF)
composed of the three Polyakov loops as the three-quark source instead of
the three-quark Wilson loop, and thus,
our results are not contaminated by systematic effects due to  the spatial Wilson lines. 
By employing the multilevel algorithm extensively,
we have then obtained remarkably accurate data on the potential for $O(200)$ sets of the
three-quark geometries,
which include not only the cases  that  three quarks are put at the 
vertices of  acute (ACT),  right (RGT), and obtuse (OBTN and OBTW) triangles,
but also the extreme cases such that  three quarks are put in line (LIN).
As a special case, we have also investigated the quark-diquark (QDQ) potential.

\par
What we have shown on the three-quark potential is summarized as follows.
\begin{enumerate}
\itemsep=0cm

\item The potentials of ACT, RGT, OBTN with $r_{\rm min}/a > 2$
plotted against  $L_{\rm str}$ can fall into one curve, which 
show the same linearly-rising behavior at long distance as in the quark-antiquark potential.

\item
The potential of QDQ is identical to the quark-antiquark 
potential as in Eq.~\eqref{eqn:fit-vqdq} except for the constant shift.

\item The string tension of the potential, 
identified as the coefficient in front of $L_{\rm str}$,
is common to that of the quark-antiquark potential
 (we have shown this without a fit procedure).

\item 
The potentials of triangle
geometries
 are clearly different from 
the half of the sum of the two-body quark-antiquark potential 
when the size of the triangle becomes larger,
which can be described by the functional form in Eq.~\eqref{eq:finalform-case1}.

\item 
The potential of LIN 
is very close to the half of the sum of the two-body quark-antiquark potentials,
which can be described by the functional form  in Eq.~\eqref{eq:finalform-case2}
(up to a tiny relative error about 0.3~\% at $\beta=6.00$).

\item  The potential, its derivative, and the remnant of the constant term, which we call the junction term,
show good scaling behaviors with respect to the lattice spacing
(although only the equilateral triangle
geometries have  been examined).

\end{enumerate}

\par
It seems that there is no unique functional form of the potential which covers 
all three-quark geometries,
which in turn implies that the potential is very sensitive
not only to the location of three quarks but also to
the nontrivial flux-tube structure spanned among the three quarks.
The functional forms that we have successfully categorized
into three types as in Eqs.~\eqref{eqn:fit-vqdq},~\eqref{eq:finalform-case1}, and~\eqref{eq:finalform-case2}
clearly indicate this feature.
Probably, the potential of OBTW, which has not been addressed fully in the present study,
may play an intermediate role between ACT-RGT-OBTN and LIN.
It must be quite important to look at the distribution of  energy density 
in the three-quark system at zero temperature by using the PLCF 
as in the  finite temperature case~\cite{Bornyakov:2004yg,Bakry:2014gea,Bakry:2015csa}.

\par
In order to apply the three-quark potential to baryon spectroscopy, 
on the other hand, 
it would be desirable to have a unique functional form.
This may be possible by introducing a kind of a form factor. 
One of the practical ideas may be to extend 
the functional form that we have found further by  using
the distance such as $h$, an averaged distance between the Fermat-Torricelli 
point and three sides of a triangle.
As  we have demonstrated in Fig.~\ref{fig:rratio_vs_h}, 
the difference between all three-quark potentials
from the half of the sum of the quark-antiquark potential
can be described by a quadratic function of $h$.

\section*{Acknowledgments}
Y.K. was partially supported by the Ministry of Education, Science, 
Sports and Culture, Japan, Grant-in-Aid for Young Scientists (B) (24740176).
Our numerical simulations were performed on supercomputers NEC SX8 
at Research Center for Nuclear Physics (RCNP), 
and NEC SX-ACE at Cybermedia Center (CMC), Osaka University.
We thank V.~Dmitra\v sinovi\' c, 
N.~Brambilla,  Ph. de~Forcrand, and the anonymous
referee of Physical Review D
for useful comments on the manuscript.

\appendix
\section{The three-quark potential data}
\label{sect:alldata}

\setlength{\LTcapwidth}{\linewidth}

\begin{longtable}[c]{ccccccccc}
\hline\hline
\endlastfoot
\caption{List of the  three-quark potential (ACT)}
\label{tbl:datatable_act}\\
\hline\hline
No. & $\vec{x}_{1}$  & $\vec{x}_{2}$ & $\vec{x}_{3}$ & $r_{1}$&  $r_{2}$&  $r_{3}$&  $\theta_{\rm max}$& $V_{3q}$ \\
\hline
  1& (1,0,0) & (0,1,0) & (0,0,1) &     1.41&    1.41&    1.41&   60&      0.9300\\
  2& (2,0,0) & (0,2,0) & (0,0,2) &     2.83&    2.83&    2.83&   60&      1.1723\\
  3& (3,0,0) & (0,3,0) & (0,0,3) &     4.24&    4.24&    4.24&   60&      1.3252\\
  4& (4,0,0) & (0,4,0) & (0,0,4) &     5.66&    5.66&    5.66&   60&      1.4538\\
  5& (5,0,0) & (0,5,0) & (0,0,5) &     7.07&    7.07&    7.07&   60&      1.5737\\
  6& (6,0,0) & (0,6,0) & (0,0,6) &     8.49&    8.49&    8.49&   60&      1.6899\\
  7& (7,0,0) & (0,7,0) & (0,0,7) &     9.90&    9.90&    9.90&   60&      1.8050\\
  8& (8,0,0) & (0,8,0) & (0,0,8) &    11.31&   11.31&   11.31&   60&      1.9210\\
  9& (2,0,0) & (0,1,0) & (0,0,1) &     1.41&    2.24&    2.24&   72&      1.0347\\
 10& (3,0,0) & (0,1,0) & (0,0,1) &     1.41&    3.16&    3.16&   77&      1.1156\\
 11& (4,0,0) & (0,1,0) & (0,0,1) &     1.41&    4.12&    4.12&   80&      1.1822\\
 12& (5,0,0) & (0,1,0) & (0,0,1) &     1.41&    5.10&    5.10&   82&      1.2412\\
 13& (6,0,0) & (0,1,0) & (0,0,1) &     1.41&    6.08&    6.08&   83&      1.2959\\
 14& (7,0,0) & (0,1,0) & (0,0,1) &     1.41&    7.07&    7.07&   84&      1.3481\\
 15& (8,0,0) & (0,1,0) & (0,0,1) &     1.41&    8.06&    8.06&   85&      1.3988\\
 16& (9,0,0) & (0,1,0) & (0,0,1) &     1.41&    9.06&    9.06&   86&      1.4483\\
 17& (10,0,0) & (0,1,0) & (0,0,1) &     1.41&   10.05&   10.05&   86&      1.4972\\
 18& (11,0,0) & (0,1,0) & (0,0,1) &     1.41&   11.05&   11.05&   86&      1.5430\\
 19& (1,0,0) & (0,2,0) & (0,0,2) &     2.83&    2.24&    2.24&   78&      1.1155\\
 20& (3,0,0) & (0,2,0) & (0,0,2) &     2.83&    3.61&    3.61&   67&      1.2310\\
 21& (4,0,0) & (0,2,0) & (0,0,2) &     2.83&    4.47&    4.47&   72&      1.2873\\
 22& (5,0,0) & (0,2,0) & (0,0,2) &     2.83&    5.39&    5.39&   75&      1.3410\\
 23& (6,0,0) & (0,2,0) & (0,0,2) &     2.83&    6.32&    6.32&   77&      1.3927\\
 24& (7,0,0) & (0,2,0) & (0,0,2) &     2.83&    7.28&    7.28&   79&      1.4432\\
 25& (8,0,0) & (0,2,0) & (0,0,2) &     2.83&    8.25&    8.25&   80&      1.4926\\
 26& (9,0,0) & (0,2,0) & (0,0,2) &     2.83&    9.22&    9.22&   81&      1.5414\\
 27& (10,0,0) & (0,2,0) & (0,0,2) &     2.83&   10.20&   10.20&   82&      1.5897\\
 28& (11,0,0) & (0,2,0) & (0,0,2) &     2.83&   11.18&   11.18&   83&      1.6348\\
 29& (1,0,0) & (0,3,0) & (0,0,3) &     4.24&    3.16&    3.16&   84&      1.2476\\
 30& (2,0,0) & (0,3,0) & (0,0,3) &     4.24&    3.61&    3.61&   72&      1.2820\\
 31& (4,0,0) & (0,3,0) & (0,0,3) &     4.24&    5.00&    5.00&   65&      1.3718\\
 32& (5,0,0) & (0,3,0) & (0,0,3) &     4.24&    5.83&    5.83&   69&      1.4196\\
 33& (6,0,0) & (0,3,0) & (0,0,3) &     4.24&    6.71&    6.71&   72&      1.4676\\
 34& (7,0,0) & (0,3,0) & (0,0,3) &     4.24&    7.62&    7.62&   74&      1.5154\\
 35& (8,0,0) & (0,3,0) & (0,0,3) &     4.24&    8.54&    8.54&   76&      1.5632\\
 36& (9,0,0) & (0,3,0) & (0,0,3) &     4.24&    9.49&    9.49&   77&      1.6107\\
 37& (10,0,0) & (0,3,0) & (0,0,3) &     4.24&   10.44&   10.44&   78&      1.6579\\
 38& (11,0,0) & (0,3,0) & (0,0,3) &     4.24&   11.40&   11.40&   79&      1.7022\\
 39& (1,0,0) & (0,4,0) & (0,0,4) &     5.66&    4.12&    4.12&   87&      1.3578\\
 40& (2,0,0) & (0,4,0) & (0,0,4) &     5.66&    4.47&    4.47&   78&      1.3814\\
 41& (3,0,0) & (0,4,0) & (0,0,4) &     5.66&    5.00&    5.00&   69&      1.4148\\
 42& (5,0,0) & (0,4,0) & (0,0,4) &     5.66&    6.40&    6.40&   64&      1.4961\\
 43& (6,0,0) & (0,4,0) & (0,0,4) &     5.66&    7.21&    7.21&   67&      1.5401\\
 44& (7,0,0) & (0,4,0) & (0,0,4) &     5.66&    8.06&    8.06&   69&      1.5852\\
 45& (8,0,0) & (0,4,0) & (0,0,4) &     5.66&    8.94&    8.94&   72&      1.6307\\
 46& (9,0,0) & (0,4,0) & (0,0,4) &     5.66&    9.85&    9.85&   73&      1.6768\\
 47& (10,0,0) & (0,4,0) & (0,0,4) &     5.66&   10.77&   10.77&   75&      1.7225\\
 48& (11,0,0) & (0,4,0) & (0,0,4) &     5.66&   11.70&   11.70&   76&      1.7655\\
 49& (1,0,0) & (0,5,0) & (0,0,5) &     7.07&    5.10&    5.10&   88&      1.4579\\
 50& (2,0,0) & (0,5,0) & (0,0,5) &     7.07&    5.39&    5.39&   82&      1.4758\\
 51& (3,0,0) & (0,5,0) & (0,0,5) &     7.07&    5.83&    5.83&   75&      1.5028\\
 52& (4,0,0) & (0,5,0) & (0,0,5) &     7.07&    6.40&    6.40&   67&      1.5361\\
 53& (6,0,0) & (0,5,0) & (0,0,5) &     7.07&    7.81&    7.81&   63&      1.6140\\
 54& (7,0,0) & (0,5,0) & (0,0,5) &     7.07&    8.60&    8.60&   66&      1.6563\\
 55& (8,0,0) & (0,5,0) & (0,0,5) &     7.07&    9.43&    9.43&   68&      1.6994\\
 56& (9,0,0) & (0,5,0) & (0,0,5) &     7.07&   10.30&   10.30&   70&      1.7439\\
 57& (10,0,0) & (0,5,0) & (0,0,5) &     7.07&   11.18&   11.18&   72&      1.7877\\
 58& (11,0,0) & (0,5,0) & (0,0,5) &     7.07&   12.08&   12.08&   73&      1.8307\\
 59& (1,0,0) & (0,6,0) & (0,0,6) &     8.49&    6.08&    6.08&   88&      1.5530\\
 60& (2,0,0) & (0,6,0) & (0,0,6) &     8.49&    6.32&    6.32&   84&      1.5676\\
 61& (3,0,0) & (0,6,0) & (0,0,6) &     8.49&    6.71&    6.71&   78&      1.5902\\
 62& (4,0,0) & (0,6,0) & (0,0,6) &     8.49&    7.21&    7.21&   72&      1.6191\\
 63& (5,0,0) & (0,6,0) & (0,0,6) &     8.49&    7.81&    7.81&   66&      1.6529\\
 64& (7,0,0) & (0,6,0) & (0,0,6) &     8.49&    9.22&    9.22&   63&      1.7295\\
 65& (8,0,0) & (0,6,0) & (0,0,6) &     8.49&   10.00&   10.00&   65&      1.7699\\
 66& (9,0,0) & (0,6,0) & (0,0,6) &     8.49&   10.82&   10.82&   67&      1.8108\\
 67& (10,0,0) & (0,6,0) & (0,0,6) &     8.49&   11.66&   11.66&   69&      1.8556\\
 68& (11,0,0) & (0,6,0) & (0,0,6) &     8.49&   12.53&   12.53&   70&      1.9051\\
\end{longtable}

\begin{longtable}[c]{ccccccccc}
\hline\hline
\endlastfoot
\caption{List of the  three-quark potential (RGT)}
\label{tbl:datatable_rgt}\\
\hline\hline
No. & $\vec{x}_{1}$  & $\vec{x}_{2}$ & $\vec{x}_{3}$ & $r_{1}$&  $r_{2}$&  $r_{3}$&  $\theta_{\rm max}$& $V_{3q}$ \\
\hline
  1& (1,0,0) & (0,1,0) & (0,0,0) &     1.00&    1.00&    1.41&   90&      0.8139\\
  2& (2,0,0) & (0,1,0) & (0,0,0) &     1.00&    2.00&    2.24&   90&      0.9588\\
  3& (3,0,0) & (0,1,0) & (0,0,0) &     1.00&    3.00&    3.16&   90&      1.0500\\
  4& (4,0,0) & (0,1,0) & (0,0,0) &     1.00&    4.00&    4.12&   90&      1.1197\\
  5& (5,0,0) & (0,1,0) & (0,0,0) &     1.00&    5.00&    5.10&   90&      1.1799\\
  6& (6,0,0) & (0,1,0) & (0,0,0) &     1.00&    6.00&    6.08&   90&      1.2352\\
  7& (2,0,0) & (0,2,0) & (0,0,0) &     2.00&    2.00&    2.83&   90&      1.0796\\
  8& (3,0,0) & (0,2,0) & (0,0,0) &     2.00&    3.00&    3.61&   90&      1.1597\\
  9& (4,0,0) & (0,2,0) & (0,0,0) &     2.00&    4.00&    4.47&   90&      1.2244\\
 10& (5,0,0) & (0,2,0) & (0,0,0) &     2.00&    5.00&    5.39&   90&      1.2820\\
 11& (7,0,0) & (0,1,0) & (0,0,0) &     1.00&    7.00&    7.07&   90&      1.2877\\
 12& (8,0,0) & (0,1,0) & (0,0,0) &     1.00&    8.00&    8.06&   90&      1.3386\\
 13& (9,0,0) & (0,1,0) & (0,0,0) &     1.00&    9.00&    9.06&   90&      1.3884\\
 14& (10,0,0) & (0,1,0) & (0,0,0) &     1.00&   10.00&   10.05&   90&      1.4373\\
 15& (11,0,0) & (0,1,0) & (0,0,0) &     1.00&   11.00&   11.05&   90&      1.4831\\
 16& (6,0,0) & (0,2,0) & (0,0,0) &     2.00&    6.00&    6.32&   90&      1.3359\\
 17& (7,0,0) & (0,2,0) & (0,0,0) &     2.00&    7.00&    7.28&   90&      1.3875\\
 18& (8,0,0) & (0,2,0) & (0,0,0) &     2.00&    8.00&    8.25&   90&      1.4377\\
 19& (9,0,0) & (0,2,0) & (0,0,0) &     2.00&    9.00&    9.22&   90&      1.4872\\
 20& (10,0,0) & (0,2,0) & (0,0,0) &     2.00&   10.00&   10.20&   90&      1.5358\\
 21& (11,0,0) & (0,2,0) & (0,0,0) &     2.00&   11.00&   11.18&   90&      1.5813\\
 22& (3,0,0) & (0,3,0) & (0,0,0) &     3.00&    3.00&    4.24&   90&      1.2322\\
 23& (4,0,0) & (0,3,0) & (0,0,0) &     3.00&    4.00&    5.00&   90&      1.2922\\
 24& (5,0,0) & (0,3,0) & (0,0,0) &     3.00&    5.00&    5.83&   90&      1.3470\\
 25& (6,0,0) & (0,3,0) & (0,0,0) &     3.00&    6.00&    6.71&   90&      1.3990\\
 26& (7,0,0) & (0,3,0) & (0,0,0) &     3.00&    7.00&    7.62&   90&      1.4495\\
 27& (8,0,0) & (0,3,0) & (0,0,0) &     3.00&    8.00&    8.54&   90&      1.4990\\
 28& (9,0,0) & (0,3,0) & (0,0,0) &     3.00&    9.00&    9.49&   90&      1.5478\\
 29& (10,0,0) & (0,3,0) & (0,0,0) &     3.00&   10.00&   10.44&   90&      1.5960\\
 30& (11,0,0) & (0,3,0) & (0,0,0) &     3.00&   11.00&   11.40&   90&      1.6412\\
 31& (4,0,0) & (0,4,0) & (0,0,0) &     4.00&    4.00&    5.66&   90&      1.3487\\
 32& (5,0,0) & (0,4,0) & (0,0,0) &     4.00&    5.00&    6.40&   90&      1.4010\\
 33& (6,0,0) & (0,4,0) & (0,0,0) &     4.00&    6.00&    7.21&   90&      1.4513\\
 34& (7,0,0) & (0,4,0) & (0,0,0) &     4.00&    7.00&    8.06&   90&      1.5005\\
 35& (8,0,0) & (0,4,0) & (0,0,0) &     4.00&    8.00&    8.94&   90&      1.5489\\
 36& (9,0,0) & (0,4,0) & (0,0,0) &     4.00&    9.00&    9.85&   90&      1.5972\\
 37& (10,0,0) & (0,4,0) & (0,0,0) &     4.00&   10.00&   10.77&   90&      1.6451\\
 38& (11,0,0) & (0,4,0) & (0,0,0) &     4.00&   11.00&   11.70&   90&      1.6896\\
 39& (5,0,0) & (0,5,0) & (0,0,0) &     5.00&    5.00&    7.07&   90&      1.4513\\
 40& (6,0,0) & (0,5,0) & (0,0,0) &     5.00&    6.00&    7.81&   90&      1.5001\\
 41& (7,0,0) & (0,5,0) & (0,0,0) &     5.00&    7.00&    8.60&   90&      1.5481\\
 42& (8,0,0) & (0,5,0) & (0,0,0) &     5.00&    8.00&    9.43&   90&      1.5958\\
 43& (9,0,0) & (0,5,0) & (0,0,0) &     5.00&    9.00&   10.30&   90&      1.6433\\
\end{longtable}

\begin{longtable}[c]{ccccccccc}
\hline\hline
\endlastfoot
\caption{List of the  three-quark potential (OBTN)}
\label{tbl:datatable_obn}\\
\hline\hline
No. & $\vec{x}_{1}$  & $\vec{x}_{2}$ & $\vec{x}_{3}$ & $r_{1}$&  $r_{2}$&  $r_{3}$&  $\theta_{\rm max}$& $V_{3q}$ \\
\hline
  1& (2,0,0) & (1,0,0) & (0,0,2) &     2.24&    2.83&    1.00&  117&      1.0049\\
  2& (3,0,0) & (1,0,0) & (0,0,2) &     2.24&    3.61&    2.00&  117&      1.1264\\
  3& (4,0,0) & (1,0,0) & (0,0,2) &     2.24&    4.47&    3.00&  117&      1.2051\\
  4& (5,0,0) & (1,0,0) & (0,0,2) &     2.24&    5.39&    4.00&  117&      1.2684\\
  5& (2,0,0) & (1,0,0) & (0,0,3) &     3.16&    3.61&    1.00&  108&      1.0742\\
  6& (3,0,0) & (1,0,0) & (0,0,3) &     3.16&    4.24&    2.00&  108&      1.1881\\
  7& (4,0,0) & (1,0,0) & (0,0,3) &     3.16&    5.00&    3.00&  108&      1.2621\\
  8& (5,0,0) & (1,0,0) & (0,0,3) &     3.16&    5.83&    4.00&  108&      1.3225\\
  9& (2,0,0) & (1,0,0) & (0,0,4) &     4.12&    4.47&    1.00&  104&      1.1352\\
 10& (3,0,0) & (1,0,0) & (0,0,4) &     4.12&    5.00&    2.00&  104&      1.2443\\
 11& (3,0,0) & (2,0,0) & (0,0,4) &     4.47&    5.00&    1.00&  117&      1.1621\\
 12& (4,0,0) & (1,0,0) & (0,0,4) &     4.12&    5.66&    3.00&  104&      1.3147\\
 13& (4,0,0) & (2,0,0) & (0,0,4) &     4.47&    5.66&    2.00&  117&      1.2738\\
 14& (5,0,0) & (1,0,0) & (0,0,4) &     4.12&    6.40&    4.00&  104&      1.3726\\
 15& (5,0,0) & (2,0,0) & (0,0,4) &     4.47&    6.40&    3.00&  117&      1.3457\\
 16& (2,0,0) & (1,0,0) & (0,0,5) &     5.10&    5.39&    1.00&  101&      1.1910\\
 17& (3,0,0) & (1,0,0) & (0,0,5) &     5.10&    5.83&    2.00&  101&      1.2971\\
 18& (3,0,0) & (2,0,0) & (0,0,5) &     5.39&    5.83&    1.00&  112&      1.2117\\
 19& (4,0,0) & (1,0,0) & (0,0,5) &     5.10&    6.40&    3.00&  101&      1.3649\\
 20& (4,0,0) & (2,0,0) & (0,0,5) &     5.39&    6.40&    2.00&  112&      1.3206\\
 21& (5,0,0) & (1,0,0) & (0,0,5) &     5.10&    7.07&    4.00&  101&      1.4207\\
 22& (5,0,0) & (2,0,0) & (0,0,5) &     5.39&    7.07&    3.00&  112&      1.3904\\
 23& (2,0,0) & (1,0,0) & (0,0,6) &     6.08&    6.32&    1.00&   99&      1.2438\\
 24& (3,0,0) & (1,0,0) & (0,0,6) &     6.08&    6.71&    2.00&   99&      1.3479\\
 25& (3,0,0) & (2,0,0) & (0,0,6) &     6.32&    6.71&    1.00&  108&      1.2604\\
 26& (4,0,0) & (1,0,0) & (0,0,6) &     6.08&    7.21&    3.00&   99&      1.4139\\
 27& (4,0,0) & (2,0,0) & (0,0,6) &     6.32&    7.21&    2.00&  108&      1.3674\\
\end{longtable}

\begin{longtable}[c]{ccccccccc}
\hline\hline
\endlastfoot
\caption{List of the  three-quark potential (OBTW)}
\label{tbl:datatable_obw}\\
\hline\hline
No. & $\vec{x}_{1}$  & $\vec{x}_{2}$ & $\vec{x}_{3}$ & $r_{1}$&  $r_{2}$&  $r_{3}$&  $\theta_{\rm max}$& $V_{3q}$ \\
\hline
  1& (2,0,0) & (1,0,0) & (0,0,1) &     1.41&    2.24&    1.00&  135&      0.9243\\
  2& (3,0,0) & (1,0,0) & (0,0,1) &     1.41&    3.16&    2.00&  135&      1.0570\\
  3& (3,0,0) & (2,0,0) & (0,0,1) &     2.24&    3.16&    1.00&  153&      1.0169\\
  4& (4,0,0) & (1,0,0) & (0,0,1) &     1.41&    4.12&    3.00&  135&      1.1410\\
  5& (4,0,0) & (2,0,0) & (0,0,1) &     2.24&    4.12&    2.00&  153&      1.1425\\
  6& (4,0,0) & (3,0,0) & (0,0,1) &     3.16&    4.12&    1.00&  162&      1.0904\\
  7& (5,0,0) & (1,0,0) & (0,0,1) &     1.41&    5.10&    4.00&  135&      1.2068\\
  8& (5,0,0) & (2,0,0) & (0,0,1) &     2.24&    5.10&    3.00&  153&      1.2226\\
  9& (5,0,0) & (3,0,0) & (0,0,1) &     3.16&    5.10&    2.00&  162&      1.2122\\
 10& (5,0,0) & (4,0,0) & (0,0,1) &     4.12&    5.10&    1.00&  166&      1.1529\\
 11& (3,0,0) & (2,0,0) & (0,0,2) &     2.83&    3.61&    1.00&  135&      1.0618\\
 12& (4,0,0) & (2,0,0) & (0,0,2) &     2.83&    4.47&    2.00&  135&      1.1821\\
 13& (4,0,0) & (3,0,0) & (0,0,2) &     3.61&    4.47&    1.00&  146&      1.1184\\
 14& (5,0,0) & (2,0,0) & (0,0,2) &     2.83&    5.39&    3.00&  135&      1.2596\\
 15& (5,0,0) & (3,0,0) & (0,0,2) &     3.61&    5.39&    2.00&  146&      1.2374\\
 16& (5,0,0) & (4,0,0) & (0,0,2) &     4.47&    5.39&    1.00&  153&      1.1725\\
 17& (3,0,0) & (2,0,0) & (0,0,3) &     3.61&    4.24&    1.00&  124&      1.1117\\
 18& (4,0,0) & (2,0,0) & (0,0,3) &     3.61&    5.00&    2.00&  124&      1.2272\\
 19& (4,0,0) & (3,0,0) & (0,0,3) &     4.24&    5.00&    1.00&  135&      1.1551\\
 20& (5,0,0) & (2,0,0) & (0,0,3) &     3.61&    5.83&    3.00&  124&      1.3017\\
 21& (5,0,0) & (3,0,0) & (0,0,3) &     4.24&    5.83&    2.00&  135&      1.2711\\
 22& (5,0,0) & (4,0,0) & (0,0,3) &     5.00&    5.83&    1.00&  143&      1.2007\\
 23& (4,0,0) & (3,0,0) & (0,0,4) &     5.00&    5.66&    1.00&  127&      1.1962\\
 24& (5,0,0) & (3,0,0) & (0,0,4) &     5.00&    6.40&    2.00&  127&      1.3094\\
 25& (5,0,0) & (4,0,0) & (0,0,4) &     5.66&    6.40&    1.00&  135&      1.2346\\
 26& (4,0,0) & (3,0,0) & (0,0,5) &     5.83&    6.40&    1.00&  121&      1.2394\\
 27& (5,0,0) & (3,0,0) & (0,0,5) &     5.83&    7.07&    2.00&  121&      1.3503\\
 28& (5,0,0) & (4,0,0) & (0,0,5) &     6.40&    7.07&    1.00&  129&      1.2721\\
\end{longtable}

\begin{longtable}[c]{ccccccccc}
\hline\hline
\endlastfoot
\caption{List of the  three-quark potential (LIN)}
\label{tbl:datatable_lin}\\
\hline\hline
No. & $\vec{x}_{1}$  & $\vec{x}_{2}$ & $\vec{x}_{3}$ & $r_{1}$&  $r_{2}$&  $r_{3}$&  $\theta_{\rm max}$& $V_{3q}$ \\
\hline
  1& (2,0,0) & (1,0,0) & (0,0,0) &     1.00&    2.00&    1.00&  180&      0.8478\\
  2& (3,0,0) & (1,0,0) & (0,0,0) &     1.00&    3.00&    2.00&  180&      0.9914\\
  3& (4,0,0) & (1,0,0) & (0,0,0) &     1.00&    4.00&    3.00&  180&      1.0786\\
  4& (4,0,0) & (2,0,0) & (0,0,0) &     2.00&    4.00&    2.00&  180&      1.1203\\
  5& (5,0,0) & (1,0,0) & (0,0,0) &     1.00&    5.00&    4.00&  180&      1.1457\\
  6& (5,0,0) & (2,0,0) & (0,0,0) &     2.00&    5.00&    3.00&  180&      1.2017\\
  7& (6,0,0) & (1,0,0) & (0,0,0) &     1.00&    6.00&    5.00&  180&      1.2044\\
  8& (6,0,0) & (2,0,0) & (0,0,0) &     2.00&    6.00&    4.00&  180&      1.2661\\
  9& (6,0,0) & (3,0,0) & (0,0,0) &     3.00&    6.00&    3.00&  180&      1.2803\\
 10& (7,0,0) & (1,0,0) & (0,0,0) &     1.00&    7.00&    6.00&  180&      1.2588\\
 11& (7,0,0) & (2,0,0) & (0,0,0) &     2.00&    7.00&    5.00&  180&      1.3233\\
 12& (7,0,0) & (3,0,0) & (0,0,0) &     3.00&    7.00&    4.00&  180&      1.3434\\
 13& (8,0,0) & (1,0,0) & (0,0,0) &     1.00&    8.00&    7.00&  180&      1.3108\\
 14& (8,0,0) & (2,0,0) & (0,0,0) &     2.00&    8.00&    6.00&  180&      1.3769\\
 15& (8,0,0) & (3,0,0) & (0,0,0) &     3.00&    8.00&    5.00&  180&      1.3998\\
 16& (8,0,0) & (4,0,0) & (0,0,0) &     4.00&    8.00&    4.00&  180&      1.4055\\
 17& (9,0,0) & (1,0,0) & (0,0,0) &     1.00&    9.00&    8.00&  180&      1.3612\\
 18& (9,0,0) & (2,0,0) & (0,0,0) &     2.00&    9.00&    7.00&  180&      1.4283\\
 19& (9,0,0) & (3,0,0) & (0,0,0) &     3.00&    9.00&    6.00&  180&      1.4527\\
 20& (9,0,0) & (4,0,0) & (0,0,0) &     4.00&    9.00&    5.00&  180&      1.4614\\
 21& (10,0,0) & (1,0,0) & (0,0,0) &     1.00&   10.00&    9.00&  180&      1.4108\\
 22& (10,0,0) & (2,0,0) & (0,0,0) &     2.00&   10.00&    8.00&  180&      1.4786\\
 23& (10,0,0) & (3,0,0) & (0,0,0) &     3.00&   10.00&    7.00&  180&      1.5040\\
 24& (10,0,0) & (4,0,0) & (0,0,0) &     4.00&   10.00&    6.00&  180&      1.5142\\
 25& (10,0,0) & (5,0,0) & (0,0,0) &     5.00&   10.00&    5.00&  180&      1.5171\\
 26& (11,0,0) & (1,0,0) & (0,0,0) &     1.00&   11.00&   10.00&  180&      1.4591\\
 27& (11,0,0) & (2,0,0) & (0,0,0) &     2.00&   11.00&    9.00&  180&      1.5279\\
 28& (11,0,0) & (3,0,0) & (0,0,0) &     3.00&   11.00&    8.00&  180&      1.5540\\
 29& (11,0,0) & (4,0,0) & (0,0,0) &     4.00&   11.00&    7.00&  180&      1.5654\\
 30& (11,0,0) & (5,0,0) & (0,0,0) &     5.00&   11.00&    6.00&  180&      1.5697\\
 31& (12,0,0) & (1,0,0) & (0,0,0) &     1.00&   12.00&   11.00&  180&      1.4987\\
 32& (12,0,0) & (2,0,0) & (0,0,0) &     2.00&   12.00&   10.00&  180&      1.5740\\
\end{longtable}

\begin{longtable}[c]{ccccccccc}
\hline\hline
\endlastfoot
\caption{List of the  three-quark potential (QDQ)}
\label{tbl:datatable_qdq}\\
\hline\hline
No. & $\vec{x}_{1}$  & $\vec{x}_{2}$ & $\vec{x}_{3}$ & $r_{1}$&  $r_{2}$&  $r_{3}$&  $\theta_{\rm max}$& $V_{3q}$ \\
\hline
  1& (1,0,0) & (0,0,0) & (0,0,0) &     0.00&    1.00&    1.00& --- &      0.5480\\
  2& (2,0,0) & (0,0,0) & (0,0,0) &     0.00&    2.00&    2.00& --- &      0.7331\\
  3& (3,0,0) & (0,0,0) & (0,0,0) &     0.00&    3.00&    3.00& --- &      0.8347\\
  4& (4,0,0) & (0,0,0) & (0,0,0) &     0.00&    4.00&    4.00& --- &      0.9075\\
  5& (5,0,0) & (0,0,0) & (0,0,0) &     0.00&    5.00&    5.00& --- &      0.9689\\
  6& (6,0,0) & (0,0,0) & (0,0,0) &     0.00&    6.00&    6.00& --- &      1.0248\\
  7& (7,0,0) & (0,0,0) & (0,0,0) &     0.00&    7.00&    7.00& --- &      1.0777\\
  8& (8,0,0) & (0,0,0) & (0,0,0) &     0.00&    8.00&    8.00& --- &      1.1287\\
  9& (9,0,0) & (0,0,0) & (0,0,0) &     0.00&    9.00&    9.00& --- &      1.1786\\
 10& (10,0,0) & (0,0,0) & (0,0,0) &     0.00&   10.00&   10.00& --- &      1.2277\\
 11& (11,0,0) & (0,0,0) & (0,0,0) &     0.00&   11.00&   11.00& --- &      1.2736\\
 12& (12,0,0) & (0,0,0) & (0,0,0) &     0.00&   12.00&   12.00& --- &      1.2979\\
 13& (1,0,0) & (1,0,0) & (0,0,1) &     1.41&    1.41&    0.00& --- &      0.6650\\
 14& (2,0,0) & (2,0,0) & (0,0,1) &     2.24&    2.24&    0.00& --- &      0.7693\\
 15& (3,0,0) & (3,0,0) & (0,0,1) &     3.16&    3.16&    0.00& --- &      0.8498\\
 16& (4,0,0) & (4,0,0) & (0,0,1) &     4.12&    4.12&    0.00& --- &      0.9161\\
 17& (5,0,0) & (5,0,0) & (0,0,1) &     5.10&    5.10&    0.00& --- &      0.9748\\
 18& (1,0,0) & (1,0,0) & (0,0,2) &     2.24&    2.24&    0.00& --- &      0.7693\\
 19& (2,0,0) & (2,0,0) & (0,0,2) &     2.83&    2.83&    0.00& --- &      0.8255\\
 20& (3,0,0) & (3,0,0) & (0,0,2) &     3.61&    3.61&    0.00& --- &      0.8832\\
 21& (4,0,0) & (4,0,0) & (0,0,2) &     4.47&    4.47&    0.00& --- &      0.9385\\
 22& (5,0,0) & (5,0,0) & (0,0,2) &     5.39&    5.39&    0.00& --- &      0.9914\\
 23& (1,0,0) & (1,0,0) & (0,0,3) &     3.16&    3.16&    0.00& --- &      0.8498\\
\end{longtable}

\newpage

\begin{minipage}{\linewidth}
\begin{longtable}[c]{ccclll}
\hline\hline
\endlastfoot
\caption{The three-quark potentials of the equilateral geometries at
$\beta=5.85$ ($N_{\rm iupd}=500000$ and $N_{\rm cnf}=8$), 
$6.00$ ($N_{\rm iupd}=500000$ and $N_{\rm cnf}=9$), and
 $6.30$ ($N_{\rm iupd}=400000$ and $N_{\rm cnf}=6$), where the three
quarks are placed at $(x,0,0)$, $(0,x,0)$ and $(0,0,x)$.
The distances $Y=\sqrt{6}x$ and $R=\sqrt{3}Y/9=\sqrt{2}x/3$ are used
when we plot the potential data.
The corresponding plot is shown in Fig.~\ref{fig:pot3q_scale}.}
\label{tbl:pot3q_regular_scale}\\
\hline\hline
 $x/a$ & $Y/a$ &$R/a$& $\beta=5.85$ & $\beta=6.00$ & $\beta=6.30$ \\
\hline
1&  2.449 &  0.471 & 1.01563(39)  &  0.93078(23)&  0.82543(12)  \\
2&   4.899 & 0.943 &  1.3374(11)  &  1.17491(76)&  0.99852(53)  \\
3&    7.348 &  1.414 &  1.5703(21) &   1.3300(15)  &   1.0886(12)  \\
4&    9.798 &  1.886 & 1.7813(34) &   1.4609(23)  &   1.1540(22)  \\ 
5&    12.247 &  2.357 &  1.9855(53) &   1.5830(33)&   1.2083(36)  \\ 
6&    14.697 & 2.828  & 2.1871(78) &   1.7013(42)&   1.2569(48)  \\ 
7&    17.146 & 3.300  &  2.385(10)  &   1.8187(52) &   1.3026(66)  \\ 
8&   19.596 &  3.771  &                       &    1.930(11)  &   1.3364(70)  \\ 
9&   22.045 &   4.243  &                      &    1.997(27)  &    1.383(18)  \\ 
10 &   24.495 & 4.714  &  &   &     1.440(23)  \\ 
11&   26.944 &  5.185  & &   &     1.442(48)  \\ 
\end{longtable}
\end{minipage}

\begin{minipage}{\linewidth}
\begin{longtable}[c]{cccccc}
\hline\hline
\endlastfoot
\caption{Fit results of the three-quark potential
in Table~\ref{tbl:pot3q_regular_scale}  to 
$V_{3q}^{\rm (Y)}=-A_{3q}^{\rm (Y)}/Y+\sigma_{3q} Y + \mu_{3q}$.}
\label{tbl:fit-pot3q-regular}\\
\hline\hline
$\beta$ & $A_{3q}^{\rm (Y)}$  & $\sigma_{3q}\,a^{2}$   & $\mu_{3q}a$  
& {\footnotesize fit range} $x/a$ ($Y/a$)   &  $\chi^{2}/N_{\rm df}$ \\
\hline
5.85  &  0.64(1)\F &     0.0776(5)   &  1.089(5) &  1 -- 7 (2.45 -- 17.1) &   0.15 \\
6.00 &   0.662(6)  &    0.0446(3)  &   1.092(3)  & 1 -- 8  (2.45 -- 19.6) &   0.09  \\
6.30  &   0.635(6)  &    0.0180(3) &      1.041(3) &   1 -- 7  (2.45 -- 17.1) &  2.1 \\
\end{longtable}
\end{minipage}

\section{The quark-antiquark potential data}
\label{sect:qqbarpot}

\begin{minipage}[h]{\linewidth}
\begin{longtable}[c]{clll}
\hline\hline
\endlastfoot
\caption{The quark-antiquark potentials at 
$\beta=5.85$ ($N_{\rm iupd}=50000$ and $N_{\rm cnf}=133$), 
$6.00$ ($N_{\rm iupd}=100000$ and $N_{\rm cnf}=20$), and 
$6.30$ ($N_{\rm iupd}=6000$ and $N_{\rm cnf}=40$),
where the quark and antiquark are separated only along  the on-axis.
For other simulation parameters, see Table~\ref{tbl:simulation}.
The data at  $\beta=5.85$ and $6.30$ were computed when we studied the 
relativistic corrections to the quark-antiquark potential~\cite{Koma:2006fw,Koma:2009ws}
and for this reason some parts remain blank, which however 
are harmless in the present analysis.
The corresponding plot is shown in Fig.~\ref{fig:potqqbar_scale}.}
\label{tbl:potqqbar_scale}\\
\hline\hline
 $r/a$ & $\beta=5.85$  & $\beta=6.00$ & $\beta=6.30$\\
 \hline
   1 &                           & 0.501511(36) &  \\ 
   2 & 0.762272(85)&  0.68631(12) &0.600853(55)  \\ 
   3 &  0.90051(16) &  0.78796(24)  &0.67114(11)  \\ 
   4 &  1.00877(25) &  0.86105(39) & 0.71569(19)  \\ 
   5 &  1.10516(35) &  0.92273(56)  &0.74970(30)  \\ 
   6 &  1.19592(46) &  0.97891(75) & 0.77837(41)  \\ 
   7 &  1.28359(58) &  1.03203(97) & 0.80403(56)  \\ 
   8&  1.36941(71) &   1.0833(12)  & 0.82733(74)  \\ 
   9 &  1.45401(84) &   1.1333(14)   &0.84850(96)  \\ 
  10 &  1.53776(99) &   1.1820(17) & 0.8667(12)  \\ 
  11 &                           &   1.2269(21) & \\ 
  12&                           &   1.2499(24) & \\ 
\end{longtable}
\end{minipage}

\begin{minipage}[h]{\linewidth}
\begin{longtable}[c]{cccccc}
\hline\hline
\endlastfoot
\caption{Fit results of the quark-antiquark potential
in Table~\ref{tbl:potqqbar_scale}  to
$V_{\qqb}(r) =-A_{\qqb}/r +\sigma_{\qqb}\,  r + \mu_{\qqb}$.}
\label{tbl:fit-potqqbar}\\
\hline\hline
$\beta$   &    $A_{\qqb}$  & $\sigma_{\qqb}\, a^{2}$  
 & $\mu_{\qqb}a$    &  {\footnotesize fit range}  $r/a$  & $\chi^{2}/N_{\rm df}$ \\
\hline
5.85    &  0.354(2)   & 0.0790(1) &  0.781(1) &  3 -- 10  &1.1  \\
6.00  &  0.340(2)   & 0.0449(2) &   0.766(1) &  2 -- 10 &0.13 \\
6.30 &  0.313(2)   & 0.0182(1) &  0.721(1)  &  3 -- 10 & 1.2    \\
\end{longtable}
\end{minipage}

\begin{minipage}[h]{\linewidth}
\begin{longtable}[c]{cllll}
\hline\hline
\endlastfoot
\caption{The quark-antiquark potential at $\beta=6.00$
 ($N_{\rm iupd}=100000$ and $N_{\rm cnf}=20$), where 
the quark and antiquark are separated along  the off-axis.
For other simulation parameters, see Table~\ref{tbl:simulation}.
The Euclidean distance for the label $(i,j,k)$ is 
$r/a=\sqrt{i^{2}+j^{2}+k^{2}}$.
These data are used when comparing the three-quark potential with 
the quark-antiquark potential at $\beta=6.00$.
All data including the on-axis data at $\beta=6.00$  in Table~\ref{tbl:potqqbar_scale} are plotted in 
Fig.~\ref{fig:qqbar-pot-reference}.}
\label{tbl:potqqbar-b600}\\
\hline\hline
 $i$  & $(i,i,0)$ & $(i,i,i)$& $(2i,i,0)$\\
 \hline
   1  & 0.618206(71)&  0.67289(10) &  0.72246(15)  \\ 
   2  &  0.77882(23)   &  0.82851(33) &0.89243(48)  \\ 
   3  &  0.87886(45)   &  0.93620(65) &1.01763(94)  \\ 
   4  &  0.96147(74)   &   1.0301(10) &  1.1316(15)  \\ 
   5  &   1.0370(11)    &   1.1185(15)   &   1.2404(23)  \\ 
   6 &   1.1091(15)     &   1.2040(20)  &   1.3173(30)  \\ 
   7 &   1.1792(20)     &   1.2879(27)  & \\ 
   8 &   1.2480(26)    &   1.3705(36)  & \\ 
   9  &   1.3155(33)  &   1.4518(55)  & \\ 
  10  &   1.3797(45)  &    1.521(10)  & \\ 
  11  &   1.4311(70)  &    1.565(27)   &\\ 
  12  &   1.4500(89)  &    1.567(31)   &\\ 
\end{longtable}
\end{minipage}

\begin{figure}[h]
\includegraphics[width=\figwidth]{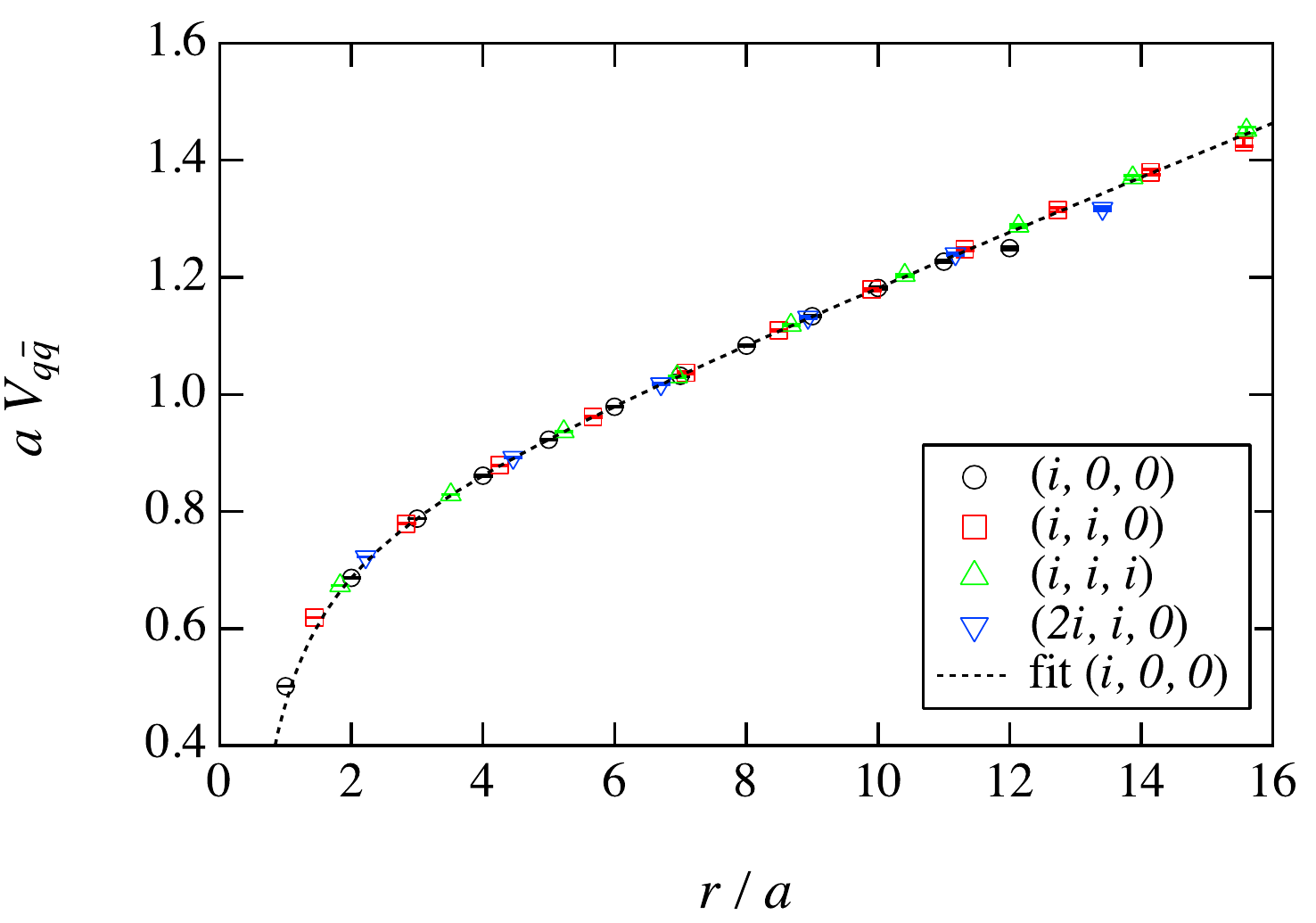}
\caption{The quark-antiquark potential at $\beta=6.00$ 
as a function of the interquark distance $r/a$.
The raw data and the fit results are summarized 
in Tables~\ref{tbl:potqqbar_scale},  \ref{tbl:fit-potqqbar},  and \ref{tbl:potqqbar-b600}.}
\label{fig:qqbar-pot-reference}
\end{figure}

\newpage


\end{document}